\documentclass[%
 reprint,
 amsmath,amssymb,
 aps,
]{revtex4-1}

\usepackage{color}

\usepackage{hyperref}
\usepackage{graphicx}
\usepackage{dcolumn}
\usepackage{bm}
\usepackage{color}

\begin{document}

\preprint{APS/123-QED}

\title{Spatial Embedding Imposes Constraints on the Network Architectures \\of Neural Systems}

\author{Jennifer Stiso$^{1,2}$}
\author{Danielle S. Bassett$^{1,3,4,5,6}$}

\affiliation{
 $^1$Department of Bioengineering, School of Engineering and Applied Sciences, University of Pennsylvania, Philadelphia, PA, 19104
}
\affiliation{
$^5$Department of Physics \& Astronomy, College of Arts and Sciences, University of Pennsylvania, Philadelphia, PA, 19104
}
\affiliation{
 $^3$Department of Electrical and Systems Engineering, School of Engineering and Applied Sciences, University of Pennsylvania, Philadelphia, PA, 19104
}
\affiliation{
 $^4$Department of Neurology, Perelman School of Medicine, University of Pennsylvania, Philadelphia, PA, 19104
}
\affiliation{
 $^2$Neuroscience Graduate Group, Perelman School of Medicine, University of Pennsylvania, Philadelphia, PA, 19104
}
\affiliation{
$^6$To whom correspondence should be addressed: dsb@seas.upenn.edu
}

\begin{abstract}
	A fundamental understanding of the network architecture of the brain is necessary for the further development of theories explicating circuit function. Recent progress has capitalized on quantitative tools from network science to parsimoniously describe and predict neural activity and connectivity across multiple spatial and temporal scales. Perhaps as a historical artifact of its origins in mathematics or perhaps as a derivative of its initial application to abstract informational systems, network science provides many methods and summary statistics that address the network’s topological characteristics with little or no thought to its physical instantiation. Yet, for embedded systems, physical laws can directly constrain processes of network growth, development, and function, and an appreciation of those physical laws is therefore critical to an understanding of the system. Recent evidence demonstrates that the constraints imposed by the physical shape and volume of the brain, and by the mechanical forces at play in its development, have marked effects on the observed network topology and function. Here, we review the rules imposed by space on the development of neural networks and show that these rules give rise to a specific set of complex topologies. We present evidence that these fundamental wiring rules affect the repertoire of neural dynamics that can emerge from the system, and thereby inform our understanding of network dysfunction in disease. We also discuss several computational tools, mathematical models, and algorithms that have proven useful in delineating the effects of spatial embedding on a given networked system and are important considerations for addressing future problems in network neuroscience. Finally, we outline several open questions regarding the network architectures that support circuit function, the answers to which will require a thorough and honest appraisal of the role of physical space in brain network anatomy and physiology.
\end{abstract}

\maketitle

\section*{Network topology versus geometry in neural systems}

In contemporary neuroscience, increasing volumes of data are being brought to bear on the question of how heterogeneous and distributed interactions between neural units might give rise to complex behaviors. Such interactions form characteristic patterns across multiple spatial scales, spanning from the relatively small scales of molecules and cells, to the relatively large scales of areas and lobes \cite{Bassett2017}. A natural language in which to describe such interactions is network science, which elegantly represents interconnected systems as sets of \emph{nodes} linked by \emph{edges} \cite{bassett2018nature}. Intuitively, nodes often represent proteins, neurons, subcortical nuclei, or large cortical areas, and edges often represent either (i) structural links in the form of chemical bonds, synapses, or white-matter tracts, or (ii) functional links in the form of statistical relations between nodal activity time series. Generally, the resultant network architecture can be fruitfully studied using tools from graph theory to obtain mechanistic insights pertinent to cognition \cite{medaglia2015cognitive,petersen2015brain}, above and beyond those provided by studies of regional activation (\textbf{Box 1}). 

In particular, several fundamental questions in neuroscience are quintessentially network questions concerning the physical relationships between functional units. How does the physical structure of a circuit affect its function? How does coordinated activity at small spatial scales give rise to emergent phenomena at large spatial scales? How might alterations in neurodevelopmental processes lead to circuit malfunction in psychiatric disorders? How might pathology spread through cortical and subcortical tissue giving rise to the well-known clinical presentations of neurological disease? These questions collectively highlight the fact that the brain --- and its multiple networks of interacting units --- is physically embedded into a fixed three-dimensional enclosure. Natural consequences of this embedding include diverse physical drivers of early connection formation and physical constraints on the resultant adult network architecture. An understanding of the system's constitution and basal dynamics therefore require not only approaches to quantify and predict network topology, but also tools, theories, and methods to quantify and predict network geometry and its role in both enabling and constraining system function.  

In this review, we provide evidence to support the notion that a consideration of the brain's physical embedding will prove critical for a holistic understanding of neural circuit function. We focus our comments on the utility of informing this consideration with emerging computational tools developed for the characterization of spatial networks. Indeed, while network science was originally developed in the context of systems devoid of clear spatial characteristics \cite{ducruet2014spatial}, the field has steadily developed tools and intuitions for spatially embedded network systems \cite{Barthelemy2011}. In the light of these recent technical developments, we begin by recounting observations from empirical studies addressing the question of how brain networks are embedded into physical space. Next, we discuss the relevance of this spatial embedding for an understanding of network function and dysfunction. We complement these empirical discussions with a more technical exposition on the relevant tools, methods, and statistical approaches to be considered when analyzing brain networks. Lastly, we close by outlining a few important future directions in methodological development and neuroscientific investigation that would benefit from an honest appraisal of the role of space in neural network architecture and dynamics.

\section*{Physical constraints on network topology and geometry}

The processes and influences that guide the formation of structural connections in neural systems are disparate and varied \cite{Bullmore2012,Chen2017}. Evidence from genetics suggests that neurons with similar functions as operationalized by similar gene expression tend to have more similar connection profiles than neurons with less similar functions \cite{Bullmore2012,French2011,Rubinov}. Of course, it is important to note that some spatial similarity of expression profiles is expected due to the influence of small scale spatial gradients in growth factors over periods of development \cite{Bullmore2012}. However, evidence from the Allen Brain Atlas suggests that interareal connectivity profiles in rodent brains are even more correlated with gene co-expression than expected simply based on such spatial relationships \cite{French2011}. This heightened correlation could be partially explained by observations in mathematical modeling studies that neurons with similar inputs (and therefore potentially performing similar functions) tend to have more similar connection profiles than neurons with dissimilar inputs \cite{Vertes}. 

Yet, while genetic coding and functional utility each play important roles, a key challenge lies in summarizing the various constraints on connection formation in a simple and intuitive theory that can guide future predictions. One particularly acclaimed candidate mechanism for such a theory is that of physical constraints on the development, maintenance, and use of connections. Metabolism related to neural architecture and function is costly, utilizing 20\% of the body's energy, despite comprising only 2\% of its volume \cite{Laughlin2002,Niven2008}. Even the development of axons alone, comprising only a small portion of cortical tissue, extorts a large material cost \cite{Bullmore2012}. The existence of these pervasive costs motivated early work to postulate that wiring minimization is a fundamental driver of connection formation. Consistent with this hypothesis, the connectomes of multiple species are predominantly comprised of wires extending over markedly short distances \cite{Budd2012,Cherniak1994,Cherniak2010,Niven2008,Raj2011,RiaErcsey-Ravasz2013,Song2014}, and this observation holds across different methods of data collection \cite{Cherniak2010,Song2014,RiaErcsey-Ravasz2013}. 

However, mounting evidence suggests that pressures for wiring minimization may compete against pressures for efficient communication \cite{Bassett2010,Kaiser2006,Zalesky2012}. Early evidence supporting the role of efficient communication came from the observation that one can fix the network architecture of inter-areal projections in the macaque cortex and then rearrange the location of areas in space to obtain a configuration with significantly (32\%) lower wiring cost than that present in the real system \cite{Kaiser2006}.  A similar method can be used to obtain a configuration of the \emph{C. elegans} neuronal connectome with 48\% lower wiring cost than that present in the real system \cite{Kaiser2006}. Interestingly, the connections whose length is decreased most also tend to be those that shorten the characteristic path length -- one of many ways to quantify how efficiently a network can communicate  \cite{Kaiser2006,Avena-Koenigsberger2018}. Consistent observations have been made in human white matter tractography \cite{Bassett2010}, the mouse inter-areal connectome \cite{Rubinov}, and dendritic arbors \cite{Budd2012,Bullmore2012}. Notably, computational models that instantiate both constraints on wiring and efficient communication produce topologies more similar to the true topologies than models that instantiate a constraint on wiring minimization alone \cite{Chen2013,Vertes}.  Moreover, models that allow for changes in this tradeoff over developmental time periods better fit observed connectome growth patterns than prior models \cite{Nicosia2013}.

It is precisely this balance between wiring minimization and communication efficiency that is thought to produce the complex network topologies observed in neural systems, along with markedly precise spatial embedding \cite{Chen2013,Kaiser2017}. To better understand this spatially embedded topology, it is useful to consider methods that can simultaneously (rather than independently) assess topology and geometry. One such method that has proven particularly useful in the study of neural systems from mice to humans is Rentian scaling, which assesses the efficiency of a network's spatial embedding \cite{Bassett2010,Sperry2017,Klimm2014,pineda2015disparate,sadovsky2014mouse}. Originally developed in the context of computer circuits, Rentian scaling describes a power-law scaling relationship between the number of nodes in a volume and the number of connections crossing the boundary of the volume \cite{Bassett2010,Bullmore2012}. The existence of such a power law relationship with an exponent known as Rent's exponent is consistent with an efficient spatial embedding of a complex topology \cite{christie2000interpretation,alcalde2017method}. In turn that efficient spatial embedding is thought to support a broad repertoire of functional dynamics. For example, tracts that bridge disparate areas of cortex to increase communication efficiency despite greater wiring cost, also critically add to the functional diversity of the brain in a manner that is distinct from that predicted by path length alone \cite{Betzel201720186}.

\section*{Reflections of physical constraints in local, mesoscale, and global network topology}

\begin{figure*}
	\begin{center}
		\centerline{\includegraphics[width=0.85\textwidth]{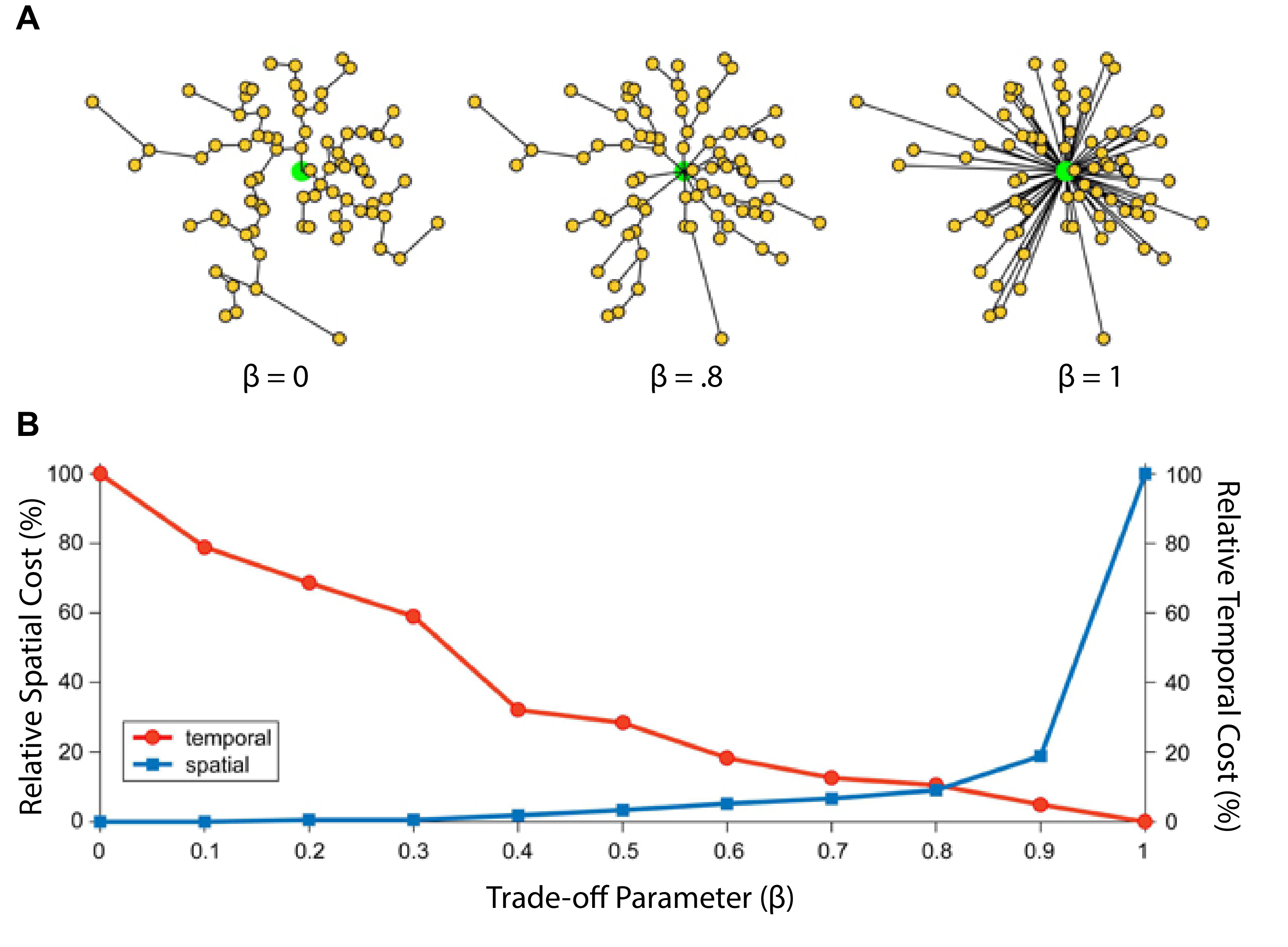}}
		\caption{\textbf{Effect of wiring minimization and communication efficiency on network topology.} Networks were generated by modulating the balance between a constraint on wiring (in the figure referred to as \emph{a spatial cost}) and a constraint on information routing efficiency (in the figure referred to as a \emph{temporal cost}). The parameter $\beta$, which ranges between 0 and 1, tunes this balance by weighting spatial cost against temporal cost. When $\beta=0$ only the spatial cost is considered, while when $\beta=1$ only the temporal cost is considered. \emph{(A)} Examples of networks at different values of $\beta$ when only the spatial constraint exists \emph{(left)}, when only the temporal constraint exists \emph{(right)}, and when the two constraints are balanced \emph{(middle)}. Root nodes are shown in green and all other nodes are shown in yellow. \emph{(B)} Spatial costs (blue) and temporal costs (red) vary as a function of $\beta$. This figure was adapted with permission from \cite{Budd2012}.}
		\label{fig1}
	\end{center}
\end{figure*}

Across species, the brain consistently exhibits a set of topological features at local, meso-, and global scales that can be relatively simply explained by spatial wiring rules \cite{Henderson2011,Park2013, Sporns2004}. At the local scale, multiple modalities have been used to demonstrate that a key conserved topological feature is the existence of hubs, or nodes of unexpectedly high degree \cite{VandenHeuvel2012,Seidlitz2018}. Such hubs emerge naturally in computational models in which the location of nodes are fixed in space, and edges between nodes are rewired to miminize average wiring length and to maximize topological efficiency by minimizing the average shortest path length (\textbf{Box 1}). Interestingly, the number and degree of hubs varies systematically with the relative importance of the two constraints \cite{Budd2012,Chen2013} (\textbf{Fig.~\ref{fig1}}). When wiring minimization is not enforced, networks become star graphs with a single giant hub \cite{Budd2012,Chen2013}; when both constraints are balanced, networks contain several hubs of varying degrees, consistent with the topology observed in brain networks \cite{Chen2013}. It is notable that such constraints can be implemented within the natural processes of development; for example, in adult \emph{C. elegans}, hub neurons have been tracked back to the earliest born neurons in the embryo, which accumulate a large number of connections along the normative growth trajectory  \cite{Kaiser2017,Varier2011}. 

At the mesoscale, a key conserved topological feature is modularity, or the existence of internally dense and externally sparse communities of nodes \cite{Henderson2011,hilgetag2016brain}. The strength of modularity in a network is commonly quantified using a modularity quality index (\textbf{Box 1}). In computational models, this index obtained under pressures of wiring minimization and communication efficiency (quantified with path length) was more similar to that empirically measured in the connectomes of the macaque and \emph{C. elegans} than to that obtained under either constraint separately \cite{Chen2013,Chen2017}.  Again it is notable that such constraints can be implemented within the natural processes of development; for example, in \emph{C. elegans}, communities form when many neurons are born in a similar temporal window, and therefore typically share a common progenitor type, spatial location, and genetic profile \cite{Kaiser2017, Perin2011}.  Spaces between modules can form cavities or cycles, or intuitively holes in the network, that can be identified with emerging tools from applied algebraic topology (\textbf{Box 2}) \cite{Sizemore2017a}. The locations, prevalence, and weight structure of these cycles differs markedly between geometric and random networks \cite{Kahle,Kahle2009}, with patterns of functional connectivity among neurons exhibiting characteristics similar to those observed in spatially constrained geometric networks \cite{Giusti}. It will be interesting in future to gain a deeper understanding of the relations between cycles and modules, and their emergence through the spatially constrained processes of development.

At the global scale, a key conserved topological feature is small-worldness, or the confluence of unexpectedly high clustering and short path length (\textbf{Box 1}) \cite{Watts1998}. Such an architecture is thought to be particularly conducive to a balance between local information processing within the clusters, and global information transmission across the topologically long distance connections \cite{shih2015connectomics}. Similar to the existence of hubs, modules, and cavities, small-world architecture in a network can naturally arise from spatial constraints on wiring \cite{kaiser2004spatial}. Intuitively, clusters tend to form in spatially nearby regions in order to minimize wiring cost, while long distance connections facilitating efficient communication tend to form only occasionally due to their elevated wiring cost \cite{Bassett2017b}. In concert with these empirical observations, computational models that account for wiring economy produce networks with small-world architecture reminiscent of that observed in real neural systems \cite{Avena-Koenigsberger2014}. Collectively, these studies demonstrate the influence of parsimonious wiring rules on complex network topology. Future work could be directed to better understand the aspects of connectome topology that remain unexplained and thus may arise from more subtle rules \cite{Chen2017}.

\section*{Relevance of network geometry for dynamics and cognition}

Pressures for wiring minimization and communication efficiency can exist alongside developmental processes that produce non-isotropically structured organs. Such processes include migration, elongation, segregation, folding, and closure that accompany neurulation resulting in the bilaterally symmetric nervous system composed of proencephalon, mesencephalon, rhombencephalon, and spinal cord as well as patterning across multiple overlapping signaling gradients. It is intuitively possible that such processes could also explain the observed differences in the network topologies of different sectors of the brain \cite{scholtens2014linking,heuvel2015bridging}, which can impinge on the functions that those sectors are optimized to perform (\textbf{Box 3}). Indeed, prior work has noted the co-existence of complex structural topologies and spatial gradients of specific function \cite{Jbabdi2013}, although it has been difficult to achieve a mechanistic understanding of exactly how the two relate to one another. One particularly promising recent line of investigation has proposed the existence of a set of primary spatial gradients that explain variance in large-scale connectivity \cite{Huntenburg2018,Margulies2016}. In both humans and macaques, the primary axis of variance is bounded on one end by the transmodal default mode system, and on the other end by the unimodal sensory systems \cite{Margulies2016}. Notably, this gradient is tightly linked to the geometry of the network, with the regions located at one end having maximal spatial distances from the regions located at the other end \cite{Margulies2016}. Additionally, the regions located at the peaks of the transmodal gradient have substantial overlap with structural hubs in human connectomes \cite{VandenHeuvel2013,petersen2015brain}. As discussed earlier, these hub could serve to optimally balance wiring cost and communication efficiency across the connectome, in addition to explaining patterns of functional connectivity. Put simply, such evidence supports the notion that the cortex is fundamentally organized along a dimension of function from concrete to abstract, and that dimension manifests clearly in the network's spatial embedding.

The diverse spatial location of cortical and subcortical areas has important implications for the patterns of neural dynamics that one would expect to observe. Consider, for example, the patterns of intrinsic activity noted consistently across species, individuals, and imaging modalities \cite{Lu2012,Margulies2016,Mitra2015c,Raichle2015}. While prior work has reported a consistent architecture of correlations between regional time series \cite{Raichle2015}, the manner in which the across-region activity pattern at one time point is related to the across-region activity pattern at other time points is not well understood. Recent work addressing this gap has posited the existence of so-called \emph{lag threads}, or spatial similarities between whole-brain activity patterns at non-zero time lags \cite{Mitra2014}. Unexplained by vasculature, the orthogonal threads are thought to reflect slow, subthreshold changes in the membrane potential of large neuronal populations \cite{MITRA2018}, a notion that is supported by subtle changes in lag thread structure across sleep and during cognitively demanding tasks \cite{Mitra2014,Mitra2015}. Notably, regions of the default mode (which coincide with peaks of the transmodal gradient discussed above) participate in lag thread motifs, where changes in the activity in one region lead to changes in the activity of another region more frequently than expected in an appropriate statistical null model \cite{Mitra2015c,Raichle2015}. The existence of these dependencies is consistent with time-invariant, possibly structural features of brain organization producing highly reproducible patterns of activity.

\begin{figure*}
	\begin{center}
		\centering
		\includegraphics[width=0.98\textwidth]{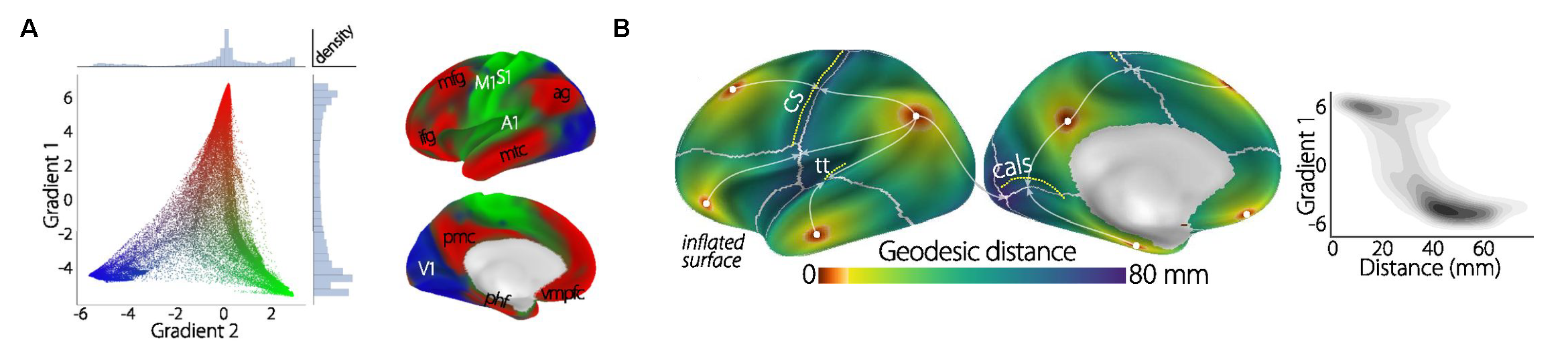}
		\caption{\textbf{Spatial distribution of intrinsic neural activity.} Principal gradients of functional connectivity calculated in the structural connections of both humans and macaques. The first two principal gradients explained approximately 40\% of the observed variance. \emph{(A)} (Left) A scatter plot of the first two principal gradients, with transmodal regions shown in red, visual regions shown in blue, and sensorimotor regions shown in green. (Right) The same colors are used to show the distribution of points visualized on a cortical surface. The pattern suggests the existence of a macroscale gradient of connectivity that reflects the systematic integration of information across different sensory modalities. \emph{(B, Left)} The minimum geodesic distance (mm) between each point on the cortical surface and the positive peaks of the first principal gradient. The peaks are shown as white circles. \emph{(B, Right)} A scatter plot depicting the relationship between distance and location on the trans- to uni-modal gradient. Put differently, transmodal regions with high values in the principal gradient are maximally distant from unimodal regions with low values in the principal gradient. This figure was adapted with permission from \cite{Margulies2016}.}
	\end{center}
\end{figure*}

\section*{Relevance of network geometry for disease}

The spatial architecture of brain networks not only impacts our understanding of dynamics and cognition, but also our understanding of neurological disease and psychiatric disorders. Mounting evidence suggests that many diseases and disorders of mental health can be thought of fruitfully as network disorders, where the anatomy and physiology of cross-regional communication can go awry \cite{braun2018from,stam2014modern}. Intuitively, spatial anisotropies of developmental processes, spatial specificity of pathology, and spatial inhomogeneities of drug targets that either lead to or accompany these disorders can also explain alterations in the spatial characteristics of brain networks \cite{bassett2018understanding}. In this section we briefly discuss this correspondence in epilepsy, a particularly common neurological disease, and in schizophrenia, a particularly devastating psychiatric disorder. 

Despite a diverse pathophysiology and a renitent unifying biological mechanism, epilepsy is characterized by altered network dynamics in the form of seizures that display spatially consistent patterns. For example, an ictal period often begins with a marked spatial decorrelation between distributed brain regions followed by a period in which abnormally synchronized activity propagates in consistent spatial patterns \cite{Jirsa2014,Wendling2003}. In addition to broad patterns of spatial decorrelation, individual siezures also show stereotyped patterns of both spiral waves and travelling waves of activity \cite{Chamberlain2011, Viventi2011,Gonzalez-Ramirez2015,Richardson2005,Ursino2006,Jirsa2014,Martinet2017}. In silico studies have demonstrated that a simple adaptive model of synaptically coupled and spatially embedded excitatory neurons can reproduce many basic features of these waveforms, including their speed and the size of the wavefront \cite{Gonzalez-Ramirez2015}. Yet, it is worth noting that travelling waves are not unique to epilepsy, but also occur in healthy human and non-human primates where they are thought to play a role in transporting task-relevant information \cite{Beggs2012,Massimini2004b,Richardson2005,Rubino2006,Takahashi2015,Takahashi2011a,Zhang2018}. However, marked differences in wave propagation in healthy and epileptic cortical tissue suggests that the precise spatial progression is important, potentially supported by distinct underlying microstructures \cite{Benucci2007}. Finally, even interictal dynamics are altered in epilepsy, as manifest by marked decreases in average functional connectivity across the brain combined with local increases in functional connectivity and efficiency in default mode areas \cite{Douw2015,Bonilha,DeSalvo2014}. These connectivity patterns have some utility in predicting seizure spread, but the guiding principles leading to these changes and how they relate to fine scale patterns of activity remains unclear \cite{Jirsa2017}.

While its pathophysiology is quite distinct from that implicated in epilepsy, schizophrenia is also a condition marked by severe network disturbances that have broad ramifications for cognitive function \cite{Bassett2008,Griffa2013,Zalesky2012}. Some of these network alterations appear to selectively affect connections of certain physical lengths, reflecting an alteration in the network's spatial embedding \cite{Alexander-Bloch2013}. Specifically, evidence suggests a reduced hierarchical structure and increased connection distance in the anatomical connectivity of multimodal cortex in patients with schizophrenia compared to healthy controls, indicative of less efficient spatial wiring \cite{Bassett2008}. Moreover, in functional brain networks, patients display longer high-weight connections, decreased clustering, and increased topological efficiency in comparison to healthy controls \cite{Alexander-Bloch2013}. The lack of strong, short distance functional connections is in line with evidence from animal studies suggesting an over-pruning of synapses in childhood onset schizophrenia \cite{Alexander-Bloch2013}. Here, the intuitions gained from a consideration of the network's spatial embedding offer important directions for future work in linking non-invasive imaging phenotypes with invasive biomarkers of neural dysfunction in disease.

\section*{Statistics, Null Models, and Generative Models}

In the previous sections, we outlined developmental rules for efficient wiring and we discussed the reflections of these rules in spatial patterns of healthy and diseased brain dynamics. Collectively, the studies that we have reviewed motivate the broader use and further development of sophisticated and easily-implementable tools for the analysis of a network's spatial embedding \cite{Kaiser2011}. Here we outline the current state of the field in developing effective network statistics, network null models, and generative network models that account for spatial embedding.~\\

\noindent \textbf{Network Statistics.} A simple way to examine network architecture in the context of spatial embedding is to incorporate the Euclidean distance of connections into local, meso-scale, and global statistics \cite{Alexander-Bloch2013,Duarte-Carvajalino2012}. Arguably the simplest local statistics that remain spatially sensitive are moments of the distribution of edge lengths in the network, including the mean, variance, skewness, and kurtosis. One can also compute graph metrics that have been extended to consider space, such as the physical network efficiency and the physical edge betweenness \cite{buhl2004efficiency}. Intuitively, both begin by defining the length of the shortest physical (as opposed to topological) path along network edges between any two nodes. The physical network efficiency then takes the inverse of the harmonic mean of this length, while the physical edge betweenness provides the fraction of shortest physical paths between all node pairs that traverse a given edge \cite{papadopoulos2018comparing}. One could also define a physical clustering coefficient in a similar manner. Finally, one can assess the system for Rentian scaling as described earlier, providing information on how efficiently the complex network topology has been embedded into the physical space \cite{Bassett2010,Sperry2017,Klimm2014,pineda2015disparate,sadovsky2014mouse}. In the context of neural systems, these spatially informed graph statistics can be used to account for the physical nature of information processing, propagation, and transmission.

Complementing local and global graph statistics is an assessment of a network's community structure, a mesoscale property frequently assessed by considering the existence and strength of network modules \cite{fortunato2016community}. From that community structure, one can determine the spatial embedding of communities, for example by assessing their laterality in bilaterally symmetric systems such as the brain  \cite{Doron2012,chai2016functional,he2018disrupted}. One of the most common ways to assess community structure is to maximize a modularity quality function, which identifies assortative modules with dense within-module connectivity and sparse between-module connectivity \cite{Newman2003} (although see \cite{Betzel} for methods to identify non-assortative communities). Mechanistically, this algorithm compares the strength of observed connections between two nodes in a community to that expected under a given a null model. The most commonly used null model in this context is the Newman-Girvan or configuration model, which preserves the strength distribution of the network \cite{Newman2003}. However, this null model operationalizes a purely topological constraint -- the strength distribution -- and does not acknowledge any spatial constraints that may exist in the system. For this reason, many investigators across scientific domains have begun developing alternative null models that account for physical laws \cite{papadopoulos2016evolution} or physical contraints \cite{Betzel2016a,Expert2010,Sarzynska2015} on their system of interest.

In the context of brain networks, it is worth considering three distinct null models for modularity maximization that incorporate information about the physical space of the network's embedding. First, one can directly incorporate the wiring minimization constraint observed in brain networks by defining a null model with a probability of connection between two nodes that decays exponentially as a function of distance \cite{Betzel2016a}. Using this model, one can detect different, and more spatially distributed modules than those obtained when one uses the configuration model \cite{Betzel2016a}. Second, one can employ gravity models \cite{Expert2010}, which account for the number of connections expected given a certain distance (typically a power law or inverse of distance), weighted by the relative importance of each location (typically a quantification of the population or size of a given location) \cite{Expert2010,Sarzynska2015}. Third, one can employ radiation models designed to capture flow of information between regions, by weighting distance functions by the flux or flow of each location \cite{Sarzynska2015}. Of course, there exists no single correct null model for community detection that will suit every question in neuroscience. However, we propose that many studies could test tighter, more targeted hypotheses about community structure in brain networks by using a null model that accounts for the brain's spatial nature.\\

\noindent \textbf{Network Null Models.} When considering a network representation of a neural system, one often computes a statistical quantitity of interest and then compares that quantity to that expected in a random network null model. If the observed quantity is significantly greater than or less than that expected, one concludes that the network under study shows meaningful architecture of potential relevance to the biology. Perhaps the most common random network null model is that which randomly permutes the locations of edges in the network while preserving the number of nodes, number of edges, and edge weight distribution. However, one may also be interested to determine whether observed statistics are different from what one would expect simply from the spatial embedding or wiring rules of the network \cite{Roberts2016,Samu2014,Wiedermann2016}. To address these questions, one can rewire the observed network by conditionally swapping two links if the swap preserves the mean wiring length of the network \cite{Samu2014}. By pairing this model with a reduced null model in which connections are only swapped if they \emph{reduce} connnection length, one can assess the role of long distance connections in the network, which will be preserved in the spatial null but not preserved in the reduced null \cite{Samu2014}. In addition to preserving the mean wiring length, one might also wish to preserve the full edge length distribution by, for example, (1) fitting a function to the relationship between the mean and variance of edge weights and their distances, (2) removing the effect of that relationship from the data, (3) randomly rewiring the network, and (4) adding the effect back into the rewired network \cite{Roberts2016}.

\begin{figure*}
	\begin{center}
		\centering
		\includegraphics[width=0.95\textwidth]{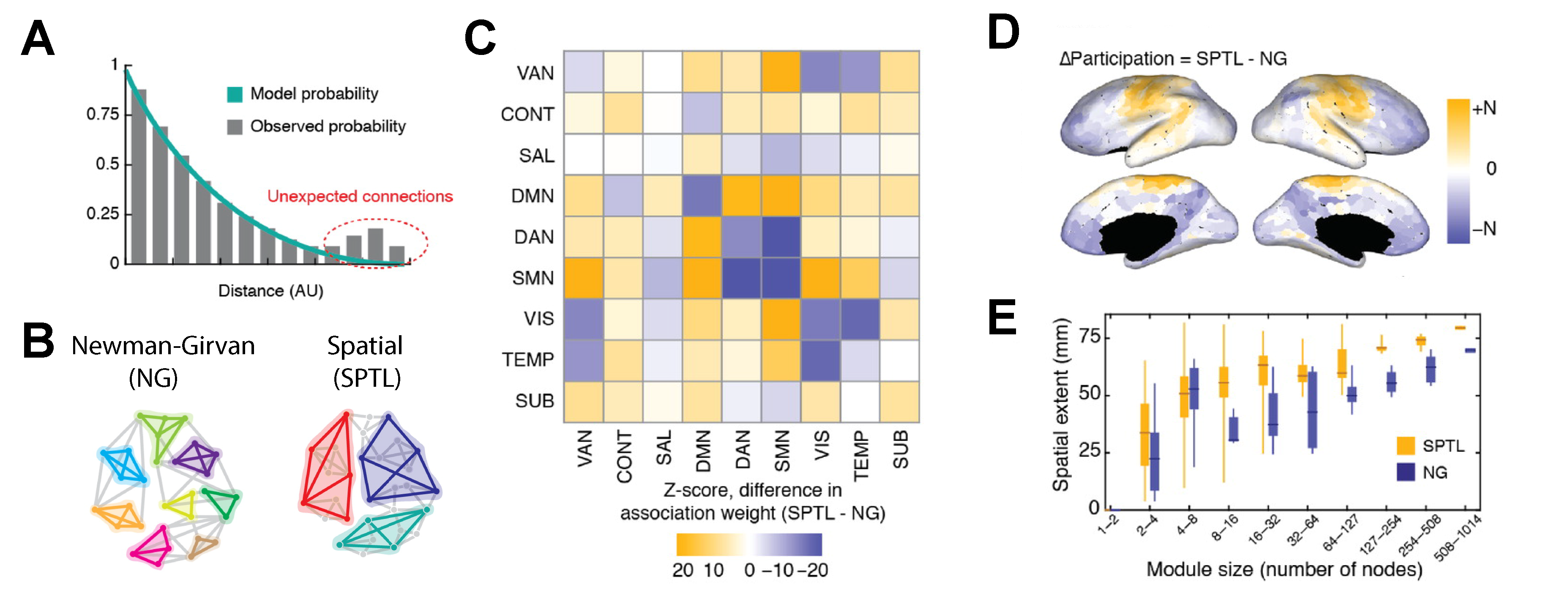}
		\caption{ \textbf{Community structure obtained with spatially embedded and non-embedded null models.} \emph{(A)} A schematic of a spatially-informed null model. The model expects fewer long distance connections than short distance connections. \emph{(B)} A schematic of the anticipated difference between the spatial null model and the Newman-Girvan (NG) null model; spatial communities will have longer distance connections and not capture clustering of spatially nearby regions. \emph{(C)} Differences in the association matrices between the two models. Positive (negative) numbers indicate when two nodes were more likely to be co-assigned to the same module under the spatial (NG) model. \emph{(D)} The difference in the participation coefficient between the spatial and NG models. The participation coefficient quantifies how diverse a node's connections are across modules. \emph{(E)} The difference in spatial spread of modules in both models; the spatially embedded model tends to produce modules that cover larger distances. This figure was adapted with permission from \cite{Betzel2016a}.}
	\end{center}
\end{figure*}

To complement insights obtained from edge swapping algorithms, one can also construct null model networks by stipulating a wiring rule \emph{a priori} while fixing the locations of nodes within the embedded system. In this vein, studies have fruitfully used null models based on minimum spanning tree and greedy triangulation methods \cite{cui2018classification,janssen2017neural,smit2016life}.  A minimum spanning tree is a graph that connects all of the nodes in a network such that the sum of the total edge weights is minimal. To extend this notion to spatial networks, one can preserve the true geographic locations of all nodes in the empirical network and compute the minimum spanning tree on the matrix of Euclidean distances between all node pairs \cite{papadopoulos2018comparing}. Representing the opposite extreme is the greedy triangulation model, which is particularly relevant for the study of empirical networks that are planar (lying along a surface) as opposed to non-planar (lying within a volume). In the context of neural systems, planar or planar-like networks are observed in vasculature, and in thinned models of cortex that either consider a single lamina or a coarse-grained model collapsing across laminae \cite{blinder2013cortical,schmid2017depth}. To construct a greedy triangulation null model, one can preserve the true geographic locations of all nodes in the empirical network and iteratively connect pairs of nodes in ascending order of their distance while ensuring that no edges cross. After constructing such minimally and maximally wired null models, one can calculate relative measures of wiring length, physical efficiency, physical betweenness centrality, and community structure by normalizing the empirical values by those expected in the two extremes \cite{papadopoulos2018comparing}.\\

\noindent \textbf{Generative Network Models.} Generative network models can be used to test hypotheses about the rules guiding network growth, development, and evolution \cite{betzel2017generative}. Often, an ensemble of generative models are constructed, and summary graph statistics from the empirical network are compared to the statistics of each of the generative models with the goal of inferring which wiring rule was most likely to have produced the observed architecture \cite{Klimm2014,Betzel2016b,Vertes}. Evidence from such studies suggests that spatially embedded models tend to more accurately reproduce network measures of large-scale neural systems than models that do not account for space \cite{Klimm2014}. One particularly influential study considered 13 generative models that all incorporated a wiring probability that increased with distance \cite{Betzel2016b}. Consistent with other work, the authors found that the model that only included the wiring minimization constraint was unable to recreate long distance connections of individual connectomes in humans \cite{Betzel2016b,Chen2017,Vertes}. Successive generative models were then added that attempted to recreate certain aspects of topology in addition to these geometric constraints \cite{Betzel2016b}. The models that performed the best were those that preserved homophilic attraction such that connections preferentially formed between nodes that had similar connection profiles \cite{Betzel2016b}. Continued advancement of generative network models, and inclusion of additional biological features such as bilateral symmetry, serves as an exciting approach to test mechanistic predictions about how network topology forms in spatially embedded neural systems.

\section*{Future Directions}

The spatial embedding of the brain is an important driver of its connectivity, which in turn directly constrains neural function and by extension behavior. Emerging tools from network science can be used to assess this spatial architecture, thereby allowing investigators to test more specific hypotheses about brain network structure and dynamics. While we envision that the use of these tools will significantly expand our understanding, it is also important to acknowledge their limitations. In particular, the majority of currently available network tools make the simplifying assumption that all of the relations of interests are strictly dyadic in nature, and exist between inherently separable components \cite{Butts2009}. In truth, however, features that arise from spatial embedding can also manifest as continuous or overlapping maps and gradients \cite{Jbabdi2013}, motivating the use of tools from applied algebraic topology that can account for non-dyadic interactions (Box 2). As the field moves forward, we envision existing and yet-to-be-developed tools for characterizing the spatial embedding of brain networks will prove critical for our understanding of network processes underlying cognition, and alterations to those processes accompanying disease.

\clearpage
\newpage
\section*{Box 1: Simple Network Statistics}
In a network representation of the brain, units ranging from neurons or neuronal ensembles to nuclei and areas are represented as network \emph{nodes} and unit-to-unit interactions ranging from physical connections to statistical similarities in activity time series are represented as network \emph{edges}. The architecture of the network can be quantitatively characterized using statistics from graph theory \cite{Rubinov2010}. Here, we mathematically define some of the topological statistics mentioned elsewhere in this paper. 
\begin{itemize}
	\item \emph{Degree and Strength}. The degree of a node is the number of connections it has. In a binary graph encoded in the adjacency matrix $\mathbf{A}$, where two regions $i$ and $j$ are connected if $A_{ij} = 1$, and not connected if $A_{ij} = 0$, then the degree $k_{i}$ is defined as $k_{i} = \sum_{i,j \in N}A_{ij}$, where $N$ is the set of all nodes. In a weighted graph, where $A_{ij}$ is the strength of the connection between nodes $i$ and $j$, then the strength $s_{i}$ is defined as $s_{i} = \sum_{i,j \in N}A_{ij}$.
	\item \emph{Path Length and Network Efficiency}. The term path length frequently refers to the average length of the shortest path in a network. The shortest path between any two nodes is given by the path requiring the fewest hops. The network efficiency is given by the inverse of the harmonic mean of the shortest path length. To be precise, we can write the path length of node $i$ as $L_{i } = \frac{1}{n}\sum_{i \in N}\frac{\sum_{i \in N, j \neq i} d_{i,j}}{n - 1}$, where $d_{i,j}$ is the shortest path length between two nodes and $n$ is the number of nodes.
	\item \emph{Clustering Coefficient}. The clustering coefficient can be used to quantify the fraction of a node's neighbors that are also neighbors with each other. Specifically, the clustering coefficient of node $i$ is given by $C_{i} = \frac{1}{n} \sum_{i \in N}\frac{2t_{i}}{k_{i}(k_{i} - 1)}$, where $t_{i}$ is the number of triangles around node $i$ \cite{Watts1998}. The clustering coefficient of the network is the average clustering coefficient of all of its nodes.
	\item \emph{Modularity}. While several modularity quality functions exist, the most common is $Q = \sum_{ij}[A_{ij} - \gamma P_{ij}]\delta (c_{i}c_{j})$, where $Q$ is the modularity quality index, $P_{ij}$ is the expected number of connections between node $i$ and node $j$ under a specified null model, $\delta()$ is the Kroenecker delta, and $c_i$ indicates the community assignment of node $i$. The tuning parameter $\gamma$ ranges from $(0,\infty)$ and can be used to tune the average community size.
\end{itemize}

\begin{figure*}
	\begin{center}
		\centerline{\includegraphics[width=0.85\textwidth]{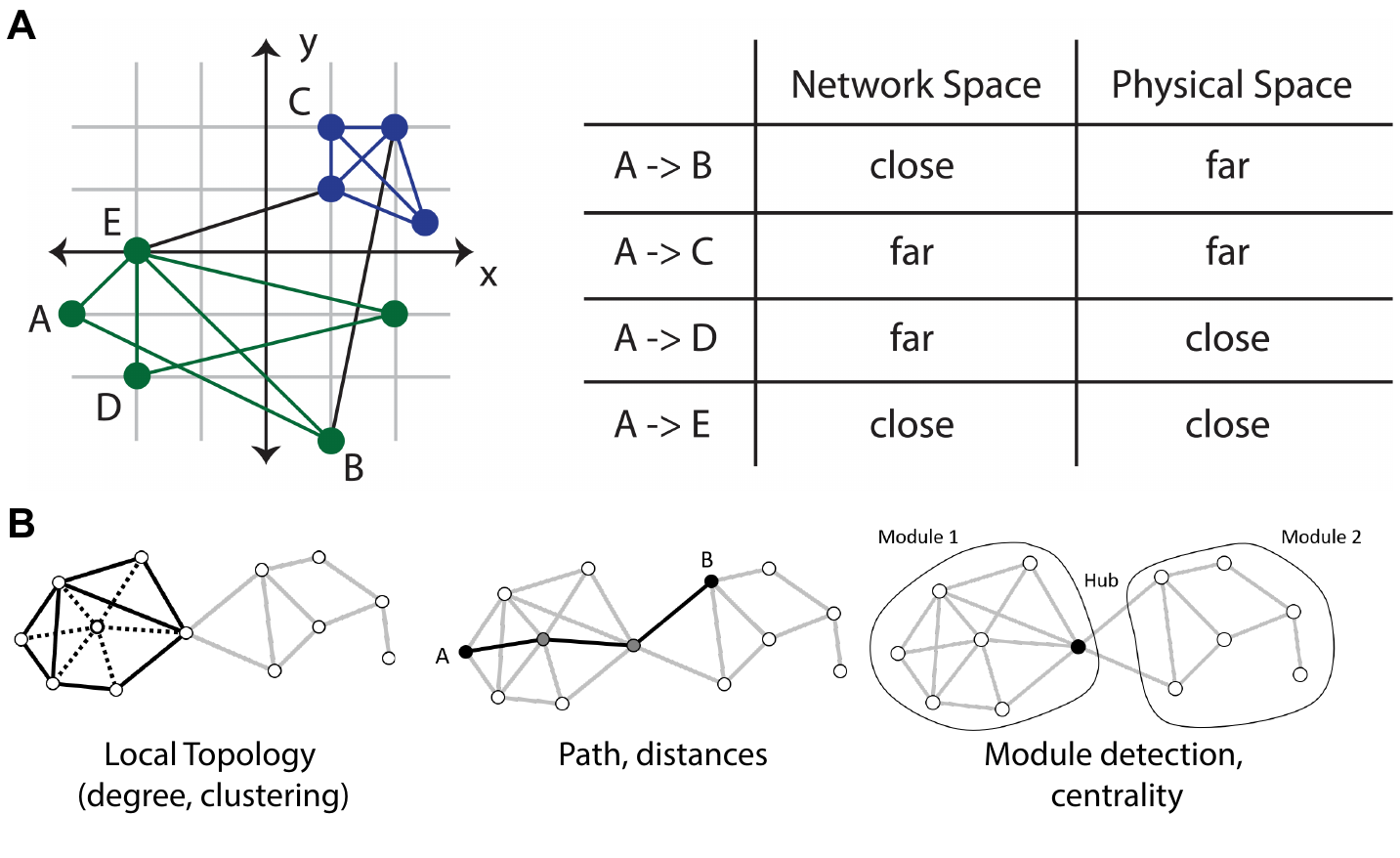}}
		\caption{\textbf{Schematic of network measures.} \emph{(A)} An illustration of network space (topology) and physical space (geometry). The network is embedded into a physical space, indicated by the $x$- and $y$-axes. The topological and physical distances between the nodes are not necessarily related. \emph{(B)} The network representation enables the calculation of local, mesoscale, and global features to describe the pattern of connections in topological space (as shown here) as well as the pattern of connections in physical space (as we describe in the main text). This figure was adapted with permission from \cite{Bassett2017} and from \cite{garcia2018applications}.}
	\end{center}
\end{figure*}

\section*{Box 2: Applied Algebraic Topology}
While graph theory is a powerful and accessible framework for analyzing complex networks, complementary information can be gained by using different mathematical formalisms. Here, we describe an alternative approach to studying structure in networks that relies on tools developed in the field of applied algebraic topology, specifically persistent homology \cite{giusti2016twos,Zomorodian2005,Carlsson2009}. Persistent homology can be used to study intrinsically mesoscale structures called \emph{cycles} and \emph{cliques} \cite{sizemore2018importance}. Cliques are all-to-all connected subsets of nodes in a network. The presence of many, large cliques indicates many highly connected units are present in the network \cite{Reimann2017}. Cycles are looped patterns of cliques which may enclose a cavity, or topological void, within the network. Cliques and cavities by definition reside within a binary graph, however one can expand a weighted network into a sequence of binary graphs via iterative thresholding \cite{Giusti,Petri2013}. Then using persistent homology one can track the birth, persistence, and death of cavities along this sequence which gives a wholistic insight into the global network (Fig. Box 2, panel A). 

In random graphs, the number of births and deaths across thresholds follows a characteristic pattern \cite{Kahle2009}. At high thresholds and low edge density, a few low dimensional cavities exist, while at low thresholds and high edge density, more high-dimensional cavities exist (Fig. Box 2, panel B) \cite{Sizemore2017, Kahle2009, Horak2009}. Interestingly, geometric graphs -- which can be used to instantiate spatial constraints on the topology -- show a markedly different distribution. There are many low dimensional cavities, and \emph{fewer} cavities with increasing dimension\cite{Sizemore2017,Kahle,Kahle2009,Bobrowski} (Fig. Box 2, panel C). This general pattern has been recapitulated in functional networks constructed from firing of hippocampal neurons, indicating a geometric rather than random nature to neuronal co-firing \cite{Giusti}. Furthermore, the persistent homology of human connectomes \cite{Sizemore2017a} and rat microcircuits \cite{Dotko2016a} is distinct from that expected in a minimally wired null model. In humans, the presence of widespread subcortical connections leads to more cavities being born at high densities \cite{Sizemore2017a}, while rat microcircuits display more high dimensional cavities in general \cite{Dotko2016a}. Further investigation into how wiring rules shape the topology of neural systems may shed light on how the brain's spatial embedding shapes connectivity across scales and species.

\begin{figure}
	\begin{center}
		\centerline{\includegraphics[width=0.45\textwidth]{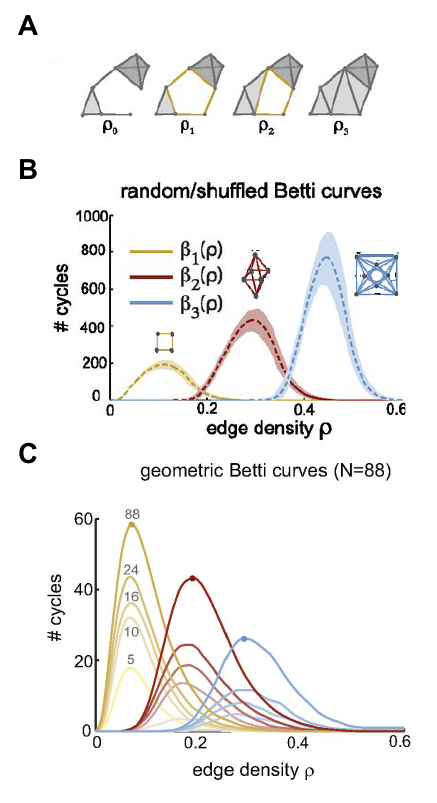}}
		\caption{\textbf{Applied algebraic topology.} \emph{(A)} An illustration of thresholding a weighted network across different densities ($\rho$). At $\rho_1$, a cavity of dimension 1 is born (shown in yellow), which then dies at $\rho_3$. \emph{(B)} The characteristic pattern of births and deaths (called Betti curves) for cycles of dimension 1 (yellow), dimension 2 (red), and dimension 3 (blue) from a random network. The cliques of each dimension are shown near each corresponding Betti curve for reference. \emph{(C)} The same pattern, but for geometric networks. Different lines of the same color indicate different dimensions of embedding. This figure was adapted with permission from \cite{Giusti}.}
	\end{center}
\end{figure}

\section*{Box 3: Control Theory}

Network control theory provides a potentially powerful approach for modeling neural dynamics \cite{Schiff2011}. Hailing from physics and engineering, network control theory characterizes a complex system as composed of nodes interconnected by edges, and then specifies a model of network dynamics to determine how external input affects the nodes' time-varying activity \cite{Liu2011}. Most studies of network control in neural systems stipulate a linear, time-invariant model of dynamics: $\dot{x}(t) = \mathbf{A}x(t) + \mathbf{B}u(t)$, where $x$ is some measure of brain state, $\mathbf{A}$ is a structural connectivity matrix, $u$ is the input into the system (exogenous stimulation, or endogenous input from other brain regions), and $\mathbf{B}$ selects the control set, or regions to provide input to \cite{Bryson1996,yan2017network}. Assuming this model of dynamics, one can calculate the control \emph{energy} required to reach specific brain states, which can be used a state dependent measure of the efficiency of control \cite{gu2017optimal,betzel2016optimally}. Control theory can also posit control metrics that quantify how efficiently a node would drive the brain to various states. Two commonly used metrics are \emph{average controllability} and \emph{modal controllability} \cite{pasqualetti2014controllability}. When every node is included in the control set, average controllability is proportional to the average energy required to drive the node to any state \cite{jeganathan2018fronto}. Conversely, modal controllability is high in nodes where a small input will result in large perturbations to all eigenmodes of the system, and is interpreted to be high in nodes that can easily drive the brain to hard-to-reach states \cite{Gu2014,Tang,Wu-Yan2017}.

If these properties are important for helping the brain transition between states, one would expect them not to be randomly distributed across the cortex, but to be clustered into spatially constrained, functionally relevant systems. More specifically, one might expect functional systems that drive the brain to many accessible states, such as the default mode system, to have high average controllability, while regions that drive the brain to hard-to-reach, cognitively demanding states (executive control areas) to have high modal controllability. Data from healthy human adults supports these two hypotheses \cite{Gu2014}. Moreover, both average and modal controllability increase across development and are correlated with cognitive performance generally \cite{Tang} and executive function specifically \cite{cornblath2018sex}. The manner in which network control tracks individual differences reflects the fact that the capacity for a network to enact control is dependent upon its topology \cite{Menara2018,Kim2017}. Further efforts are needed to distill exactly how spatial embedding and wiring constraints impinge on that control capacity, and how it is altered in psychiatric disorders \cite{jeganathan2018fronto} and neurological disease \cite{taylor2015optimal}.

\section*{Box 4: Outstanding Questions}
\begin{itemize}
	\item How do spatially constrained developmental processes constrain the formation of cycles in brain networks?
	\item What are the aspects of connectome topology that remain unexplained by wiring minimization or communication efficiency and thus may arise from more subtle rules?
	\item How do the structural topologies that arise from physical growth rules support functional gradients?
	\item What is the precise relationship between invariant features of brain activity and the underlying anatomical structure?
	\item How does the development of the connectome determine spatial progression of activity through the cortex in health and disease?
	\item To what extent can we link macro-scale structural topology with small scale developmental rules?
	\item Can a deeper understanding of connectome development be used to help identify new biomarkers for network diseases?
	\item What additional rules can be incorporated into generative models of the brain to recapitulate its topology?
	\item How can frontiers in network science help characterize non-dyadic relationships in the brain?
\end{itemize}

\section*{Glossary}
\begin{itemize}
	\item \textbf{adjacency matrix}: The adjacency matrix of a graph is an $N \times N$ matrix, where $N$ is the number of nodes. Each element $A_{ij}$ of the matrix gives the strength of the edge between nodes $i$ and $j$.
	\item \textbf{cycle}: In applied algebraic topology, a cycle is an empty space (or lack of edges) in a graph surrounded by all-to-all connected subgraphs of the same dimension. The dimension here refers to the number of nodes included in each all-to-all connected subgraph.
    \item \textbf{edge}: From the perspective of graph theory, an edge is a connection between nodes. From the perspective of neuroscience, an edge is a statistical dependency (functional) or estimated physical connection (structural) between nodes.
    \item \textbf{geometry}: The geometry of a network reflects features of a graph in the context of physical space.
    \item \textbf{hub}: A central node in the network, typically having many connections.
	\item \textbf{node}: From the perspective of graph theory, a node is the unit where edges connect in the graph. From the perspective of neuroscience, a node is a brain region, neuron, or protein whose interactions one wishes to understand. 
	\item \textbf{topology}: The quantification of features of a graph in the context of space defined by the graph itself, without respect to any physical embedding.
\end{itemize}

\section*{Highlights}
\begin{itemize}
	\item The physical embedding of neural systems imposes constraints on the possible patterns of connections that form, and in turn on the possible repertoire of functional motifs. Utilizing new tools from network science that account for these constraints can help researchers understand fundamentals of brain dynamics in health and disease.
	\item Prominent, competing rules guiding the formation of brain networks include the minimization of wiring cost, and the maximization of communication efficiency. These competing mechanisms lead to networks with high local clustering with sparse long distance connections.
	\item Recent work suggests that intrinsic functional connectivity varies along dimensions that are tightly linked to the spatial embedding of the brain and the topological properties that arise in the presence of spatial constraints. Similarly, these topological properties show widespread changes in the context of the network diseases such as epilepsy and schizophrenia.
	\item A rich, and continually growing repertoire of statistics, null models, and generative models exist to aid researchers in testing focused hypotheses about the role of physical embedding in observed neural phenomena.
\end{itemize}

\section*{Acknowledgments}
We would like to thank Ann Sizemore for her help with Box 2: Applied Algebraic Topology. D.S.B. and J.S. acknowledge support from the John D. and Catherine T. MacArthur Foundation, the Alfred P. Sloan Foundation, the ISI Foundation, the Paul Allen Foundation, the Army Research Laboratory (W911NF-10-2-0022), the Army Research Office (Bassett-W911NF-14-1-0679, Grafton-W911NF-16-1-0474, DCIST-W911NF-17-2-0181), the Office of Naval Research, the National Institute of Mental Health (2-R01-DC-009209-11, R01-MH112847, R01-MH107235, R21-M MH-106799), the National Institute of Child Health and Human Development (1R01HD086888-01), National Institute of Neurological Disorders and Stroke (R01 NS099348), and the National Science Foundation (BCS-1441502, BCS-1430087, NSF PHY-1554488 and BCS-1631550). The content is solely the responsibility of the authors and does not necessarily represent the official views of any of the funding agencies.\\


%


\begin{thebibliography}{139}%
	\makeatletter
	\providecommand \@ifxundefined [1]{%
		\@ifx{#1\undefined}
	}%
	\providecommand \@ifnum [1]{%
		\ifnum #1\expandafter \@firstoftwo
		\else \expandafter \@secondoftwo
		\fi
	}%
	\providecommand \@ifx [1]{%
		\ifx #1\expandafter \@firstoftwo
		\else \expandafter \@secondoftwo
		\fi
	}%
	\providecommand \natexlab [1]{#1}%
	\providecommand \enquote  [1]{``#1''}%
	\providecommand \bibnamefont  [1]{#1}%
	\providecommand \bibfnamefont [1]{#1}%
	\providecommand \citenamefont [1]{#1}%
	\providecommand \href@noop [0]{\@secondoftwo}%
	\providecommand \href [0]{\begingroup \@sanitize@url \@href}%
	\providecommand \@href[1]{\@@startlink{#1}\@@href}%
	\providecommand \@@href[1]{\endgroup#1\@@endlink}%
	\providecommand \@sanitize@url [0]{\catcode `\\12\catcode `\$12\catcode
		`\&12\catcode `\#12\catcode `\^12\catcode `\_12\catcode `\%12\relax}%
	\providecommand \@@startlink[1]{}%
	\providecommand \@@endlink[0]{}%
	\providecommand \url  [0]{\begingroup\@sanitize@url \@url }%
	\providecommand \@url [1]{\endgroup\@href {#1}{\urlprefix }}%
	\providecommand \urlprefix  [0]{URL }%
	\providecommand \Eprint [0]{\href }%
	\providecommand \doibase [0]{http://dx.doi.org/}%
	\providecommand \selectlanguage [0]{\@gobble}%
	\providecommand \bibinfo  [0]{\@secondoftwo}%
	\providecommand \bibfield  [0]{\@secondoftwo}%
	\providecommand \translation [1]{[#1]}%
	\providecommand \BibitemOpen [0]{}%
	\providecommand \bibitemStop [0]{}%
	\providecommand \bibitemNoStop [0]{.\EOS\space}%
	\providecommand \EOS [0]{\spacefactor3000\relax}%
	\providecommand \BibitemShut  [1]{\csname bibitem#1\endcsname}%
	\let\auto@bib@innerbib\@empty
	\bibitem [{\citenamefont {Bassett}\ and\ \citenamefont
		{Sporns}(2017)}]{Bassett2017}%
	\BibitemOpen
	\bibfield  {author} {\bibinfo {author} {\bibfnamefont {D.~S.}\ \bibnamefont
			{Bassett}}\ and\ \bibinfo {author} {\bibfnamefont {O.}~\bibnamefont
			{Sporns}},\ }\href {\doibase 10.1038/nn.4502} {\bibfield  {journal} {\bibinfo
			{journal} {Nature Neuroscience}\ }\textbf {\bibinfo {volume} {20}},\
		\bibinfo {pages} {353} (\bibinfo {year} {2017})},\ \Eprint
	{http://arxiv.org/abs/0106096v1} {arXiv:0106096v1 [arXiv:cond-mat]}
	\BibitemShut {NoStop}%
	\bibitem [{\citenamefont {Bassett}\ \emph
		{et~al.}(2018{\natexlab{a}})\citenamefont {Bassett}, \citenamefont {Zurn},\
		and\ \citenamefont {Gold}}]{bassett2018nature}%
	\BibitemOpen
	\bibfield  {author} {\bibinfo {author} {\bibfnamefont {D.~S.}\ \bibnamefont
			{Bassett}}, \bibinfo {author} {\bibfnamefont {P.}~\bibnamefont {Zurn}}, \
		and\ \bibinfo {author} {\bibfnamefont {J.~I.}\ \bibnamefont {Gold}},\
	}\href@noop {} {\bibfield  {journal} {\bibinfo  {journal} {Nature Reviews
				Neuroscience}\ }\textbf {\bibinfo {volume} {In Press}} (\bibinfo {year}
		{2018}{\natexlab{a}})}\BibitemShut {NoStop}%
	\bibitem [{\citenamefont {Medaglia}\ \emph {et~al.}(2015)\citenamefont
		{Medaglia}, \citenamefont {Lynall},\ and\ \citenamefont
		{Bassett}}]{medaglia2015cognitive}%
	\BibitemOpen
	\bibfield  {author} {\bibinfo {author} {\bibfnamefont {J.~D.}\ \bibnamefont
			{Medaglia}}, \bibinfo {author} {\bibfnamefont {M.~E.}\ \bibnamefont
			{Lynall}}, \ and\ \bibinfo {author} {\bibfnamefont {D.~S.}\ \bibnamefont
			{Bassett}},\ }\href@noop {} {\bibfield  {journal} {\bibinfo  {journal} {J
				Cogn Neurosci}\ }\textbf {\bibinfo {volume} {27}},\ \bibinfo {pages} {1471}
		(\bibinfo {year} {2015})}\BibitemShut {NoStop}%
	\bibitem [{\citenamefont {Petersen}\ and\ \citenamefont
		{Sporns}(2015)}]{petersen2015brain}%
	\BibitemOpen
	\bibfield  {author} {\bibinfo {author} {\bibfnamefont {S.~E.}\ \bibnamefont
			{Petersen}}\ and\ \bibinfo {author} {\bibfnamefont {O.}~\bibnamefont
			{Sporns}},\ }\href@noop {} {\bibfield  {journal} {\bibinfo  {journal}
			{Neuron}\ }\textbf {\bibinfo {volume} {88}},\ \bibinfo {pages} {207}
		(\bibinfo {year} {2015})}\BibitemShut {NoStop}%
	\bibitem [{\citenamefont {Ducruet}\ and\ \citenamefont
		{Beauguitte}(2014)}]{ducruet2014spatial}%
	\BibitemOpen
	\bibfield  {author} {\bibinfo {author} {\bibfnamefont {C.}~\bibnamefont
			{Ducruet}}\ and\ \bibinfo {author} {\bibfnamefont {L.}~\bibnamefont
			{Beauguitte}},\ }\href@noop {} {\bibfield  {journal} {\bibinfo  {journal}
			{Networks and Spatial Economics}\ }\textbf {\bibinfo {volume} {14}},\
		\bibinfo {pages} {297} (\bibinfo {year} {2014})}\BibitemShut {NoStop}%
	\bibitem [{\citenamefont {Barth{\'{e}}lemy}(2011)}]{Barthelemy2011}%
	\BibitemOpen
	\bibfield  {author} {\bibinfo {author} {\bibfnamefont {M.}~\bibnamefont
			{Barth{\'{e}}lemy}},\ }\href {\doibase 10.1016/j.physrep.2010.11.002}
	{\bibfield  {journal} {\bibinfo  {journal} {Physics Reports}\ }\textbf
		{\bibinfo {volume} {499}},\ \bibinfo {pages} {1} (\bibinfo {year}
		{2011})}\BibitemShut {NoStop}%
	\bibitem [{\citenamefont {Bullmore}\ and\ \citenamefont
		{Sporns}(2012)}]{Bullmore2012}%
	\BibitemOpen
	\bibfield  {author} {\bibinfo {author} {\bibfnamefont {E.}~\bibnamefont
			{Bullmore}}\ and\ \bibinfo {author} {\bibfnamefont {O.}~\bibnamefont
			{Sporns}},\ }\href {\doibase 10.1038/nrn3214} {\bibfield  {journal} {\bibinfo
			{journal} {Nature Reviews Neuroscience}\ }\textbf {\bibinfo {volume} {13}},\
		\bibinfo {pages} {336} (\bibinfo {year} {2012})},\ \Eprint
	{http://arxiv.org/abs/NIHMS150003} {arXiv:NIHMS150003} \BibitemShut {NoStop}%
	\bibitem [{\citenamefont {Chen}\ \emph {et~al.}(2017)\citenamefont {Chen},
		\citenamefont {Wang}, \citenamefont {Hilgetag},\ and\ \citenamefont
		{Zhou}}]{Chen2017}%
	\BibitemOpen
	\bibfield  {author} {\bibinfo {author} {\bibfnamefont {Y.}~\bibnamefont
			{Chen}}, \bibinfo {author} {\bibfnamefont {S.}~\bibnamefont {Wang}}, \bibinfo
		{author} {\bibfnamefont {C.~C.}\ \bibnamefont {Hilgetag}}, \ and\ \bibinfo
		{author} {\bibfnamefont {C.}~\bibnamefont {Zhou}},\ }\href {\doibase
		10.1371/journal.pcbi.1005776} {\bibfield  {journal} {\bibinfo  {journal}
			{PLoS computational biology}\ }\textbf {\bibinfo {volume} {13}},\ \bibinfo
		{pages} {e1005776} (\bibinfo {year} {2017})}\BibitemShut {NoStop}%
	\bibitem [{\citenamefont {French}\ and\ \citenamefont
		{Pavlidis}(2011)}]{French2011}%
	\BibitemOpen
	\bibfield  {author} {\bibinfo {author} {\bibfnamefont {L.}~\bibnamefont
			{French}}\ and\ \bibinfo {author} {\bibfnamefont {P.}~\bibnamefont
			{Pavlidis}},\ }\href {\doibase 10.1371/journal.pcbi.1001049} {\bibfield
		{journal} {\bibinfo  {journal} {PLoS Comput Biol}\ }\textbf {\bibinfo
			{volume} {7}} (\bibinfo {year} {2011}),\
		10.1371/journal.pcbi.1001049}\BibitemShut {NoStop}%
	\bibitem [{\citenamefont {Rubinov}\ \emph {et~al.}(2015)\citenamefont
		{Rubinov}, \citenamefont {Ypma}, \citenamefont {Watson},\ and\ \citenamefont
		{Bullmore}}]{Rubinov}%
	\BibitemOpen
	\bibfield  {author} {\bibinfo {author} {\bibfnamefont {M.}~\bibnamefont
			{Rubinov}}, \bibinfo {author} {\bibfnamefont {R.~J.~F.}\ \bibnamefont
			{Ypma}}, \bibinfo {author} {\bibfnamefont {C.}~\bibnamefont {Watson}}, \ and\
		\bibinfo {author} {\bibfnamefont {E.~T.}\ \bibnamefont {Bullmore}},\ }\href
	{\doibase 10.1073/pnas.1420315112} {\bibfield  {journal} {\bibinfo  {journal}
			{Proceedings of the National Academy of Sciences}\ }\textbf {\bibinfo
			{volume} {112}},\ \bibinfo {pages} {10032} (\bibinfo {year}
		{2015})}\BibitemShut {NoStop}%
	\bibitem [{\citenamefont {Vertes}\ \emph {et~al.}(2012)\citenamefont {Vertes},
		\citenamefont {Alexander-Bloch}, \citenamefont {Gogtay}, \citenamefont
		{Giedd}, \citenamefont {Rapoport},\ and\ \citenamefont {Bullmore}}]{Vertes}%
	\BibitemOpen
	\bibfield  {author} {\bibinfo {author} {\bibfnamefont {P.~E.}\ \bibnamefont
			{Vertes}}, \bibinfo {author} {\bibfnamefont {A.~F.}\ \bibnamefont
			{Alexander-Bloch}}, \bibinfo {author} {\bibfnamefont {N.}~\bibnamefont
			{Gogtay}}, \bibinfo {author} {\bibfnamefont {J.~N.}\ \bibnamefont {Giedd}},
		\bibinfo {author} {\bibfnamefont {J.~L.}\ \bibnamefont {Rapoport}}, \ and\
		\bibinfo {author} {\bibfnamefont {E.~T.}\ \bibnamefont {Bullmore}},\ }\href
	{\doibase 10.1073/pnas.1111738109} {\bibfield  {journal} {\bibinfo  {journal}
			{Proceedings of the National Academy of Sciences}\ }\textbf {\bibinfo
			{volume} {109}},\ \bibinfo {pages} {5868} (\bibinfo {year}
		{2012})}\BibitemShut {NoStop}%
	\bibitem [{\citenamefont {Laughlin}\ and\ \citenamefont
		{Sejnowski}(2002)}]{Laughlin2002}%
	\BibitemOpen
	\bibfield  {author} {\bibinfo {author} {\bibfnamefont {S.~B.}\ \bibnamefont
			{Laughlin}}\ and\ \bibinfo {author} {\bibfnamefont {T.~J.}\ \bibnamefont
			{Sejnowski}},\ }\href
	{http://science.sciencemag.org/content/sci/301/5641/1870.full.pdf} {\bibfield
		{journal} {\bibinfo  {journal} {Science}\ }\textbf {\bibinfo {volume}
			{108}},\ \bibinfo {pages} {22} (\bibinfo {year} {2002})}\BibitemShut
	{NoStop}%
	\bibitem [{\citenamefont {Niven}\ and\ \citenamefont
		{Laughlin}(2008)}]{Niven2008}%
	\BibitemOpen
	\bibfield  {author} {\bibinfo {author} {\bibfnamefont {J.~E.}\ \bibnamefont
			{Niven}}\ and\ \bibinfo {author} {\bibfnamefont {S.~B.}\ \bibnamefont
			{Laughlin}},\ }\href {\doibase 10.1242/jeb.017574} {\bibfield  {journal}
		{\bibinfo  {journal} {Journal of Experimental Biology}\ } (\bibinfo {year}
		{2008}),\ 10.1242/jeb.017574}\BibitemShut {NoStop}%
	\bibitem [{\citenamefont {Budd}\ and\ \citenamefont
		{Kisv{\'{a}}rday}(2012)}]{Budd2012}%
	\BibitemOpen
	\bibfield  {author} {\bibinfo {author} {\bibfnamefont {J.~M.~L.}\
			\bibnamefont {Budd}}\ and\ \bibinfo {author} {\bibfnamefont {Z.~F.}\
			\bibnamefont {Kisv{\'{a}}rday}},\ }\href {\doibase 10.3389/fnana.2012.00042}
	{\bibfield  {journal} {\bibinfo  {journal} {Frontiers in Neuroanatomy}\
		}\textbf {\bibinfo {volume} {6}} (\bibinfo {year} {2012}),\
		10.3389/fnana.2012.00042}\BibitemShut {NoStop}%
	\bibitem [{\citenamefont {Cherniak}(1994)}]{Cherniak1994}%
	\BibitemOpen
	\bibfield  {author} {\bibinfo {author} {\bibfnamefont {C.}~\bibnamefont
			{Cherniak}},\ }\href
	{http://www.jneurosci.org/content/jneuro/14/4/2418.full.pdf} {\bibfield
		{journal} {\bibinfo  {journal} {The Journal of Neuroscience}\ }\textbf
		{\bibinfo {volume} {14}},\ \bibinfo {pages} {2418} (\bibinfo {year}
		{1994})}\BibitemShut {NoStop}%
	\bibitem [{\citenamefont {Cherniak}\ \emph {et~al.}(2010)\citenamefont
		{Cherniak}, \citenamefont {Mokhtarzada},\ and\ \citenamefont
		{Rodriguez-Esteban}}]{Cherniak2010}%
	\BibitemOpen
	\bibfield  {author} {\bibinfo {author} {\bibfnamefont {C.}~\bibnamefont
			{Cherniak}}, \bibinfo {author} {\bibfnamefont {Z.}~\bibnamefont
			{Mokhtarzada}}, \ and\ \bibinfo {author} {\bibfnamefont {R.}~\bibnamefont
			{Rodriguez-Esteban}},\ }\href {\doibase 10.1016/B0-12-370878-8/00100-2}
	{\bibfield  {journal} {\bibinfo  {journal} {Evolution of Nervous Systems}\
		}\textbf {\bibinfo {volume} {1}},\ \bibinfo {pages} {269} (\bibinfo {year}
		{2010})}\BibitemShut {NoStop}%
	\bibitem [{\citenamefont {Raj}\ and\ \citenamefont {Chen}(2011)}]{Raj2011}%
	\BibitemOpen
	\bibfield  {author} {\bibinfo {author} {\bibfnamefont {A.}~\bibnamefont
			{Raj}}\ and\ \bibinfo {author} {\bibfnamefont {Y.-H.}\ \bibnamefont {Chen}},\
	}\href {\doibase 10.1371/} {\bibfield  {journal} {\bibinfo  {journal} {PLoS
				ONE}\ }\textbf {\bibinfo {volume} {6}} (\bibinfo {year} {2011}),\
		10.1371/}\BibitemShut {NoStop}%
	\bibitem [{\citenamefont {{Ria Ercsey-Ravasz}}\ \emph
		{et~al.}(2013)\citenamefont {{Ria Ercsey-Ravasz}}, \citenamefont {Markov},
		\citenamefont {Lamy}, \citenamefont {{Van Essen}}, \citenamefont {Knoblauch},
		\citenamefont {Toroczkai},\ and\ \citenamefont
		{Kennedy}}]{RiaErcsey-Ravasz2013}%
	\BibitemOpen
	\bibfield  {author} {\bibinfo {author} {\bibfnamefont {M.}~\bibnamefont {{Ria
					Ercsey-Ravasz}}}, \bibinfo {author} {\bibfnamefont {N.~T.}\ \bibnamefont
			{Markov}}, \bibinfo {author} {\bibfnamefont {C.}~\bibnamefont {Lamy}},
		\bibinfo {author} {\bibfnamefont {D.~C.}\ \bibnamefont {{Van Essen}}},
		\bibinfo {author} {\bibfnamefont {K.}~\bibnamefont {Knoblauch}}, \bibinfo
		{author} {\bibfnamefont {Z.~N.}\ \bibnamefont {Toroczkai}}, \ and\ \bibinfo
		{author} {\bibfnamefont {H.}~\bibnamefont {Kennedy}},\ }\href {\doibase
		10.1016/j.neuron.2013.07.036} {\bibfield  {journal} {\bibinfo  {journal}
			{Neuron}\ }\textbf {\bibinfo {volume} {80}},\ \bibinfo {pages} {184}
		(\bibinfo {year} {2013})}\BibitemShut {NoStop}%
	\bibitem [{\citenamefont {Song}\ \emph {et~al.}(2014)\citenamefont {Song},
		\citenamefont {Kennedy},\ and\ \citenamefont {Wang}}]{Song2014}%
	\BibitemOpen
	\bibfield  {author} {\bibinfo {author} {\bibfnamefont {H.~F.}\ \bibnamefont
			{Song}}, \bibinfo {author} {\bibfnamefont {H.}~\bibnamefont {Kennedy}}, \
		and\ \bibinfo {author} {\bibfnamefont {X.-J.}\ \bibnamefont {Wang}},\ }\href
	{\doibase 10.1073/pnas.1414153111} {\bibfield  {journal} {\bibinfo  {journal}
			{Proceedings of the National Academy of Sciences}\ }\textbf {\bibinfo
			{volume} {111}},\ \bibinfo {pages} {16580} (\bibinfo {year}
		{2014})}\BibitemShut {NoStop}%
	\bibitem [{\citenamefont {Bassett}\ \emph {et~al.}(2010)\citenamefont
		{Bassett}, \citenamefont {Greenfield}, \citenamefont {Meyer-Lindenberg},
		\citenamefont {Weinberger}, \citenamefont {Moore},\ and\ \citenamefont
		{Bullmore}}]{Bassett2010}%
	\BibitemOpen
	\bibfield  {author} {\bibinfo {author} {\bibfnamefont {D.~S.}\ \bibnamefont
			{Bassett}}, \bibinfo {author} {\bibfnamefont {D.~L.}\ \bibnamefont
			{Greenfield}}, \bibinfo {author} {\bibfnamefont {A.}~\bibnamefont
			{Meyer-Lindenberg}}, \bibinfo {author} {\bibfnamefont {D.~R.}\ \bibnamefont
			{Weinberger}}, \bibinfo {author} {\bibfnamefont {S.~W.}\ \bibnamefont
			{Moore}}, \ and\ \bibinfo {author} {\bibfnamefont {E.~T.}\ \bibnamefont
			{Bullmore}},\ }\href {\doibase 10.1371/journal.pcbi.1000748} {\bibfield
		{journal} {\bibinfo  {journal} {PLoS Comput Biol}\ }\textbf {\bibinfo
			{volume} {6}} (\bibinfo {year} {2010}),\
		10.1371/journal.pcbi.1000748}\BibitemShut {NoStop}%
	\bibitem [{\citenamefont {Kaiser}\ and\ \citenamefont
		{Hilgetag}(2006)}]{Kaiser2006}%
	\BibitemOpen
	\bibfield  {author} {\bibinfo {author} {\bibfnamefont {M.}~\bibnamefont
			{Kaiser}}\ and\ \bibinfo {author} {\bibfnamefont {C.~C.}\ \bibnamefont
			{Hilgetag}},\ }\href {\doibase 10.1371/journal.pcbi.0020095} {\bibfield
		{journal} {\bibinfo  {journal} {PLoS Computational Biology}\ }\textbf
		{\bibinfo {volume} {2}},\ \bibinfo {pages} {0805} (\bibinfo {year} {2006})},\
	\Eprint {http://arxiv.org/abs/0607034} {arXiv:0607034 [q-bio]} \BibitemShut
	{NoStop}%
	\bibitem [{\citenamefont {Zalesky}\ \emph {et~al.}(2012)\citenamefont
		{Zalesky}, \citenamefont {Fornito}, \citenamefont {Egan}, \citenamefont
		{Pantelis},\ and\ \citenamefont {Bullmore}}]{Zalesky2012}%
	\BibitemOpen
	\bibfield  {author} {\bibinfo {author} {\bibfnamefont {A.}~\bibnamefont
			{Zalesky}}, \bibinfo {author} {\bibfnamefont {A.}~\bibnamefont {Fornito}},
		\bibinfo {author} {\bibfnamefont {G.~F.}\ \bibnamefont {Egan}}, \bibinfo
		{author} {\bibfnamefont {C.}~\bibnamefont {Pantelis}}, \ and\ \bibinfo
		{author} {\bibfnamefont {E.~T.}\ \bibnamefont {Bullmore}},\ }\href {\doibase
		10.1002/hbm.21379} {\bibfield  {journal} {\bibinfo  {journal} {Human Brain
				Mapping}\ }\textbf {\bibinfo {volume} {33}},\ \bibinfo {pages} {2535}
		(\bibinfo {year} {2012})},\ \Eprint {http://arxiv.org/abs/arXiv:1011.1669v3}
	{arXiv:arXiv:1011.1669v3} \BibitemShut {NoStop}%
	\bibitem [{\citenamefont {Avena-Koenigsberger}\ \emph
		{et~al.}(2018)\citenamefont {Avena-Koenigsberger}, \citenamefont {Misic},\
		and\ \citenamefont {Sporns}}]{Avena-Koenigsberger2018}%
	\BibitemOpen
	\bibfield  {author} {\bibinfo {author} {\bibfnamefont {A.}~\bibnamefont
			{Avena-Koenigsberger}}, \bibinfo {author} {\bibfnamefont {B.}~\bibnamefont
			{Misic}}, \ and\ \bibinfo {author} {\bibfnamefont {O.}~\bibnamefont
			{Sporns}},\ }\href {\doibase 10.1038/nrn.2017.149} {\bibfield  {journal}
		{\bibinfo  {journal} {Nature Publishing Group}\ }\textbf {\bibinfo {volume}
			{19}} (\bibinfo {year} {2018}),\ 10.1038/nrn.2017.149}\BibitemShut {NoStop}%
	\bibitem [{\citenamefont {Chen}\ \emph {et~al.}(2013)\citenamefont {Chen},
		\citenamefont {Wang}, \citenamefont {Hilgetag},\ and\ \citenamefont
		{Zhou}}]{Chen2013}%
	\BibitemOpen
	\bibfield  {author} {\bibinfo {author} {\bibfnamefont {Y.}~\bibnamefont
			{Chen}}, \bibinfo {author} {\bibfnamefont {S.}~\bibnamefont {Wang}}, \bibinfo
		{author} {\bibfnamefont {C.~C.}\ \bibnamefont {Hilgetag}}, \ and\ \bibinfo
		{author} {\bibfnamefont {C.}~\bibnamefont {Zhou}},\ }\href {\doibase
		10.1371/journal.pcbi.1002937} {\bibfield  {journal} {\bibinfo  {journal}
			{PLoS Comput Biol}\ }\textbf {\bibinfo {volume} {9}} (\bibinfo {year}
		{2013}),\ 10.1371/journal.pcbi.1002937}\BibitemShut {NoStop}%
	\bibitem [{\citenamefont {Nicosia}\ \emph {et~al.}(2013)\citenamefont
		{Nicosia}, \citenamefont {Vertes}, \citenamefont {Schafer}, \citenamefont
		{Latora},\ and\ \citenamefont {Bullmore}}]{Nicosia2013}%
	\BibitemOpen
	\bibfield  {author} {\bibinfo {author} {\bibfnamefont {V.}~\bibnamefont
			{Nicosia}}, \bibinfo {author} {\bibfnamefont {P.~E.}\ \bibnamefont {Vertes}},
		\bibinfo {author} {\bibfnamefont {W.~R.}\ \bibnamefont {Schafer}}, \bibinfo
		{author} {\bibfnamefont {V.}~\bibnamefont {Latora}}, \ and\ \bibinfo {author}
		{\bibfnamefont {E.~T.}\ \bibnamefont {Bullmore}},\ }\href {\doibase
		10.1073/pnas.1300753110} {\bibfield  {journal} {\bibinfo  {journal}
			{Proceedings of the National Academy of Sciences}\ }\textbf {\bibinfo
			{volume} {110}},\ \bibinfo {pages} {7880} (\bibinfo {year} {2013})},\ \Eprint
	{http://arxiv.org/abs/1304.7364} {arXiv:1304.7364} \BibitemShut {NoStop}%
	\bibitem [{\citenamefont {Kaiser}(2017)}]{Kaiser2017}%
	\BibitemOpen
	\bibfield  {author} {\bibinfo {author} {\bibfnamefont {M.}~\bibnamefont
			{Kaiser}},\ }\href {\doibase 10.1016/j.tics.2017.05.010} {\bibfield
		{journal} {\bibinfo  {journal} {Trends in Cognitive Sciences}\ }\textbf
		{\bibinfo {volume} {21}},\ \bibinfo {pages} {703} (\bibinfo {year}
		{2017})}\BibitemShut {NoStop}%
	\bibitem [{\citenamefont {Sperry}\ \emph {et~al.}(2017)\citenamefont {Sperry},
		\citenamefont {Telesford}, \citenamefont {Klimm},\ and\ \citenamefont
		{Bassett}}]{Sperry2017}%
	\BibitemOpen
	\bibfield  {author} {\bibinfo {author} {\bibfnamefont {M.~M.}\ \bibnamefont
			{Sperry}}, \bibinfo {author} {\bibfnamefont {Q.~K.}\ \bibnamefont
			{Telesford}}, \bibinfo {author} {\bibfnamefont {F.}~\bibnamefont {Klimm}}, \
		and\ \bibinfo {author} {\bibfnamefont {D.~S.}\ \bibnamefont {Bassett}},\
	}\href {\doibase 10.1093/comnet/cnw010} {\bibfield  {journal} {\bibinfo
			{journal} {Journal of Complex Networks}\ }\textbf {\bibinfo {volume} {5}},\
		\bibinfo {pages} {199} (\bibinfo {year} {2017})}\BibitemShut {NoStop}%
	\bibitem [{\citenamefont {Klimm}\ \emph {et~al.}(2014)\citenamefont {Klimm},
		\citenamefont {Bassett}, \citenamefont {Carlson}, \citenamefont {Mucha},\
		and\ \citenamefont {Hilgetag}}]{Klimm2014}%
	\BibitemOpen
	\bibfield  {author} {\bibinfo {author} {\bibfnamefont {F.}~\bibnamefont
			{Klimm}}, \bibinfo {author} {\bibfnamefont {D.~S.}\ \bibnamefont {Bassett}},
		\bibinfo {author} {\bibfnamefont {J.~M.}\ \bibnamefont {Carlson}}, \bibinfo
		{author} {\bibfnamefont {P.~J.}\ \bibnamefont {Mucha}}, \ and\ \bibinfo
		{author} {\bibfnamefont {C.~C.}\ \bibnamefont {Hilgetag}},\ }\href {\doibase
		10.1371/journal.pcbi.1003491} {\bibfield  {journal} {\bibinfo  {journal}
			{PLoS Comput Biol}\ }\textbf {\bibinfo {volume} {10}} (\bibinfo {year}
		{2014}),\ 10.1371/journal.pcbi.1003491}\BibitemShut {NoStop}%
	\bibitem [{\citenamefont {Pineda-Pardo}\ \emph {et~al.}(2015)\citenamefont
		{Pineda-Pardo}, \citenamefont {Martinez}, \citenamefont {Solana},
		\citenamefont {Hernandez-Tamames}, \citenamefont {Colom},\ and\ \citenamefont
		{del Pozo}}]{pineda2015disparate}%
	\BibitemOpen
	\bibfield  {author} {\bibinfo {author} {\bibfnamefont {J.~A.}\ \bibnamefont
			{Pineda-Pardo}}, \bibinfo {author} {\bibfnamefont {K.}~\bibnamefont
			{Martinez}}, \bibinfo {author} {\bibfnamefont {A.~B.}\ \bibnamefont
			{Solana}}, \bibinfo {author} {\bibfnamefont {J.~A.}\ \bibnamefont
			{Hernandez-Tamames}}, \bibinfo {author} {\bibfnamefont {R.}~\bibnamefont
			{Colom}}, \ and\ \bibinfo {author} {\bibfnamefont {F.}~\bibnamefont {del
				Pozo}},\ }\href@noop {} {\bibfield  {journal} {\bibinfo  {journal} {Brain
				Topogr}\ }\textbf {\bibinfo {volume} {28}},\ \bibinfo {pages} {187} (\bibinfo
		{year} {2015})}\BibitemShut {NoStop}%
	\bibitem [{\citenamefont {Sadovsky}\ and\ \citenamefont
		{MacLean}(2014)}]{sadovsky2014mouse}%
	\BibitemOpen
	\bibfield  {author} {\bibinfo {author} {\bibfnamefont {A.~J.}\ \bibnamefont
			{Sadovsky}}\ and\ \bibinfo {author} {\bibfnamefont {J.~N.}\ \bibnamefont
			{MacLean}},\ }\href@noop {} {\bibfield  {journal} {\bibinfo  {journal} {J
				Neurosci}\ }\textbf {\bibinfo {volume} {34}},\ \bibinfo {pages} {7769}
		(\bibinfo {year} {2014})}\BibitemShut {NoStop}%
	\bibitem [{\citenamefont {Christie}\ and\ \citenamefont
		{Stroobandt}(2000)}]{christie2000interpretation}%
	\BibitemOpen
	\bibfield  {author} {\bibinfo {author} {\bibfnamefont {P.}~\bibnamefont
			{Christie}}\ and\ \bibinfo {author} {\bibfnamefont {D.}~\bibnamefont
			{Stroobandt}},\ }\href@noop {} {\bibfield  {journal} {\bibinfo  {journal}
			{IEEE Trans. on VLSI Systems}\ }\textbf {\bibinfo {volume} {8}},\ \bibinfo
		{pages} {639} (\bibinfo {year} {2000})}\BibitemShut {NoStop}%
	\bibitem [{\citenamefont {Alcalde~Cuesta}\ \emph {et~al.}(2017)\citenamefont
		{Alcalde~Cuesta}, \citenamefont {Gonzalez~Sequeiros},\ and\ \citenamefont
		{Lozano~Rojo}}]{alcalde2017method}%
	\BibitemOpen
	\bibfield  {author} {\bibinfo {author} {\bibfnamefont {F.}~\bibnamefont
			{Alcalde~Cuesta}}, \bibinfo {author} {\bibfnamefont {P.}~\bibnamefont
			{Gonzalez~Sequeiros}}, \ and\ \bibinfo {author} {\bibfnamefont
			{A.}~\bibnamefont {Lozano~Rojo}},\ }\href@noop {} {\bibfield  {journal}
		{\bibinfo  {journal} {Sci Rep}\ }\textbf {\bibinfo {volume} {7}},\ \bibinfo
		{pages} {5378} (\bibinfo {year} {2017})}\BibitemShut {NoStop}%
	\bibitem [{\citenamefont {Betzel}\ and\ \citenamefont
		{Bassett}(2018)}]{Betzel201720186}%
	\BibitemOpen
	\bibfield  {author} {\bibinfo {author} {\bibfnamefont {R.~F.}\ \bibnamefont
			{Betzel}}\ and\ \bibinfo {author} {\bibfnamefont {D.~S.}\ \bibnamefont
			{Bassett}},\ }\href {\doibase 10.1073/pnas.1720186115} {\bibfield  {journal}
		{\bibinfo  {journal} {Proceedings of the National Academy of Sciences}\ }
		(\bibinfo {year} {2018}),\ 10.1073/pnas.1720186115},\ \Eprint
	{http://arxiv.org/abs/http://www.pnas.org/content/early/2018/05/07/1720186115.full.pdf}
	{http://www.pnas.org/content/early/2018/05/07/1720186115.full.pdf}
	\BibitemShut {NoStop}%
	\bibitem [{\citenamefont {Henderson}\ and\ \citenamefont
		{Robinson}(2011)}]{Henderson2011}%
	\BibitemOpen
	\bibfield  {author} {\bibinfo {author} {\bibfnamefont {J.~A.}\ \bibnamefont
			{Henderson}}\ and\ \bibinfo {author} {\bibfnamefont {P.~A.}\ \bibnamefont
			{Robinson}},\ }\href {\doibase 10.1103/PhysRevLett.107.018102} {\bibfield
		{journal} {\bibinfo  {journal} {Physical Review Letters}\ }\textbf {\bibinfo
			{volume} {107}} (\bibinfo {year} {2011}),\
		10.1103/PhysRevLett.107.018102}\BibitemShut {NoStop}%
	\bibitem [{\citenamefont {Park}\ and\ \citenamefont
		{Friston}(2013)}]{Park2013}%
	\BibitemOpen
	\bibfield  {author} {\bibinfo {author} {\bibfnamefont {H.~J.}\ \bibnamefont
			{Park}}\ and\ \bibinfo {author} {\bibfnamefont {K.}~\bibnamefont {Friston}},\
	}\href {\doibase 10.1126/science.1238411} {\bibfield  {journal} {\bibinfo
			{journal} {Science}\ }\textbf {\bibinfo {volume} {342}} (\bibinfo {year}
		{2013}),\ 10.1126/science.1238411}\BibitemShut {NoStop}%
	\bibitem [{\citenamefont {Sporns}\ \emph {et~al.}(2004)\citenamefont {Sporns},
		\citenamefont {Chialvo}, \citenamefont {Kaiser},\ and\ \citenamefont
		{Hilgetag}}]{Sporns2004}%
	\BibitemOpen
	\bibfield  {author} {\bibinfo {author} {\bibfnamefont {O.}~\bibnamefont
			{Sporns}}, \bibinfo {author} {\bibfnamefont {D.~R.}\ \bibnamefont {Chialvo}},
		\bibinfo {author} {\bibfnamefont {M.}~\bibnamefont {Kaiser}}, \ and\ \bibinfo
		{author} {\bibfnamefont {C.~C.}\ \bibnamefont {Hilgetag}},\ }\href {\doibase
		10.1016/j.tics.2004.07.008} {\bibfield  {journal} {\bibinfo  {journal}
			{Trends in Cognitive Sciences}\ }\textbf {\bibinfo {volume} {8}},\ \bibinfo
		{pages} {418} (\bibinfo {year} {2004})}\BibitemShut {NoStop}%
	\bibitem [{\citenamefont {van~den Heuvel}\ \emph {et~al.}(2012)\citenamefont
		{van~den Heuvel}, \citenamefont {Kahn}, \citenamefont {Goni},\ and\
		\citenamefont {Sporns}}]{VandenHeuvel2012}%
	\BibitemOpen
	\bibfield  {author} {\bibinfo {author} {\bibfnamefont {M.~P.}\ \bibnamefont
			{van~den Heuvel}}, \bibinfo {author} {\bibfnamefont {R.~S.}\ \bibnamefont
			{Kahn}}, \bibinfo {author} {\bibfnamefont {J.}~\bibnamefont {Goni}}, \ and\
		\bibinfo {author} {\bibfnamefont {O.}~\bibnamefont {Sporns}},\ }\href
	{\doibase 10.1073/pnas.1203593109} {\bibfield  {journal} {\bibinfo  {journal}
			{Proceedings of the National Academy of Sciences}\ }\textbf {\bibinfo
			{volume} {109}},\ \bibinfo {pages} {11372} (\bibinfo {year}
		{2012})}\BibitemShut {NoStop}%
	\bibitem [{\citenamefont {Seidlitz}\ \emph {et~al.}(2018)\citenamefont
		{Seidlitz}, \citenamefont {V{\'{a}}{\v{s}}a}, \citenamefont {Shinn},
		\citenamefont {Romero-Garcia}, \citenamefont {Whitaker}, \citenamefont
		{V{\'{e}}rtes}, \citenamefont {Wagstyl}, \citenamefont {{Kirkpatrick
				Reardon}}, \citenamefont {Clasen}, \citenamefont {Liu}, \citenamefont
		{Messinger}, \citenamefont {Leopold}, \citenamefont {Fonagy}, \citenamefont
		{Dolan}, \citenamefont {Jones}, \citenamefont {Goodyer}, \citenamefont
		{Raznahan},\ and\ \citenamefont {Bullmore}}]{Seidlitz2018}%
	\BibitemOpen
	\bibfield  {author} {\bibinfo {author} {\bibfnamefont {J.}~\bibnamefont
			{Seidlitz}}, \bibinfo {author} {\bibfnamefont {F.}~\bibnamefont
			{V{\'{a}}{\v{s}}a}}, \bibinfo {author} {\bibfnamefont {M.}~\bibnamefont
			{Shinn}}, \bibinfo {author} {\bibfnamefont {R.}~\bibnamefont
			{Romero-Garcia}}, \bibinfo {author} {\bibfnamefont {K.~J.}\ \bibnamefont
			{Whitaker}}, \bibinfo {author} {\bibfnamefont {P.~E.}\ \bibnamefont
			{V{\'{e}}rtes}}, \bibinfo {author} {\bibfnamefont {K.}~\bibnamefont
			{Wagstyl}}, \bibinfo {author} {\bibfnamefont {P.}~\bibnamefont {{Kirkpatrick
					Reardon}}}, \bibinfo {author} {\bibfnamefont {L.}~\bibnamefont {Clasen}},
		\bibinfo {author} {\bibfnamefont {S.}~\bibnamefont {Liu}}, \bibinfo {author}
		{\bibfnamefont {A.}~\bibnamefont {Messinger}}, \bibinfo {author}
		{\bibfnamefont {D.~A.}\ \bibnamefont {Leopold}}, \bibinfo {author}
		{\bibfnamefont {P.}~\bibnamefont {Fonagy}}, \bibinfo {author} {\bibfnamefont
			{R.~J.}\ \bibnamefont {Dolan}}, \bibinfo {author} {\bibfnamefont {P.~B.}\
			\bibnamefont {Jones}}, \bibinfo {author} {\bibfnamefont {I.~M.}\ \bibnamefont
			{Goodyer}}, \bibinfo {author} {\bibfnamefont {A.}~\bibnamefont {Raznahan}}, \
		and\ \bibinfo {author} {\bibfnamefont {E.~T.}\ \bibnamefont {Bullmore}},\
	}\href {\doibase 10.1016/j.neuron.2017.11.039} {\bibfield  {journal}
		{\bibinfo  {journal} {Neuron}\ }\textbf {\bibinfo {volume} {97}},\ \bibinfo
		{pages} {231} (\bibinfo {year} {2018})}\BibitemShut {NoStop}%
	\bibitem [{\citenamefont {Varier}\ and\ \citenamefont
		{Kaiser}(2011)}]{Varier2011}%
	\BibitemOpen
	\bibfield  {author} {\bibinfo {author} {\bibfnamefont {S.}~\bibnamefont
			{Varier}}\ and\ \bibinfo {author} {\bibfnamefont {M.}~\bibnamefont
			{Kaiser}},\ }\href {\doibase 10.1371/journal.pcbi.1001044} {\bibfield
		{journal} {\bibinfo  {journal} {PLoS Comput Biol}\ }\textbf {\bibinfo
			{volume} {7}} (\bibinfo {year} {2011}),\
		10.1371/journal.pcbi.1001044}\BibitemShut {NoStop}%
	\bibitem [{\citenamefont {Hilgetag}\ and\ \citenamefont
		{Goulas}(2016)}]{hilgetag2016brain}%
	\BibitemOpen
	\bibfield  {author} {\bibinfo {author} {\bibfnamefont {C.~C.}\ \bibnamefont
			{Hilgetag}}\ and\ \bibinfo {author} {\bibfnamefont {A.}~\bibnamefont
			{Goulas}},\ }\href@noop {} {\bibfield  {journal} {\bibinfo  {journal} {Brain
				Struct Funct}\ }\textbf {\bibinfo {volume} {221}},\ \bibinfo {pages} {2361}
		(\bibinfo {year} {2016})}\BibitemShut {NoStop}%
	\bibitem [{\citenamefont {Perin}\ \emph {et~al.}(2011)\citenamefont {Perin},
		\citenamefont {Berger},\ and\ \citenamefont {Markram}}]{Perin2011}%
	\BibitemOpen
	\bibfield  {author} {\bibinfo {author} {\bibfnamefont {R.}~\bibnamefont
			{Perin}}, \bibinfo {author} {\bibfnamefont {T.~K.}\ \bibnamefont {Berger}}, \
		and\ \bibinfo {author} {\bibfnamefont {H.}~\bibnamefont {Markram}},\ }\href
	{\doibase 10.1073/pnas.1016051108} {\bibfield  {journal} {\bibinfo  {journal}
			{Proceedings of the National Academy of Sciences}\ }\textbf {\bibinfo
			{volume} {108}},\ \bibinfo {pages} {5419} (\bibinfo {year} {2011})},\ \Eprint
	{http://arxiv.org/abs/arXiv:1408.1149} {arXiv:arXiv:1408.1149} \BibitemShut
	{NoStop}%
	\bibitem [{\citenamefont {Sizemore}\ \emph
		{et~al.}(2017{\natexlab{a}})\citenamefont {Sizemore}, \citenamefont {Giusti},
		\citenamefont {Kahn}, \citenamefont {Vettel}, \citenamefont {Betzel},\ and\
		\citenamefont {Bassett}}]{Sizemore2017a}%
	\BibitemOpen
	\bibfield  {author} {\bibinfo {author} {\bibfnamefont {A.~E.}\ \bibnamefont
			{Sizemore}}, \bibinfo {author} {\bibfnamefont {C.}~\bibnamefont {Giusti}},
		\bibinfo {author} {\bibfnamefont {A.}~\bibnamefont {Kahn}}, \bibinfo {author}
		{\bibfnamefont {J.~M.}\ \bibnamefont {Vettel}}, \bibinfo {author}
		{\bibfnamefont {R.~F.}\ \bibnamefont {Betzel}}, \ and\ \bibinfo {author}
		{\bibfnamefont {D.~S.}\ \bibnamefont {Bassett}},\ }\href {\doibase
		10.1007/s10827-017-0672-6} {\bibfield  {journal} {\bibinfo  {journal}
			{Journal of Computational Neuroscience}\ }\textbf {\bibinfo {volume} {44}},\
		\bibinfo {pages} {1} (\bibinfo {year} {2017}{\natexlab{a}})},\ \Eprint
	{http://arxiv.org/abs/1608.03520} {arXiv:1608.03520} \BibitemShut {NoStop}%
	\bibitem [{\citenamefont {Kahle}(2018)}]{Kahle}%
	\BibitemOpen
	\bibfield  {author} {\bibinfo {author} {\bibfnamefont {M.}~\bibnamefont
			{Kahle}},\ }\href {http://math.stanford.edu/{~}mkahle/geometric.pdf}
	{\enquote {\bibinfo {title} {{Random Geometric Complexes}},}\ } (\bibinfo
	{year} {2018})\BibitemShut {NoStop}%
	\bibitem [{\citenamefont {Kahle}(2009)}]{Kahle2009}%
	\BibitemOpen
	\bibfield  {author} {\bibinfo {author} {\bibfnamefont {M.}~\bibnamefont
			{Kahle}},\ }\href {\doibase 10.1016/j.disc.2008.02.037} {\bibfield  {journal}
		{\bibinfo  {journal} {Discrete Mathematics}\ }\textbf {\bibinfo {volume}
			{309}},\ \bibinfo {pages} {1658} (\bibinfo {year} {2009})},\ \Eprint
	{http://arxiv.org/abs/0605536} {arXiv:0605536 [math]} \BibitemShut {NoStop}%
	\bibitem [{\citenamefont {Giusti}\ \emph {et~al.}(2015)\citenamefont {Giusti},
		\citenamefont {Pastalkova}, \citenamefont {Curto},\ and\ \citenamefont
		{Itskov}}]{Giusti}%
	\BibitemOpen
	\bibfield  {author} {\bibinfo {author} {\bibfnamefont {C.}~\bibnamefont
			{Giusti}}, \bibinfo {author} {\bibfnamefont {E.}~\bibnamefont {Pastalkova}},
		\bibinfo {author} {\bibfnamefont {C.}~\bibnamefont {Curto}}, \ and\ \bibinfo
		{author} {\bibfnamefont {V.}~\bibnamefont {Itskov}},\ }\href {\doibase
		10.1073/pnas.1506407112} {\bibfield  {journal} {\bibinfo  {journal} {Proc
				Natl Acad Sci U S A}\ }\textbf {\bibinfo {volume} {112}},\ \bibinfo {pages}
		{13455} (\bibinfo {year} {2015})},\ \Eprint {http://arxiv.org/abs/1502.06172}
	{arXiv:1502.06172} \BibitemShut {NoStop}%
	\bibitem [{\citenamefont {Watts}\ and\ \citenamefont
		{Strogatz}(1998)}]{Watts1998}%
	\BibitemOpen
	\bibfield  {author} {\bibinfo {author} {\bibfnamefont {D.~J.}\ \bibnamefont
			{Watts}}\ and\ \bibinfo {author} {\bibfnamefont {S.~H.}\ \bibnamefont
			{Strogatz}},\ }\href {\doibase 10.1038/30918} {\bibfield  {journal} {\bibinfo
			{journal} {Nature}\ }\textbf {\bibinfo {volume} {393}},\ \bibinfo {pages}
		{440} (\bibinfo {year} {1998})},\ \Eprint {http://arxiv.org/abs/0803.0939v1}
	{arXiv:0803.0939v1} \BibitemShut {NoStop}%
	\bibitem [{\citenamefont {Shih}\ \emph {et~al.}(2015)\citenamefont {Shih},
		\citenamefont {Sporns}, \citenamefont {Yuan}, \citenamefont {Su},
		\citenamefont {Lin}, \citenamefont {Chuang}, \citenamefont {Wang},
		\citenamefont {Lo}, \citenamefont {Greenspan},\ and\ \citenamefont
		{Chiang}}]{shih2015connectomics}%
	\BibitemOpen
	\bibfield  {author} {\bibinfo {author} {\bibfnamefont {C.~T.}\ \bibnamefont
			{Shih}}, \bibinfo {author} {\bibfnamefont {O.}~\bibnamefont {Sporns}},
		\bibinfo {author} {\bibfnamefont {S.~L.}\ \bibnamefont {Yuan}}, \bibinfo
		{author} {\bibfnamefont {T.~S.}\ \bibnamefont {Su}}, \bibinfo {author}
		{\bibfnamefont {Y.~J.}\ \bibnamefont {Lin}}, \bibinfo {author} {\bibfnamefont
			{C.~C.}\ \bibnamefont {Chuang}}, \bibinfo {author} {\bibfnamefont {T.~Y.}\
			\bibnamefont {Wang}}, \bibinfo {author} {\bibfnamefont {C.~C.}\ \bibnamefont
			{Lo}}, \bibinfo {author} {\bibfnamefont {R.~J.}\ \bibnamefont {Greenspan}}, \
		and\ \bibinfo {author} {\bibfnamefont {A.~S.}\ \bibnamefont {Chiang}},\
	}\href@noop {} {\bibfield  {journal} {\bibinfo  {journal} {Curr Biol}\
		}\textbf {\bibinfo {volume} {25}},\ \bibinfo {pages} {1249} (\bibinfo {year}
		{2015})}\BibitemShut {NoStop}%
	\bibitem [{\citenamefont {Kaiser}\ and\ \citenamefont
		{Hilgetag}(2004)}]{kaiser2004spatial}%
	\BibitemOpen
	\bibfield  {author} {\bibinfo {author} {\bibfnamefont {M.}~\bibnamefont
			{Kaiser}}\ and\ \bibinfo {author} {\bibfnamefont {C.~C.}\ \bibnamefont
			{Hilgetag}},\ }\href@noop {} {\bibfield  {journal} {\bibinfo  {journal} {Phys
				Rev E Stat Nonlin Soft Matter Phys}\ }\textbf {\bibinfo {volume} {69}},\
		\bibinfo {pages} {036103} (\bibinfo {year} {2004})}\BibitemShut {NoStop}%
	\bibitem [{\citenamefont {Bassett}\ and\ \citenamefont
		{Bullmore}(2017)}]{Bassett2017b}%
	\BibitemOpen
	\bibfield  {author} {\bibinfo {author} {\bibfnamefont {D.~S.}\ \bibnamefont
			{Bassett}}\ and\ \bibinfo {author} {\bibfnamefont {E.~T.}\ \bibnamefont
			{Bullmore}},\ }\href {\doibase 10.1177/1073858416667720} {\bibfield
		{journal} {\bibinfo  {journal} {Neuroscientist}\ }\textbf {\bibinfo {volume}
			{23}},\ \bibinfo {pages} {499} (\bibinfo {year} {2017})},\ \Eprint
	{http://arxiv.org/abs/1608.05665} {arXiv:1608.05665} \BibitemShut {NoStop}%
	\bibitem [{\citenamefont {Avena-Koenigsberger}\ \emph
		{et~al.}(2014)\citenamefont {Avena-Koenigsberger}, \citenamefont {Goni},
		\citenamefont {Sole},\ and\ \citenamefont
		{Sporns}}]{Avena-Koenigsberger2014}%
	\BibitemOpen
	\bibfield  {author} {\bibinfo {author} {\bibfnamefont {A.}~\bibnamefont
			{Avena-Koenigsberger}}, \bibinfo {author} {\bibfnamefont {J.}~\bibnamefont
			{Goni}}, \bibinfo {author} {\bibfnamefont {R.}~\bibnamefont {Sole}}, \ and\
		\bibinfo {author} {\bibfnamefont {O.}~\bibnamefont {Sporns}},\ }\href
	{\doibase 10.1098/rsif.2014.0881} {\bibfield  {journal} {\bibinfo  {journal}
			{Journal of The Royal Society Interface}\ }\textbf {\bibinfo {volume} {12}},\
		\bibinfo {pages} {20140881} (\bibinfo {year} {2014})}\BibitemShut {NoStop}%
	\bibitem [{\citenamefont {Scholtens}\ \emph {et~al.}(2014)\citenamefont
		{Scholtens}, \citenamefont {Schmidt}, \citenamefont {de~Reus},\ and\
		\citenamefont {van~den Heuvel}}]{scholtens2014linking}%
	\BibitemOpen
	\bibfield  {author} {\bibinfo {author} {\bibfnamefont {L.~H.}\ \bibnamefont
			{Scholtens}}, \bibinfo {author} {\bibfnamefont {R.}~\bibnamefont {Schmidt}},
		\bibinfo {author} {\bibfnamefont {M.~A.}\ \bibnamefont {de~Reus}}, \ and\
		\bibinfo {author} {\bibfnamefont {M.~P.}\ \bibnamefont {van~den Heuvel}},\
	}\href@noop {} {\bibfield  {journal} {\bibinfo  {journal} {J Neurosci}\
		}\textbf {\bibinfo {volume} {34}},\ \bibinfo {pages} {12192} (\bibinfo {year}
		{2014})}\BibitemShut {NoStop}%
	\bibitem [{\citenamefont {van~den Heuvel}\ \emph {et~al.}(2015)\citenamefont
		{van~den Heuvel}, \citenamefont {Scholtens}, \citenamefont {Feldman~Barrett},
		\citenamefont {Hilgetag},\ and\ \citenamefont
		{de~Reus}}]{heuvel2015bridging}%
	\BibitemOpen
	\bibfield  {author} {\bibinfo {author} {\bibfnamefont {M.~P.}\ \bibnamefont
			{van~den Heuvel}}, \bibinfo {author} {\bibfnamefont {L.~H.}\ \bibnamefont
			{Scholtens}}, \bibinfo {author} {\bibfnamefont {L.}~\bibnamefont
			{Feldman~Barrett}}, \bibinfo {author} {\bibfnamefont {C.~C.}\ \bibnamefont
			{Hilgetag}}, \ and\ \bibinfo {author} {\bibfnamefont {M.~A.}\ \bibnamefont
			{de~Reus}},\ }\href@noop {} {\bibfield  {journal} {\bibinfo  {journal} {J
				Neurosci}\ }\textbf {\bibinfo {volume} {35}},\ \bibinfo {pages} {13943}
		(\bibinfo {year} {2015})}\BibitemShut {NoStop}%
	\bibitem [{\citenamefont {Jbabdi}\ \emph {et~al.}(2013)\citenamefont {Jbabdi},
		\citenamefont {Sotiropoulos},\ and\ \citenamefont {Behrens}}]{Jbabdi2013}%
	\BibitemOpen
	\bibfield  {author} {\bibinfo {author} {\bibfnamefont {S.}~\bibnamefont
			{Jbabdi}}, \bibinfo {author} {\bibfnamefont {S.~N.}\ \bibnamefont
			{Sotiropoulos}}, \ and\ \bibinfo {author} {\bibfnamefont {T.~E.}\
			\bibnamefont {Behrens}},\ }\href {\doibase 10.1016/j.conb.2012.12.004}
	{\bibfield  {journal} {\bibinfo  {journal} {Current Opinion in Neurobiology}\
		}\textbf {\bibinfo {volume} {23}},\ \bibinfo {pages} {207} (\bibinfo {year}
		{2013})},\ \Eprint {http://arxiv.org/abs/NIHMS150003} {arXiv:NIHMS150003}
	\BibitemShut {NoStop}%
	\bibitem [{\citenamefont {Huntenburg}\ \emph {et~al.}(2018)\citenamefont
		{Huntenburg}, \citenamefont {Bazin},\ and\ \citenamefont
		{Margulies}}]{Huntenburg2018}%
	\BibitemOpen
	\bibfield  {author} {\bibinfo {author} {\bibfnamefont {J.~M.}\ \bibnamefont
			{Huntenburg}}, \bibinfo {author} {\bibfnamefont {P.~L.}\ \bibnamefont
			{Bazin}}, \ and\ \bibinfo {author} {\bibfnamefont {D.~S.}\ \bibnamefont
			{Margulies}},\ }\href {\doibase 10.1016/j.tics.2017.11.002} {\bibfield
		{journal} {\bibinfo  {journal} {Trends in Cognitive Sciences}\ }\textbf
		{\bibinfo {volume} {22}},\ \bibinfo {pages} {21} (\bibinfo {year}
		{2018})}\BibitemShut {NoStop}%
	\bibitem [{\citenamefont {Margulies}\ \emph {et~al.}(2016)\citenamefont
		{Margulies}, \citenamefont {Ghosh}, \citenamefont {Goulas}, \citenamefont
		{Falkiewicz}, \citenamefont {Huntenburg}, \citenamefont {Langs},
		\citenamefont {Bezgin}, \citenamefont {Eickhoff}, \citenamefont
		{Castellanos}, \citenamefont {Petrides}, \citenamefont {Jefferies},\ and\
		\citenamefont {Smallwood}}]{Margulies2016}%
	\BibitemOpen
	\bibfield  {author} {\bibinfo {author} {\bibfnamefont {D.~S.}\ \bibnamefont
			{Margulies}}, \bibinfo {author} {\bibfnamefont {S.~S.}\ \bibnamefont
			{Ghosh}}, \bibinfo {author} {\bibfnamefont {A.}~\bibnamefont {Goulas}},
		\bibinfo {author} {\bibfnamefont {M.}~\bibnamefont {Falkiewicz}}, \bibinfo
		{author} {\bibfnamefont {J.~M.}\ \bibnamefont {Huntenburg}}, \bibinfo
		{author} {\bibfnamefont {G.}~\bibnamefont {Langs}}, \bibinfo {author}
		{\bibfnamefont {G.}~\bibnamefont {Bezgin}}, \bibinfo {author} {\bibfnamefont
			{S.~B.}\ \bibnamefont {Eickhoff}}, \bibinfo {author} {\bibfnamefont {F.~X.}\
			\bibnamefont {Castellanos}}, \bibinfo {author} {\bibfnamefont
			{M.}~\bibnamefont {Petrides}}, \bibinfo {author} {\bibfnamefont
			{E.}~\bibnamefont {Jefferies}}, \ and\ \bibinfo {author} {\bibfnamefont
			{J.}~\bibnamefont {Smallwood}},\ }\href {\doibase 10.1073/pnas.1608282113}
	{\bibfield  {journal} {\bibinfo  {journal} {Proceedings of the National
				Academy of Sciences}\ }\textbf {\bibinfo {volume} {113}},\ \bibinfo {pages}
		{12574} (\bibinfo {year} {2016})},\ \Eprint
	{http://arxiv.org/abs/arXiv:1408.1149} {arXiv:arXiv:1408.1149} \BibitemShut
	{NoStop}%
	\bibitem [{\citenamefont {van~den Heuvel}\ and\ \citenamefont
		{Sporns}(2013)}]{VandenHeuvel2013}%
	\BibitemOpen
	\bibfield  {author} {\bibinfo {author} {\bibfnamefont {M.~P.}\ \bibnamefont
			{van~den Heuvel}}\ and\ \bibinfo {author} {\bibfnamefont {O.}~\bibnamefont
			{Sporns}},\ }\href {\doibase 10.1016/j.tics.2013.09.012} {\bibfield
		{journal} {\bibinfo  {journal} {Trends in Cognitive Sciences}\ }\textbf
		{\bibinfo {volume} {17}},\ \bibinfo {pages} {683} (\bibinfo {year}
		{2013})}\BibitemShut {NoStop}%
	\bibitem [{\citenamefont {Lu}\ \emph {et~al.}(2012)\citenamefont {Lu},
		\citenamefont {Zou}, \citenamefont {Gu}, \citenamefont {Raichle},
		\citenamefont {Stein},\ and\ \citenamefont {Yang}}]{Lu2012}%
	\BibitemOpen
	\bibfield  {author} {\bibinfo {author} {\bibfnamefont {H.}~\bibnamefont
			{Lu}}, \bibinfo {author} {\bibfnamefont {Q.}~\bibnamefont {Zou}}, \bibinfo
		{author} {\bibfnamefont {H.}~\bibnamefont {Gu}}, \bibinfo {author}
		{\bibfnamefont {M.~E.}\ \bibnamefont {Raichle}}, \bibinfo {author}
		{\bibfnamefont {E.~A.}\ \bibnamefont {Stein}}, \ and\ \bibinfo {author}
		{\bibfnamefont {Y.}~\bibnamefont {Yang}},\ }\href {\doibase
		10.1073/pnas.1200506109} {\bibfield  {journal} {\bibinfo  {journal}
			{Proceedings of the National Academy of Sciences}\ }\textbf {\bibinfo
			{volume} {109}},\ \bibinfo {pages} {3979} (\bibinfo {year}
		{2012})}\BibitemShut {NoStop}%
	\bibitem [{\citenamefont {Mitra}\ \emph
		{et~al.}(2015{\natexlab{a}})\citenamefont {Mitra}, \citenamefont {Snyder},
		\citenamefont {Blazey},\ and\ \citenamefont {Raichle}}]{Mitra2015c}%
	\BibitemOpen
	\bibfield  {author} {\bibinfo {author} {\bibfnamefont {A.}~\bibnamefont
			{Mitra}}, \bibinfo {author} {\bibfnamefont {A.~Z.}\ \bibnamefont {Snyder}},
		\bibinfo {author} {\bibfnamefont {T.}~\bibnamefont {Blazey}}, \ and\ \bibinfo
		{author} {\bibfnamefont {M.~E.}\ \bibnamefont {Raichle}},\ }\href
	{www.pnas.org/cgi/doi/10.1073/pnas.1523893113} {\bibfield  {journal}
		{\bibinfo  {journal} {Proc Natl Acad Sci}\ }\textbf {\bibinfo {volume}
			{17112}},\ \bibinfo {pages} {2235} (\bibinfo {year}
		{2015}{\natexlab{a}})}\BibitemShut {NoStop}%
	\bibitem [{\citenamefont {Raichle}(2015)}]{Raichle2015}%
	\BibitemOpen
	\bibfield  {author} {\bibinfo {author} {\bibfnamefont {M.~E.}\ \bibnamefont
			{Raichle}},\ }\href {\doibase 10.1146/annurev-neuro-071013-014030} {\bibfield
		{journal} {\bibinfo  {journal} {Annual Review of Neuroscience}\ }\textbf
		{\bibinfo {volume} {38}},\ \bibinfo {pages} {433} (\bibinfo {year} {2015})},\
	\Eprint {http://arxiv.org/abs/0402594v3} {arXiv:0402594v3 [arXiv:cond-mat]}
	\BibitemShut {NoStop}%
	\bibitem [{\citenamefont {Mitra}\ \emph {et~al.}(2014)\citenamefont {Mitra},
		\citenamefont {Snyder}, \citenamefont {Hacker},\ and\ \citenamefont
		{Raichle}}]{Mitra2014}%
	\BibitemOpen
	\bibfield  {author} {\bibinfo {author} {\bibfnamefont {A.}~\bibnamefont
			{Mitra}}, \bibinfo {author} {\bibfnamefont {A.~Z.}\ \bibnamefont {Snyder}},
		\bibinfo {author} {\bibfnamefont {C.~D.}\ \bibnamefont {Hacker}}, \ and\
		\bibinfo {author} {\bibfnamefont {M.~E.}\ \bibnamefont {Raichle}},\ }\href
	{\doibase 10.1152/jn.00804.2013} {\bibfield  {journal} {\bibinfo  {journal}
			{Journal of Neurophysiology}\ }\textbf {\bibinfo {volume} {111}},\ \bibinfo
		{pages} {2374} (\bibinfo {year} {2014})}\BibitemShut {NoStop}%
	\bibitem [{\citenamefont {Mitra}\ \emph {et~al.}(2018)\citenamefont {Mitra},
		\citenamefont {Kraft}, \citenamefont {Wright}, \citenamefont {Acland},
		\citenamefont {Snyder}, \citenamefont {Rosenthal}, \citenamefont
		{Czerniewski}, \citenamefont {Bauer}, \citenamefont {Snyder}, \citenamefont
		{Culver}, \citenamefont {Lee},\ and\ \citenamefont {Raichle}}]{MITRA2018}%
	\BibitemOpen
	\bibfield  {author} {\bibinfo {author} {\bibfnamefont {A.}~\bibnamefont
			{Mitra}}, \bibinfo {author} {\bibfnamefont {A.}~\bibnamefont {Kraft}},
		\bibinfo {author} {\bibfnamefont {P.}~\bibnamefont {Wright}}, \bibinfo
		{author} {\bibfnamefont {B.}~\bibnamefont {Acland}}, \bibinfo {author}
		{\bibfnamefont {A.~Z.}\ \bibnamefont {Snyder}}, \bibinfo {author}
		{\bibfnamefont {Z.}~\bibnamefont {Rosenthal}}, \bibinfo {author}
		{\bibfnamefont {L.}~\bibnamefont {Czerniewski}}, \bibinfo {author}
		{\bibfnamefont {A.}~\bibnamefont {Bauer}}, \bibinfo {author} {\bibfnamefont
			{L.}~\bibnamefont {Snyder}}, \bibinfo {author} {\bibfnamefont
			{J.}~\bibnamefont {Culver}}, \bibinfo {author} {\bibfnamefont {J.-M.}\
			\bibnamefont {Lee}}, \ and\ \bibinfo {author} {\bibfnamefont {M.~E.}\
			\bibnamefont {Raichle}},\ }\href {\doibase
		https://doi.org/10.1016/j.neuron.2018.03.015} {\bibfield  {journal} {\bibinfo
			{journal} {Neuron}\ } (\bibinfo {year} {2018}),\
		https://doi.org/10.1016/j.neuron.2018.03.015}\BibitemShut {NoStop}%
	\bibitem [{\citenamefont {Mitra}\ \emph
		{et~al.}(2015{\natexlab{b}})\citenamefont {Mitra}, \citenamefont {Snyder},
		\citenamefont {Tagliazucchi}, \citenamefont {Laufs},\ and\ \citenamefont
		{Raichle}}]{Mitra2015}%
	\BibitemOpen
	\bibfield  {author} {\bibinfo {author} {\bibfnamefont {A.}~\bibnamefont
			{Mitra}}, \bibinfo {author} {\bibfnamefont {A.~Z.}\ \bibnamefont {Snyder}},
		\bibinfo {author} {\bibfnamefont {E.}~\bibnamefont {Tagliazucchi}}, \bibinfo
		{author} {\bibfnamefont {H.}~\bibnamefont {Laufs}}, \ and\ \bibinfo {author}
		{\bibfnamefont {M.~E.}\ \bibnamefont {Raichle}},\ }\href {\doibase
		10.7554/eLife.10781.001} {\bibfield  {journal} {\bibinfo  {journal} {eLife}\
		}\textbf {\bibinfo {volume} {4}},\ \bibinfo {pages} {1} (\bibinfo {year}
		{2015}{\natexlab{b}})}\BibitemShut {NoStop}%
	\bibitem [{\citenamefont {Braun}\ \emph {et~al.}(2018)\citenamefont {Braun},
		\citenamefont {Schaefer}, \citenamefont {Betzel}, \citenamefont {Tost},
		\citenamefont {Meyer-Lindenberg},\ and\ \citenamefont
		{Bassett}}]{braun2018from}%
	\BibitemOpen
	\bibfield  {author} {\bibinfo {author} {\bibfnamefont {U.}~\bibnamefont
			{Braun}}, \bibinfo {author} {\bibfnamefont {A.}~\bibnamefont {Schaefer}},
		\bibinfo {author} {\bibfnamefont {R.~F.}\ \bibnamefont {Betzel}}, \bibinfo
		{author} {\bibfnamefont {H.}~\bibnamefont {Tost}}, \bibinfo {author}
		{\bibfnamefont {A.}~\bibnamefont {Meyer-Lindenberg}}, \ and\ \bibinfo
		{author} {\bibfnamefont {D.~S.}\ \bibnamefont {Bassett}},\ }\href@noop {}
	{\bibfield  {journal} {\bibinfo  {journal} {Neuron}\ }\textbf {\bibinfo
			{volume} {97}},\ \bibinfo {pages} {14} (\bibinfo {year} {2018})}\BibitemShut
	{NoStop}%
	\bibitem [{\citenamefont {Stam}(2014)}]{stam2014modern}%
	\BibitemOpen
	\bibfield  {author} {\bibinfo {author} {\bibfnamefont {C.~J.}\ \bibnamefont
			{Stam}},\ }\href@noop {} {\bibfield  {journal} {\bibinfo  {journal} {Nat Rev
				Neurosci}\ }\textbf {\bibinfo {volume} {15}},\ \bibinfo {pages} {683}
		(\bibinfo {year} {2014})}\BibitemShut {NoStop}%
	\bibitem [{\citenamefont {Bassett}\ \emph
		{et~al.}(2018{\natexlab{b}})\citenamefont {Bassett}, \citenamefont {Xia},\
		and\ \citenamefont {Satterthwaite}}]{bassett2018understanding}%
	\BibitemOpen
	\bibfield  {author} {\bibinfo {author} {\bibfnamefont {D.~S.}\ \bibnamefont
			{Bassett}}, \bibinfo {author} {\bibfnamefont {C.~H.}\ \bibnamefont {Xia}}, \
		and\ \bibinfo {author} {\bibfnamefont {T.~D.}\ \bibnamefont
			{Satterthwaite}},\ }\href@noop {} {\bibfield  {journal} {\bibinfo  {journal}
			{Biol Psychiatry Cogn Neurosci Neuroimaging}\ }\textbf {\bibinfo {volume}
			{S2451-9022}},\ \bibinfo {pages} {30079} (\bibinfo {year}
		{2018}{\natexlab{b}})}\BibitemShut {NoStop}%
	\bibitem [{\citenamefont {Jirsa}\ \emph {et~al.}(2014)\citenamefont {Jirsa},
		\citenamefont {Stacey}, \citenamefont {Quilichini}, \citenamefont {Ivanov},\
		and\ \citenamefont {Bernard}}]{Jirsa2014}%
	\BibitemOpen
	\bibfield  {author} {\bibinfo {author} {\bibfnamefont {V.~K.}\ \bibnamefont
			{Jirsa}}, \bibinfo {author} {\bibfnamefont {W.~C.}\ \bibnamefont {Stacey}},
		\bibinfo {author} {\bibfnamefont {P.~P.}\ \bibnamefont {Quilichini}},
		\bibinfo {author} {\bibfnamefont {A.~I.}\ \bibnamefont {Ivanov}}, \ and\
		\bibinfo {author} {\bibfnamefont {C.}~\bibnamefont {Bernard}},\ }\href
	{\doibase 10.1093/brain/awu133} {\bibfield  {journal} {\bibinfo  {journal}
			{Brain}\ }\textbf {\bibinfo {volume} {137}},\ \bibinfo {pages} {2210}
		(\bibinfo {year} {2014})},\ \Eprint {http://arxiv.org/abs/arXiv:1011.1669v3}
	{arXiv:arXiv:1011.1669v3} \BibitemShut {NoStop}%
	\bibitem [{\citenamefont {Wendling}\ \emph {et~al.}(2003)\citenamefont
		{Wendling}, \citenamefont {Bartolomei}, \citenamefont {Bellanger},
		\citenamefont {Bourien},\ and\ \citenamefont {Chauvel}}]{Wendling2003}%
	\BibitemOpen
	\bibfield  {author} {\bibinfo {author} {\bibfnamefont {F.}~\bibnamefont
			{Wendling}}, \bibinfo {author} {\bibfnamefont {F.}~\bibnamefont
			{Bartolomei}}, \bibinfo {author} {\bibfnamefont {J.~J.}\ \bibnamefont
			{Bellanger}}, \bibinfo {author} {\bibfnamefont {J.}~\bibnamefont {Bourien}},
		\ and\ \bibinfo {author} {\bibfnamefont {P.}~\bibnamefont {Chauvel}},\ }\href
	{\doibase 10.1093/brain/awg144} {\bibfield  {journal} {\bibinfo  {journal}
			{Brain}\ }\textbf {\bibinfo {volume} {126}},\ \bibinfo {pages} {1449}
		(\bibinfo {year} {2003})}\BibitemShut {NoStop}%
	\bibitem [{\citenamefont {Chamberlain}\ \emph {et~al.}(2011)\citenamefont
		{Chamberlain}, \citenamefont {Viventi}, \citenamefont {Blanco}, \citenamefont
		{Kim}, \citenamefont {Rogers},\ and\ \citenamefont {Litt}}]{Chamberlain2011}%
	\BibitemOpen
	\bibfield  {author} {\bibinfo {author} {\bibfnamefont {A.~C.}\ \bibnamefont
			{Chamberlain}}, \bibinfo {author} {\bibfnamefont {J.}~\bibnamefont
			{Viventi}}, \bibinfo {author} {\bibfnamefont {J.~A.}\ \bibnamefont {Blanco}},
		\bibinfo {author} {\bibfnamefont {D.~H.}\ \bibnamefont {Kim}}, \bibinfo
		{author} {\bibfnamefont {J.~A.}\ \bibnamefont {Rogers}}, \ and\ \bibinfo
		{author} {\bibfnamefont {B.}~\bibnamefont {Litt}},\ }\href {\doibase
		10.1109/IEMBS.2011.6090174} {\bibfield  {journal} {\bibinfo  {journal}
			{Proceedings of the Annual International Conference of the IEEE Engineering
				in Medicine and Biology Society, EMBS}\ ,\ \bibinfo {pages} {761}} (\bibinfo
		{year} {2011})}\BibitemShut {NoStop}%
	\bibitem [{\citenamefont {Viventi}\ \emph {et~al.}(2011)\citenamefont
		{Viventi}, \citenamefont {Kim}, \citenamefont {Vigeland}, \citenamefont
		{Frechette}, \citenamefont {Blanco}, \citenamefont {Kim}, \citenamefont
		{Avrin}, \citenamefont {Tiruvadi}, \citenamefont {Hwang}, \citenamefont
		{Vanleer}, \citenamefont {Wulsin}, \citenamefont {Davis}, \citenamefont
		{Gelber}, \citenamefont {Palmer}, \citenamefont {{Van Der Spiegel}},
		\citenamefont {Wu}, \citenamefont {Xiao}, \citenamefont {Huang},
		\citenamefont {Contreras}, \citenamefont {Rogers},\ and\ \citenamefont
		{Litt}}]{Viventi2011}%
	\BibitemOpen
	\bibfield  {author} {\bibinfo {author} {\bibfnamefont {J.}~\bibnamefont
			{Viventi}}, \bibinfo {author} {\bibfnamefont {D.~H.}\ \bibnamefont {Kim}},
		\bibinfo {author} {\bibfnamefont {L.}~\bibnamefont {Vigeland}}, \bibinfo
		{author} {\bibfnamefont {E.~S.}\ \bibnamefont {Frechette}}, \bibinfo {author}
		{\bibfnamefont {J.~A.}\ \bibnamefont {Blanco}}, \bibinfo {author}
		{\bibfnamefont {Y.~S.}\ \bibnamefont {Kim}}, \bibinfo {author} {\bibfnamefont
			{A.~E.}\ \bibnamefont {Avrin}}, \bibinfo {author} {\bibfnamefont {V.~R.}\
			\bibnamefont {Tiruvadi}}, \bibinfo {author} {\bibfnamefont {S.~W.}\
			\bibnamefont {Hwang}}, \bibinfo {author} {\bibfnamefont {A.~C.}\ \bibnamefont
			{Vanleer}}, \bibinfo {author} {\bibfnamefont {D.~F.}\ \bibnamefont {Wulsin}},
		\bibinfo {author} {\bibfnamefont {K.}~\bibnamefont {Davis}}, \bibinfo
		{author} {\bibfnamefont {C.~E.}\ \bibnamefont {Gelber}}, \bibinfo {author}
		{\bibfnamefont {L.}~\bibnamefont {Palmer}}, \bibinfo {author} {\bibfnamefont
			{J.}~\bibnamefont {{Van Der Spiegel}}}, \bibinfo {author} {\bibfnamefont
			{J.}~\bibnamefont {Wu}}, \bibinfo {author} {\bibfnamefont {J.}~\bibnamefont
			{Xiao}}, \bibinfo {author} {\bibfnamefont {Y.}~\bibnamefont {Huang}},
		\bibinfo {author} {\bibfnamefont {D.}~\bibnamefont {Contreras}}, \bibinfo
		{author} {\bibfnamefont {J.~A.}\ \bibnamefont {Rogers}}, \ and\ \bibinfo
		{author} {\bibfnamefont {B.}~\bibnamefont {Litt}},\ }\href {\doibase
		10.1038/nn.2973} {\bibfield  {journal} {\bibinfo  {journal} {Nature
				Neuroscience}\ }\textbf {\bibinfo {volume} {14}},\ \bibinfo {pages} {1599}
		(\bibinfo {year} {2011})}\BibitemShut {NoStop}%
	\bibitem [{\citenamefont {Gonz{\'{a}}lez-Ram{\'{i}}rez}\ \emph
		{et~al.}(2015)\citenamefont {Gonz{\'{a}}lez-Ram{\'{i}}rez}, \citenamefont
		{Ahmed}, \citenamefont {Cash}, \citenamefont {Wayne},\ and\ \citenamefont
		{Kramer}}]{Gonzalez-Ramirez2015}%
	\BibitemOpen
	\bibfield  {author} {\bibinfo {author} {\bibfnamefont {L.~R.}\ \bibnamefont
			{Gonz{\'{a}}lez-Ram{\'{i}}rez}}, \bibinfo {author} {\bibfnamefont {O.~J.}\
			\bibnamefont {Ahmed}}, \bibinfo {author} {\bibfnamefont {S.~S.}\ \bibnamefont
			{Cash}}, \bibinfo {author} {\bibfnamefont {C.~E.}\ \bibnamefont {Wayne}}, \
		and\ \bibinfo {author} {\bibfnamefont {M.~A.}\ \bibnamefont {Kramer}},\
	}\href {\doibase 10.1371/journal.pcbi.1004065} {\bibfield  {journal}
		{\bibinfo  {journal} {PLoS Computational Biology}\ }\textbf {\bibinfo
			{volume} {11}} (\bibinfo {year} {2015}),\
		10.1371/journal.pcbi.1004065}\BibitemShut {NoStop}%
	\bibitem [{\citenamefont {Richardson}\ \emph {et~al.}(2005)\citenamefont
		{Richardson}, \citenamefont {Schiff},\ and\ \citenamefont
		{Gluckman}}]{Richardson2005}%
	\BibitemOpen
	\bibfield  {author} {\bibinfo {author} {\bibfnamefont {K.~A.}\ \bibnamefont
			{Richardson}}, \bibinfo {author} {\bibfnamefont {S.~J.}\ \bibnamefont
			{Schiff}}, \ and\ \bibinfo {author} {\bibfnamefont {B.~J.}\ \bibnamefont
			{Gluckman}},\ }\href {\doibase 10.1103/PhysRevLett.94.028103} {\bibfield
		{journal} {\bibinfo  {journal} {Physical Review Letters}\ }\textbf {\bibinfo
			{volume} {94}} (\bibinfo {year} {2005}),\
		10.1103/PhysRevLett.94.028103}\BibitemShut {NoStop}%
	\bibitem [{\citenamefont {Ursino}\ and\ \citenamefont {{La
				Cara}}(2006)}]{Ursino2006}%
	\BibitemOpen
	\bibfield  {author} {\bibinfo {author} {\bibfnamefont {M.}~\bibnamefont
			{Ursino}}\ and\ \bibinfo {author} {\bibfnamefont {G.~E.}\ \bibnamefont {{La
					Cara}}},\ }\href {\doibase 10.1016/j.jtbi.2006.02.012} {\bibfield  {journal}
		{\bibinfo  {journal} {Journal of Theoretical Biology}\ }\textbf {\bibinfo
			{volume} {242}},\ \bibinfo {pages} {171} (\bibinfo {year}
		{2006})}\BibitemShut {NoStop}%
	\bibitem [{\citenamefont {Martinet}\ \emph {et~al.}(2017)\citenamefont
		{Martinet}, \citenamefont {Fiddyment}, \citenamefont {Madsen}, \citenamefont
		{Eskandar}, \citenamefont {Truccolo}, \citenamefont {Eden}, \citenamefont
		{Cash},\ and\ \citenamefont {Kramer}}]{Martinet2017}%
	\BibitemOpen
	\bibfield  {author} {\bibinfo {author} {\bibfnamefont {L.~E.}\ \bibnamefont
			{Martinet}}, \bibinfo {author} {\bibfnamefont {G.}~\bibnamefont {Fiddyment}},
		\bibinfo {author} {\bibfnamefont {J.~R.}\ \bibnamefont {Madsen}}, \bibinfo
		{author} {\bibfnamefont {E.~N.}\ \bibnamefont {Eskandar}}, \bibinfo {author}
		{\bibfnamefont {W.}~\bibnamefont {Truccolo}}, \bibinfo {author}
		{\bibfnamefont {U.~T.}\ \bibnamefont {Eden}}, \bibinfo {author}
		{\bibfnamefont {S.~S.}\ \bibnamefont {Cash}}, \ and\ \bibinfo {author}
		{\bibfnamefont {M.~A.}\ \bibnamefont {Kramer}},\ }\href {\doibase
		10.1038/ncomms14896} {\bibfield  {journal} {\bibinfo  {journal} {Nature
				Communications}\ }\textbf {\bibinfo {volume} {8}} (\bibinfo {year} {2017}),\
		10.1038/ncomms14896}\BibitemShut {NoStop}%
	\bibitem [{\citenamefont {Beggs}\ and\ \citenamefont
		{Timme}(2012)}]{Beggs2012}%
	\BibitemOpen
	\bibfield  {author} {\bibinfo {author} {\bibfnamefont {J.~M.}\ \bibnamefont
			{Beggs}}\ and\ \bibinfo {author} {\bibfnamefont {N.}~\bibnamefont {Timme}},\
	}\href {\doibase 10.3389/fphys.2012.00163} {\bibfield  {journal} {\bibinfo
			{journal} {Frontiers in Physiology}\ }\textbf {\bibinfo {volume} {3 JUN}},\
		\bibinfo {pages} {1} (\bibinfo {year} {2012})}\BibitemShut {NoStop}%
	\bibitem [{\citenamefont {Massimini}(2004)}]{Massimini2004b}%
	\BibitemOpen
	\bibfield  {author} {\bibinfo {author} {\bibfnamefont {M.}~\bibnamefont
			{Massimini}},\ }\href {\doibase 10.1523/JNEUROSCI.1318-04.2004} {\bibfield
		{journal} {\bibinfo  {journal} {Journal of Neuroscience}\ }\textbf {\bibinfo
			{volume} {24}},\ \bibinfo {pages} {6862} (\bibinfo {year}
		{2004})}\BibitemShut {NoStop}%
	\bibitem [{\citenamefont {Rubino}\ \emph {et~al.}(2006)\citenamefont {Rubino},
		\citenamefont {Robbins},\ and\ \citenamefont {Hatsopoulos}}]{Rubino2006}%
	\BibitemOpen
	\bibfield  {author} {\bibinfo {author} {\bibfnamefont {D.}~\bibnamefont
			{Rubino}}, \bibinfo {author} {\bibfnamefont {K.~A.}\ \bibnamefont {Robbins}},
		\ and\ \bibinfo {author} {\bibfnamefont {N.~G.}\ \bibnamefont
			{Hatsopoulos}},\ }\href {\doibase 10.1038/nn1802} {\bibfield  {journal}
		{\bibinfo  {journal} {Nature Neuroscience}\ }\textbf {\bibinfo {volume}
			{9}},\ \bibinfo {pages} {1549} (\bibinfo {year} {2006})}\BibitemShut
	{NoStop}%
	\bibitem [{\citenamefont {Takahashi}\ \emph {et~al.}(2015)\citenamefont
		{Takahashi}, \citenamefont {Kim}, \citenamefont {Coleman}, \citenamefont
		{Brown}, \citenamefont {Suminski}, \citenamefont {Best},\ and\ \citenamefont
		{Hatsopoulos}}]{Takahashi2015}%
	\BibitemOpen
	\bibfield  {author} {\bibinfo {author} {\bibfnamefont {K.}~\bibnamefont
			{Takahashi}}, \bibinfo {author} {\bibfnamefont {S.}~\bibnamefont {Kim}},
		\bibinfo {author} {\bibfnamefont {T.~P.}\ \bibnamefont {Coleman}}, \bibinfo
		{author} {\bibfnamefont {K.~A.}\ \bibnamefont {Brown}}, \bibinfo {author}
		{\bibfnamefont {A.~J.}\ \bibnamefont {Suminski}}, \bibinfo {author}
		{\bibfnamefont {M.~D.}\ \bibnamefont {Best}}, \ and\ \bibinfo {author}
		{\bibfnamefont {N.~G.}\ \bibnamefont {Hatsopoulos}},\ }\href {\doibase
		10.1038/ncomms8169} {\bibfield  {journal} {\bibinfo  {journal} {Nature
				Communications}\ }\textbf {\bibinfo {volume} {6}} (\bibinfo {year} {2015}),\
		10.1038/ncomms8169}\BibitemShut {NoStop}%
	\bibitem [{\citenamefont {Takahashi}\ \emph {et~al.}(2011)\citenamefont
		{Takahashi}, \citenamefont {Saleh}, \citenamefont {Penn},\ and\ \citenamefont
		{Hatsopoulos}}]{Takahashi2011a}%
	\BibitemOpen
	\bibfield  {author} {\bibinfo {author} {\bibfnamefont {K.}~\bibnamefont
			{Takahashi}}, \bibinfo {author} {\bibfnamefont {M.}~\bibnamefont {Saleh}},
		\bibinfo {author} {\bibfnamefont {R.~D.}\ \bibnamefont {Penn}}, \ and\
		\bibinfo {author} {\bibfnamefont {N.~G.}\ \bibnamefont {Hatsopoulos}},\
	}\href {\doibase 10.3389/fnhum.2011.00040} {\bibfield  {journal} {\bibinfo
			{journal} {Frontiers Human Neuroscience}\ }\textbf {\bibinfo {volume} {5}},\
		\bibinfo {pages} {40} (\bibinfo {year} {2011})}\BibitemShut {NoStop}%
	\bibitem [{\citenamefont {Zhang}\ \emph {et~al.}(2018)\citenamefont {Zhang},
		\citenamefont {Watrous}, \citenamefont {Patel},\ and\ \citenamefont
		{Jacobs}}]{Zhang2018}%
	\BibitemOpen
	\bibfield  {author} {\bibinfo {author} {\bibfnamefont {H.}~\bibnamefont
			{Zhang}}, \bibinfo {author} {\bibfnamefont {A.~J.}\ \bibnamefont {Watrous}},
		\bibinfo {author} {\bibfnamefont {A.}~\bibnamefont {Patel}}, \ and\ \bibinfo
		{author} {\bibfnamefont {J.}~\bibnamefont {Jacobs}},\ }\href {\doibase
		10.1016/j.neuron.2018.05.019} {\bibfield  {journal} {\bibinfo  {journal}
			{Neuron}\ }\textbf {\bibinfo {volume} {98}},\ \bibinfo {pages} {1269}
		(\bibinfo {year} {2018})}\BibitemShut {NoStop}%
	\bibitem [{\citenamefont {Benucci}\ \emph {et~al.}(2007)\citenamefont
		{Benucci}, \citenamefont {Frazor},\ and\ \citenamefont
		{Carandini}}]{Benucci2007}%
	\BibitemOpen
	\bibfield  {author} {\bibinfo {author} {\bibfnamefont {A.}~\bibnamefont
			{Benucci}}, \bibinfo {author} {\bibfnamefont {R.~A.}\ \bibnamefont {Frazor}},
		\ and\ \bibinfo {author} {\bibfnamefont {M.}~\bibnamefont {Carandini}},\
	}\href {\doibase 10.1016/j.neuron.2007.06.017} {\bibfield  {journal}
		{\bibinfo  {journal} {Neuron}\ }\textbf {\bibinfo {volume} {55}},\ \bibinfo
		{pages} {103} (\bibinfo {year} {2007})}\BibitemShut {NoStop}%
	\bibitem [{\citenamefont {Douw}\ \emph {et~al.}(2015)\citenamefont {Douw},
		\citenamefont {DeSalvo}, \citenamefont {Tanaka}, \citenamefont {Cole},
		\citenamefont {Liu}, \citenamefont {Reinsberger},\ and\ \citenamefont
		{Stufflebeam}}]{Douw2015}%
	\BibitemOpen
	\bibfield  {author} {\bibinfo {author} {\bibfnamefont {L.}~\bibnamefont
			{Douw}}, \bibinfo {author} {\bibfnamefont {M.~N.}\ \bibnamefont {DeSalvo}},
		\bibinfo {author} {\bibfnamefont {N.}~\bibnamefont {Tanaka}}, \bibinfo
		{author} {\bibfnamefont {A.~J.}\ \bibnamefont {Cole}}, \bibinfo {author}
		{\bibfnamefont {H.}~\bibnamefont {Liu}}, \bibinfo {author} {\bibfnamefont
			{C.}~\bibnamefont {Reinsberger}}, \ and\ \bibinfo {author} {\bibfnamefont
			{S.~M.}\ \bibnamefont {Stufflebeam}},\ }\href {\doibase 10.1002/acn3.173}
	{\bibfield  {journal} {\bibinfo  {journal} {Annals of Clinical and
				Translational Neurology}\ }\textbf {\bibinfo {volume} {2}},\ \bibinfo {pages}
		{338} (\bibinfo {year} {2015})}\BibitemShut {NoStop}%
	\bibitem [{\citenamefont {Bonilha}\ \emph {et~al.}(2012)\citenamefont
		{Bonilha}, \citenamefont {Nesland}, \citenamefont {Martz}, \citenamefont
		{Joseph}, \citenamefont {Spampinato}, \citenamefont {Edwards},\ and\
		\citenamefont {Tabesh}}]{Bonilha}%
	\BibitemOpen
	\bibfield  {author} {\bibinfo {author} {\bibfnamefont {L.}~\bibnamefont
			{Bonilha}}, \bibinfo {author} {\bibfnamefont {T.}~\bibnamefont {Nesland}},
		\bibinfo {author} {\bibfnamefont {G.~U.}\ \bibnamefont {Martz}}, \bibinfo
		{author} {\bibfnamefont {J.~E.}\ \bibnamefont {Joseph}}, \bibinfo {author}
		{\bibfnamefont {M.~V.}\ \bibnamefont {Spampinato}}, \bibinfo {author}
		{\bibfnamefont {J.~C.}\ \bibnamefont {Edwards}}, \ and\ \bibinfo {author}
		{\bibfnamefont {A.}~\bibnamefont {Tabesh}},\ }\href {\doibase
		10.1136/jnnp-2012-302476} {\bibfield  {journal} {\bibinfo  {journal} {J
				Neurol Neurosurg Psychiatry}\ }\textbf {\bibinfo {volume} {83}},\ \bibinfo
		{pages} {903} (\bibinfo {year} {2012})}\BibitemShut {NoStop}%
	\bibitem [{\citenamefont {DeSalvo}\ \emph {et~al.}(2014)\citenamefont
		{DeSalvo}, \citenamefont {Douw}, \citenamefont {Tanaka}, \citenamefont
		{Reinsberger},\ and\ \citenamefont {Stufflebeam}}]{DeSalvo2014}%
	\BibitemOpen
	\bibfield  {author} {\bibinfo {author} {\bibfnamefont {M.~N.}\ \bibnamefont
			{DeSalvo}}, \bibinfo {author} {\bibfnamefont {L.}~\bibnamefont {Douw}},
		\bibinfo {author} {\bibfnamefont {N.}~\bibnamefont {Tanaka}}, \bibinfo
		{author} {\bibfnamefont {C.}~\bibnamefont {Reinsberger}}, \ and\ \bibinfo
		{author} {\bibfnamefont {S.~M.}\ \bibnamefont {Stufflebeam}},\ }\href
	{\doibase 10.1148/radiol.13131044} {\bibfield  {journal} {\bibinfo  {journal}
			{Radiology}\ } (\bibinfo {year} {2014}),\
		10.1148/radiol.13131044}\BibitemShut {NoStop}%
	\bibitem [{\citenamefont {Jirsa}\ \emph {et~al.}(2017)\citenamefont {Jirsa},
		\citenamefont {Proix}, \citenamefont {Perdikis}, \citenamefont {Woodman},
		\citenamefont {Wang}, \citenamefont {Bernard}, \citenamefont
		{B{\"{i}}¿½nar}, \citenamefont {Chauvel}, \citenamefont {Bartolomei},
		\citenamefont {Bartolomei}, \citenamefont {Guye}, \citenamefont
		{Gonzalez-Martinez},\ and\ \citenamefont {Chauvel}}]{Jirsa2017}%
	\BibitemOpen
	\bibfield  {author} {\bibinfo {author} {\bibfnamefont {V.~K.}\ \bibnamefont
			{Jirsa}}, \bibinfo {author} {\bibfnamefont {T.}~\bibnamefont {Proix}},
		\bibinfo {author} {\bibfnamefont {D.}~\bibnamefont {Perdikis}}, \bibinfo
		{author} {\bibfnamefont {M.~M.}\ \bibnamefont {Woodman}}, \bibinfo {author}
		{\bibfnamefont {H.}~\bibnamefont {Wang}}, \bibinfo {author} {\bibfnamefont
			{C.}~\bibnamefont {Bernard}}, \bibinfo {author} {\bibfnamefont
			{C.}~\bibnamefont {B{\"{i}}¿½nar}}, \bibinfo {author} {\bibfnamefont
			{P.}~\bibnamefont {Chauvel}}, \bibinfo {author} {\bibfnamefont
			{F.}~\bibnamefont {Bartolomei}}, \bibinfo {author} {\bibfnamefont
			{F.}~\bibnamefont {Bartolomei}}, \bibinfo {author} {\bibfnamefont
			{M.}~\bibnamefont {Guye}}, \bibinfo {author} {\bibfnamefont {J.}~\bibnamefont
			{Gonzalez-Martinez}}, \ and\ \bibinfo {author} {\bibfnamefont
			{P.}~\bibnamefont {Chauvel}},\ }\href {\doibase
		10.1016/j.neuroimage.2016.04.049} {\bibfield  {journal} {\bibinfo  {journal}
			{NeuroImage}\ }\textbf {\bibinfo {volume} {145}},\ \bibinfo {pages} {377}
		(\bibinfo {year} {2017})}\BibitemShut {NoStop}%
	\bibitem [{\citenamefont {Bassett}\ \emph {et~al.}(2008)\citenamefont
		{Bassett}, \citenamefont {Bullmore}, \citenamefont {Verchinski},
		\citenamefont {Mattay}, \citenamefont {Weinberger},\ and\ \citenamefont
		{Meyer-Lindenberg}}]{Bassett2008}%
	\BibitemOpen
	\bibfield  {author} {\bibinfo {author} {\bibfnamefont {D.~S.}\ \bibnamefont
			{Bassett}}, \bibinfo {author} {\bibfnamefont {E.}~\bibnamefont {Bullmore}},
		\bibinfo {author} {\bibfnamefont {B.~A.}\ \bibnamefont {Verchinski}},
		\bibinfo {author} {\bibfnamefont {V.~S.}\ \bibnamefont {Mattay}}, \bibinfo
		{author} {\bibfnamefont {D.~R.}\ \bibnamefont {Weinberger}}, \ and\ \bibinfo
		{author} {\bibfnamefont {A.}~\bibnamefont {Meyer-Lindenberg}},\ }\href
	{\doibase 10.1523/JNEUROSCI.1929-08.2008} {\bibfield  {journal} {\bibinfo
			{journal} {Journal of Neuroscience}\ }\textbf {\bibinfo {volume} {28}},\
		\bibinfo {pages} {9239} (\bibinfo {year} {2008})}\BibitemShut {NoStop}%
	\bibitem [{\citenamefont {Griffa}\ \emph {et~al.}(2013)\citenamefont {Griffa},
		\citenamefont {Baumann}, \citenamefont {Thiran},\ and\ \citenamefont
		{Hagmann}}]{Griffa2013}%
	\BibitemOpen
	\bibfield  {author} {\bibinfo {author} {\bibfnamefont {A.}~\bibnamefont
			{Griffa}}, \bibinfo {author} {\bibfnamefont {P.~S.}\ \bibnamefont {Baumann}},
		\bibinfo {author} {\bibfnamefont {J.~P.}\ \bibnamefont {Thiran}}, \ and\
		\bibinfo {author} {\bibfnamefont {P.}~\bibnamefont {Hagmann}},\ }\href
	{\doibase 10.1016/j.neuroimage.2013.04.056} {\bibfield  {journal} {\bibinfo
			{journal} {NeuroImage}\ }\textbf {\bibinfo {volume} {80}},\ \bibinfo {pages}
		{515} (\bibinfo {year} {2013})}\BibitemShut {NoStop}%
	\bibitem [{\citenamefont {Alexander-Bloch}\ \emph {et~al.}(2013)\citenamefont
		{Alexander-Bloch}, \citenamefont {V{\'{e}}rtes}, \citenamefont {Stidd},
		\citenamefont {Lalonde}, \citenamefont {Clasen}, \citenamefont {Rapoport},
		\citenamefont {Giedd}, \citenamefont {Bullmore},\ and\ \citenamefont
		{Gogtay}}]{Alexander-Bloch2013}%
	\BibitemOpen
	\bibfield  {author} {\bibinfo {author} {\bibfnamefont {A.~F.}\ \bibnamefont
			{Alexander-Bloch}}, \bibinfo {author} {\bibfnamefont {P.~E.}\ \bibnamefont
			{V{\'{e}}rtes}}, \bibinfo {author} {\bibfnamefont {R.}~\bibnamefont {Stidd}},
		\bibinfo {author} {\bibfnamefont {F.}~\bibnamefont {Lalonde}}, \bibinfo
		{author} {\bibfnamefont {L.}~\bibnamefont {Clasen}}, \bibinfo {author}
		{\bibfnamefont {J.}~\bibnamefont {Rapoport}}, \bibinfo {author}
		{\bibfnamefont {J.}~\bibnamefont {Giedd}}, \bibinfo {author} {\bibfnamefont
			{E.~T.}\ \bibnamefont {Bullmore}}, \ and\ \bibinfo {author} {\bibfnamefont
			{N.}~\bibnamefont {Gogtay}},\ }\href {\doibase 10.1093/cercor/bhr388}
	{\bibfield  {journal} {\bibinfo  {journal} {Cerebral Cortex}\ }\textbf
		{\bibinfo {volume} {23}},\ \bibinfo {pages} {127} (\bibinfo {year}
		{2013})}\BibitemShut {NoStop}%
	\bibitem [{\citenamefont {Kaiser}(2011)}]{Kaiser2011}%
	\BibitemOpen
	\bibfield  {author} {\bibinfo {author} {\bibfnamefont {M.}~\bibnamefont
			{Kaiser}},\ }\href {\doibase 10.1016/j.neuroimage.2011.05.025} {\bibfield
		{journal} {\bibinfo  {journal} {NeuroImage}\ }\textbf {\bibinfo {volume}
			{57}},\ \bibinfo {pages} {892} (\bibinfo {year} {2011})}\BibitemShut
	{NoStop}%
	\bibitem [{\citenamefont {Duarte-Carvajalino}\ \emph
		{et~al.}(2012)\citenamefont {Duarte-Carvajalino}, \citenamefont {Jahanshad},
		\citenamefont {Lenglet}, \citenamefont {McMahon}, \citenamefont {{De
				Zubicaray}}, \citenamefont {Martin}, \citenamefont {Wright}, \citenamefont
		{Thompson},\ and\ \citenamefont {Sapiro}}]{Duarte-Carvajalino2012}%
	\BibitemOpen
	\bibfield  {author} {\bibinfo {author} {\bibfnamefont {J.~M.}\ \bibnamefont
			{Duarte-Carvajalino}}, \bibinfo {author} {\bibfnamefont {N.}~\bibnamefont
			{Jahanshad}}, \bibinfo {author} {\bibfnamefont {C.}~\bibnamefont {Lenglet}},
		\bibinfo {author} {\bibfnamefont {K.~L.}\ \bibnamefont {McMahon}}, \bibinfo
		{author} {\bibfnamefont {G.~I.}\ \bibnamefont {{De Zubicaray}}}, \bibinfo
		{author} {\bibfnamefont {N.~G.}\ \bibnamefont {Martin}}, \bibinfo {author}
		{\bibfnamefont {M.~J.}\ \bibnamefont {Wright}}, \bibinfo {author}
		{\bibfnamefont {P.~M.}\ \bibnamefont {Thompson}}, \ and\ \bibinfo {author}
		{\bibfnamefont {G.}~\bibnamefont {Sapiro}},\ }\href {\doibase
		10.1016/j.neuroimage.2011.10.096} {\bibfield  {journal} {\bibinfo  {journal}
			{NeuroImage}\ }\textbf {\bibinfo {volume} {59}},\ \bibinfo {pages} {3784}
		(\bibinfo {year} {2012})},\ \Eprint {http://arxiv.org/abs/NIHMS150003}
	{arXiv:NIHMS150003} \BibitemShut {NoStop}%
	\bibitem [{\citenamefont {Buhl}\ \emph {et~al.}(2004)\citenamefont {Buhl},
		\citenamefont {Gautrais}, \citenamefont {Sole}, \citenamefont {Kuntz},
		\citenamefont {Valverde}, \citenamefont {Deneubourg},\ and\ \citenamefont
		{Theraulaz}}]{buhl2004efficiency}%
	\BibitemOpen
	\bibfield  {author} {\bibinfo {author} {\bibfnamefont {J.}~\bibnamefont
			{Buhl}}, \bibinfo {author} {\bibfnamefont {J.}~\bibnamefont {Gautrais}},
		\bibinfo {author} {\bibfnamefont {R.~V.}\ \bibnamefont {Sole}}, \bibinfo
		{author} {\bibfnamefont {P.}~\bibnamefont {Kuntz}}, \bibinfo {author}
		{\bibfnamefont {S.}~\bibnamefont {Valverde}}, \bibinfo {author}
		{\bibfnamefont {J.~L.}\ \bibnamefont {Deneubourg}}, \ and\ \bibinfo {author}
		{\bibfnamefont {G.}~\bibnamefont {Theraulaz}},\ }\href@noop {} {\bibfield
		{journal} {\bibinfo  {journal} {The European Physical Journal B}\ }\textbf
		{\bibinfo {volume} {42}},\ \bibinfo {pages} {123} (\bibinfo {year}
		{2004})}\BibitemShut {NoStop}%
	\bibitem [{\citenamefont {Papadopoulos}\ \emph {et~al.}(2018)\citenamefont
		{Papadopoulos}, \citenamefont {Blinder}, \citenamefont {Ronellenfitsch},
		\citenamefont {Klimm}, \citenamefont {Katifori}, \citenamefont {Kleinfeld},\
		and\ \citenamefont {Bassett}}]{papadopoulos2018comparing}%
	\BibitemOpen
	\bibfield  {author} {\bibinfo {author} {\bibfnamefont {L.}~\bibnamefont
			{Papadopoulos}}, \bibinfo {author} {\bibfnamefont {P.}~\bibnamefont
			{Blinder}}, \bibinfo {author} {\bibfnamefont {H.}~\bibnamefont
			{Ronellenfitsch}}, \bibinfo {author} {\bibfnamefont {F.}~\bibnamefont
			{Klimm}}, \bibinfo {author} {\bibfnamefont {E.}~\bibnamefont {Katifori}},
		\bibinfo {author} {\bibfnamefont {D.}~\bibnamefont {Kleinfeld}}, \ and\
		\bibinfo {author} {\bibfnamefont {D.~S.}\ \bibnamefont {Bassett}},\
	}\href@noop {} {\bibfield  {journal} {\bibinfo  {journal} {arxiv}\ }\textbf
		{\bibinfo {volume} {1612}},\ \bibinfo {pages} {08058} (\bibinfo {year}
		{2018})}\BibitemShut {NoStop}%
	\bibitem [{\citenamefont {Fortunato}\ and\ \citenamefont
		{Hric}(2016)}]{fortunato2016community}%
	\BibitemOpen
	\bibfield  {author} {\bibinfo {author} {\bibfnamefont {S.}~\bibnamefont
			{Fortunato}}\ and\ \bibinfo {author} {\bibfnamefont {D.}~\bibnamefont
			{Hric}},\ }\href@noop {} {\bibfield  {journal} {\bibinfo  {journal} {Physics
				Reports}\ }\textbf {\bibinfo {volume} {659}},\ \bibinfo {pages} {1} (\bibinfo
		{year} {2016})}\BibitemShut {NoStop}%
	\bibitem [{\citenamefont {Doron}\ \emph {et~al.}(2012)\citenamefont {Doron},
		\citenamefont {Bassett},\ and\ \citenamefont {Gazzaniga}}]{Doron2012}%
	\BibitemOpen
	\bibfield  {author} {\bibinfo {author} {\bibfnamefont {K.~W.}\ \bibnamefont
			{Doron}}, \bibinfo {author} {\bibfnamefont {D.~S.}\ \bibnamefont {Bassett}},
		\ and\ \bibinfo {author} {\bibfnamefont {M.~S.}\ \bibnamefont {Gazzaniga}},\
	}\href {\doibase 10.1073/pnas.1216402109} {\bibfield  {journal} {\bibinfo
			{journal} {Proceedings of the National Academy of Sciences}\ }\textbf
		{\bibinfo {volume} {109}},\ \bibinfo {pages} {18661} (\bibinfo {year}
		{2012})}\BibitemShut {NoStop}%
	\bibitem [{\citenamefont {Chai}\ \emph {et~al.}(2016)\citenamefont {Chai},
		\citenamefont {Mattar}, \citenamefont {Blank}, \citenamefont {Fedorenko},\
		and\ \citenamefont {Bassett}}]{chai2016functional}%
	\BibitemOpen
	\bibfield  {author} {\bibinfo {author} {\bibfnamefont {L.~R.}\ \bibnamefont
			{Chai}}, \bibinfo {author} {\bibfnamefont {M.~G.}\ \bibnamefont {Mattar}},
		\bibinfo {author} {\bibfnamefont {I.~A.}\ \bibnamefont {Blank}}, \bibinfo
		{author} {\bibfnamefont {E.}~\bibnamefont {Fedorenko}}, \ and\ \bibinfo
		{author} {\bibfnamefont {D.~S.}\ \bibnamefont {Bassett}},\ }\href@noop {}
	{\bibfield  {journal} {\bibinfo  {journal} {Cereb Cortex}\ }\textbf {\bibinfo
			{volume} {26}},\ \bibinfo {pages} {4148} (\bibinfo {year}
		{2016})}\BibitemShut {NoStop}%
	\bibitem [{\citenamefont {He}\ \emph {et~al.}(2018)\citenamefont {He},
		\citenamefont {Bassett}, \citenamefont {Chaitanya}, \citenamefont {Sperling},
		\citenamefont {Kozlowski},\ and\ \citenamefont {Tracy}}]{he2018disrupted}%
	\BibitemOpen
	\bibfield  {author} {\bibinfo {author} {\bibfnamefont {X.}~\bibnamefont
			{He}}, \bibinfo {author} {\bibfnamefont {D.~S.}\ \bibnamefont {Bassett}},
		\bibinfo {author} {\bibfnamefont {G.}~\bibnamefont {Chaitanya}}, \bibinfo
		{author} {\bibfnamefont {M.~R.}\ \bibnamefont {Sperling}}, \bibinfo {author}
		{\bibfnamefont {L.}~\bibnamefont {Kozlowski}}, \ and\ \bibinfo {author}
		{\bibfnamefont {J.~I.}\ \bibnamefont {Tracy}},\ }\href@noop {} {\bibfield
		{journal} {\bibinfo  {journal} {Brain}\ ,\ \bibinfo {pages} {141}} (\bibinfo
		{year} {2018})}\BibitemShut {NoStop}%
	\bibitem [{\citenamefont {Newman}\ and\ \citenamefont
		{Girvan}(2003)}]{Newman2003}%
	\BibitemOpen
	\bibfield  {author} {\bibinfo {author} {\bibfnamefont {M.~E.~J.}\
			\bibnamefont {Newman}}\ and\ \bibinfo {author} {\bibfnamefont
			{M.}~\bibnamefont {Girvan}},\ }\href {\doibase 10.1103/PhysRevE.69.026113}
	{\bibfield  {journal} {\bibinfo  {journal} {Physical Review E}\ } (\bibinfo
		{year} {2003}),\ 10.1103/PhysRevE.69.026113},\ \Eprint
	{http://arxiv.org/abs/0308217} {arXiv:0308217 [cond-mat]} \BibitemShut
	{NoStop}%
	\bibitem [{\citenamefont {Betzel}\ \emph {et~al.}(2018)\citenamefont {Betzel},
		\citenamefont {Medaglia},\ and\ \citenamefont {Bassett}}]{Betzel}%
	\BibitemOpen
	\bibfield  {author} {\bibinfo {author} {\bibfnamefont {R.~F.}\ \bibnamefont
			{Betzel}}, \bibinfo {author} {\bibfnamefont {J.~D.}\ \bibnamefont
			{Medaglia}}, \ and\ \bibinfo {author} {\bibfnamefont {D.~S.}\ \bibnamefont
			{Bassett}},\ }\href {\doibase 10.1038/s41467-017-02681-z} {\bibfield
		{journal} {\bibinfo  {journal} {Nat Commun}\ }\textbf {\bibinfo {volume}
			{9}},\ \bibinfo {pages} {346} (\bibinfo {year} {2018})}\BibitemShut {NoStop}%
	\bibitem [{\citenamefont {Papadopoulos}\ \emph {et~al.}(2016)\citenamefont
		{Papadopoulos}, \citenamefont {Puckett}, \citenamefont {Daniels},\ and\
		\citenamefont {Bassett}}]{papadopoulos2016evolution}%
	\BibitemOpen
	\bibfield  {author} {\bibinfo {author} {\bibfnamefont {L.}~\bibnamefont
			{Papadopoulos}}, \bibinfo {author} {\bibfnamefont {J.~G.}\ \bibnamefont
			{Puckett}}, \bibinfo {author} {\bibfnamefont {K.~E.}\ \bibnamefont
			{Daniels}}, \ and\ \bibinfo {author} {\bibfnamefont {D.~S.}\ \bibnamefont
			{Bassett}},\ }\href@noop {} {\bibfield  {journal} {\bibinfo  {journal} {Phys
				Rev E}\ }\textbf {\bibinfo {volume} {94}},\ \bibinfo {pages} {032908}
		(\bibinfo {year} {2016})}\BibitemShut {NoStop}%
	\bibitem [{\citenamefont {Betzel}\ \emph
		{et~al.}(2016{\natexlab{a}})\citenamefont {Betzel}, \citenamefont {Medaglia},
		\citenamefont {Papadopoulos}, \citenamefont {Baum}, \citenamefont {Gur},
		\citenamefont {Gur}, \citenamefont {Roalf}, \citenamefont {Satterthwaite},\
		and\ \citenamefont {Bassett}}]{Betzel2016a}%
	\BibitemOpen
	\bibfield  {author} {\bibinfo {author} {\bibfnamefont {R.~F.}\ \bibnamefont
			{Betzel}}, \bibinfo {author} {\bibfnamefont {J.~D.}\ \bibnamefont
			{Medaglia}}, \bibinfo {author} {\bibfnamefont {L.}~\bibnamefont
			{Papadopoulos}}, \bibinfo {author} {\bibfnamefont {G.}~\bibnamefont {Baum}},
		\bibinfo {author} {\bibfnamefont {R.}~\bibnamefont {Gur}}, \bibinfo {author}
		{\bibfnamefont {R.}~\bibnamefont {Gur}}, \bibinfo {author} {\bibfnamefont
			{D.}~\bibnamefont {Roalf}}, \bibinfo {author} {\bibfnamefont {T.~D.}\
			\bibnamefont {Satterthwaite}}, \ and\ \bibinfo {author} {\bibfnamefont
			{D.~S.}\ \bibnamefont {Bassett}},\ }\href {http://arxiv.org/abs/1608.01161}
	{\bibfield  {journal} {\bibinfo  {journal} {Network Neuroscience}\ }
		(\bibinfo {year} {2016}{\natexlab{a}})},\ \Eprint
	{http://arxiv.org/abs/1608.01161} {arXiv:1608.01161} \BibitemShut {NoStop}%
	\bibitem [{\citenamefont {Expert}\ \emph {et~al.}(2010)\citenamefont {Expert},
		\citenamefont {Evans}, \citenamefont {Blondel},\ and\ \citenamefont
		{Lambiotte}}]{Expert2010}%
	\BibitemOpen
	\bibfield  {author} {\bibinfo {author} {\bibfnamefont {P.}~\bibnamefont
			{Expert}}, \bibinfo {author} {\bibfnamefont {T.}~\bibnamefont {Evans}},
		\bibinfo {author} {\bibfnamefont {V.~D.}\ \bibnamefont {Blondel}}, \ and\
		\bibinfo {author} {\bibfnamefont {R.}~\bibnamefont {Lambiotte}},\ }\href
	{\doibase 10.1073/pnas.1018962108} {\bibfield  {journal} {\bibinfo  {journal}
			{Proceedings of the National Academy of Sciences}\ } (\bibinfo {year}
		{2010}),\ 10.1073/pnas.1018962108},\ \Eprint {http://arxiv.org/abs/1012.3409}
	{arXiv:1012.3409} \BibitemShut {NoStop}%
	\bibitem [{\citenamefont {Sarzynska}\ \emph {et~al.}(2015)\citenamefont
		{Sarzynska}, \citenamefont {Leicht}, \citenamefont {Chowell}, \citenamefont
		{Porter}, \citenamefont {Bassett},\ and\ \citenamefont
		{Sarzynska}}]{Sarzynska2015}%
	\BibitemOpen
	\bibfield  {author} {\bibinfo {author} {\bibfnamefont {M.}~\bibnamefont
			{Sarzynska}}, \bibinfo {author} {\bibfnamefont {E.~A.}\ \bibnamefont
			{Leicht}}, \bibinfo {author} {\bibfnamefont {G.}~\bibnamefont {Chowell}},
		\bibinfo {author} {\bibfnamefont {M.~A.}\ \bibnamefont {Porter}}, \bibinfo
		{author} {\bibfnamefont {D.}~\bibnamefont {Bassett}}, \ and\ \bibinfo
		{author} {\bibfnamefont {M.}~\bibnamefont {Sarzynska}},\ }\href {\doibase
		10.1093/comnet/cnv027} {\bibfield  {journal} {\bibinfo  {journal} {Journal of
				Complex Networks}\ } (\bibinfo {year} {2015}),\
		10.1093/comnet/cnv027}\BibitemShut {NoStop}%
	\bibitem [{\citenamefont {Roberts}\ \emph {et~al.}(2016)\citenamefont
		{Roberts}, \citenamefont {Perry}, \citenamefont {Lord}, \citenamefont
		{Roberts}, \citenamefont {Mitchell}, \citenamefont {Smith}, \citenamefont
		{Calamante},\ and\ \citenamefont {Breakspear}}]{Roberts2016}%
	\BibitemOpen
	\bibfield  {author} {\bibinfo {author} {\bibfnamefont {J.~A.}\ \bibnamefont
			{Roberts}}, \bibinfo {author} {\bibfnamefont {A.}~\bibnamefont {Perry}},
		\bibinfo {author} {\bibfnamefont {A.~R.}\ \bibnamefont {Lord}}, \bibinfo
		{author} {\bibfnamefont {G.}~\bibnamefont {Roberts}}, \bibinfo {author}
		{\bibfnamefont {P.~B.}\ \bibnamefont {Mitchell}}, \bibinfo {author}
		{\bibfnamefont {R.~E.}\ \bibnamefont {Smith}}, \bibinfo {author}
		{\bibfnamefont {F.}~\bibnamefont {Calamante}}, \ and\ \bibinfo {author}
		{\bibfnamefont {M.}~\bibnamefont {Breakspear}},\ }\href {\doibase
		10.1016/j.neuroimage.2015.09.009} {\bibfield  {journal} {\bibinfo  {journal}
			{NeuroImage}\ }\textbf {\bibinfo {volume} {124}},\ \bibinfo {pages} {379}
		(\bibinfo {year} {2016})}\BibitemShut {NoStop}%
	\bibitem [{\citenamefont {Samu}\ \emph {et~al.}(2014)\citenamefont {Samu},
		\citenamefont {Seth},\ and\ \citenamefont {Nowotny}}]{Samu2014}%
	\BibitemOpen
	\bibfield  {author} {\bibinfo {author} {\bibfnamefont {D.}~\bibnamefont
			{Samu}}, \bibinfo {author} {\bibfnamefont {A.~K.}\ \bibnamefont {Seth}}, \
		and\ \bibinfo {author} {\bibfnamefont {T.}~\bibnamefont {Nowotny}},\ }\href
	{\doibase 10.1371/journal.pcbi.1003557} {\bibfield  {journal} {\bibinfo
			{journal} {PLoS Computational Biology}\ }\textbf {\bibinfo {volume} {10}}
		(\bibinfo {year} {2014}),\ 10.1371/journal.pcbi.1003557}\BibitemShut
	{NoStop}%
	\bibitem [{\citenamefont {Wiedermann}\ \emph {et~al.}(2016)\citenamefont
		{Wiedermann}, \citenamefont {Donges}, \citenamefont {Kurths},\ and\
		\citenamefont {Donner}}]{Wiedermann2016}%
	\BibitemOpen
	\bibfield  {author} {\bibinfo {author} {\bibfnamefont {M.}~\bibnamefont
			{Wiedermann}}, \bibinfo {author} {\bibfnamefont {J.~F.}\ \bibnamefont
			{Donges}}, \bibinfo {author} {\bibfnamefont {J.}~\bibnamefont {Kurths}}, \
		and\ \bibinfo {author} {\bibfnamefont {R.~V.}\ \bibnamefont {Donner}},\
	}\href {\doibase 10.1103/PhysRevE.93.042308} {\bibfield  {journal} {\bibinfo
			{journal} {Physical Review E}\ }\textbf {\bibinfo {volume} {93}} (\bibinfo
		{year} {2016}),\ 10.1103/PhysRevE.93.042308},\ \Eprint
	{http://arxiv.org/abs/1509.09293} {arXiv:1509.09293} \BibitemShut {NoStop}%
	\bibitem [{\citenamefont {Cui}\ \emph {et~al.}(2018)\citenamefont {Cui},
		\citenamefont {Xiang}, \citenamefont {Guo}, \citenamefont {Yin},
		\citenamefont {Zhang}, \citenamefont {Lan},\ and\ \citenamefont
		{Chen}}]{cui2018classification}%
	\BibitemOpen
	\bibfield  {author} {\bibinfo {author} {\bibfnamefont {X.}~\bibnamefont
			{Cui}}, \bibinfo {author} {\bibfnamefont {J.}~\bibnamefont {Xiang}}, \bibinfo
		{author} {\bibfnamefont {H.}~\bibnamefont {Guo}}, \bibinfo {author}
		{\bibfnamefont {G.}~\bibnamefont {Yin}}, \bibinfo {author} {\bibfnamefont
			{H.}~\bibnamefont {Zhang}}, \bibinfo {author} {\bibfnamefont
			{F.}~\bibnamefont {Lan}}, \ and\ \bibinfo {author} {\bibfnamefont
			{J.}~\bibnamefont {Chen}},\ }\href@noop {} {\bibfield  {journal} {\bibinfo
			{journal} {Front Comput Neurosci}\ }\textbf {\bibinfo {volume} {12}},\
		\bibinfo {pages} {31} (\bibinfo {year} {2018})}\BibitemShut {NoStop}%
	\bibitem [{\citenamefont {Janssen}\ \emph {et~al.}(2017)\citenamefont
		{Janssen}, \citenamefont {Hillebrand}, \citenamefont {Gouw}, \citenamefont
		{Gelade}, \citenamefont {Van~Mourik}, \citenamefont {Maras},\ and\
		\citenamefont {Oosterlaan}}]{janssen2017neural}%
	\BibitemOpen
	\bibfield  {author} {\bibinfo {author} {\bibfnamefont {T.~W.~P.}\
			\bibnamefont {Janssen}}, \bibinfo {author} {\bibfnamefont {A.}~\bibnamefont
			{Hillebrand}}, \bibinfo {author} {\bibfnamefont {A.}~\bibnamefont {Gouw}},
		\bibinfo {author} {\bibfnamefont {K.}~\bibnamefont {Gelade}}, \bibinfo
		{author} {\bibfnamefont {R.}~\bibnamefont {Van~Mourik}}, \bibinfo {author}
		{\bibfnamefont {A.}~\bibnamefont {Maras}}, \ and\ \bibinfo {author}
		{\bibfnamefont {J.}~\bibnamefont {Oosterlaan}},\ }\href@noop {} {\bibfield
		{journal} {\bibinfo  {journal} {Clin Neurophysiol}\ }\textbf {\bibinfo
			{volume} {128}},\ \bibinfo {pages} {2258} (\bibinfo {year}
		{2017})}\BibitemShut {NoStop}%
	\bibitem [{\citenamefont {Smit}\ \emph {et~al.}(2016)\citenamefont {Smit},
		\citenamefont {de~Geus}, \citenamefont {Boersma}, \citenamefont {Boomsma},\
		and\ \citenamefont {Stam}}]{smit2016life}%
	\BibitemOpen
	\bibfield  {author} {\bibinfo {author} {\bibfnamefont {D.~J.}\ \bibnamefont
			{Smit}}, \bibinfo {author} {\bibfnamefont {E.~J.}\ \bibnamefont {de~Geus}},
		\bibinfo {author} {\bibfnamefont {M.}~\bibnamefont {Boersma}}, \bibinfo
		{author} {\bibfnamefont {D.~I.}\ \bibnamefont {Boomsma}}, \ and\ \bibinfo
		{author} {\bibfnamefont {C.~J.}\ \bibnamefont {Stam}},\ }\href@noop {}
	{\bibfield  {journal} {\bibinfo  {journal} {Brain Connect}\ }\textbf
		{\bibinfo {volume} {6}},\ \bibinfo {pages} {312} (\bibinfo {year}
		{2016})}\BibitemShut {NoStop}%
	\bibitem [{\citenamefont {Blinder}\ \emph {et~al.}(2013)\citenamefont
		{Blinder}, \citenamefont {Tsai}, \citenamefont {Kaufhold}, \citenamefont
		{Knutsen}, \citenamefont {Suhl},\ and\ \citenamefont
		{Kleinfeld}}]{blinder2013cortical}%
	\BibitemOpen
	\bibfield  {author} {\bibinfo {author} {\bibfnamefont {P.}~\bibnamefont
			{Blinder}}, \bibinfo {author} {\bibfnamefont {P.~S.}\ \bibnamefont {Tsai}},
		\bibinfo {author} {\bibfnamefont {J.~P.}\ \bibnamefont {Kaufhold}}, \bibinfo
		{author} {\bibfnamefont {P.~M.}\ \bibnamefont {Knutsen}}, \bibinfo {author}
		{\bibfnamefont {H.}~\bibnamefont {Suhl}}, \ and\ \bibinfo {author}
		{\bibfnamefont {D.}~\bibnamefont {Kleinfeld}},\ }\href@noop {} {\bibfield
		{journal} {\bibinfo  {journal} {Nat Neurosci}\ }\textbf {\bibinfo {volume}
			{16}},\ \bibinfo {pages} {889} (\bibinfo {year} {2013})}\BibitemShut
	{NoStop}%
	\bibitem [{\citenamefont {Schmid}\ \emph {et~al.}(2017)\citenamefont {Schmid},
		\citenamefont {Tsai}, \citenamefont {Kleinfeld}, \citenamefont {Jenny},\ and\
		\citenamefont {Weber}}]{schmid2017depth}%
	\BibitemOpen
	\bibfield  {author} {\bibinfo {author} {\bibfnamefont {F.}~\bibnamefont
			{Schmid}}, \bibinfo {author} {\bibfnamefont {P.~S.}\ \bibnamefont {Tsai}},
		\bibinfo {author} {\bibfnamefont {D.}~\bibnamefont {Kleinfeld}}, \bibinfo
		{author} {\bibfnamefont {P.}~\bibnamefont {Jenny}}, \ and\ \bibinfo {author}
		{\bibfnamefont {B.}~\bibnamefont {Weber}},\ }\href@noop {} {\bibfield
		{journal} {\bibinfo  {journal} {PLoS Comput Biol}\ }\textbf {\bibinfo
			{volume} {13}},\ \bibinfo {pages} {e1005392} (\bibinfo {year}
		{2017})}\BibitemShut {NoStop}%
	\bibitem [{\citenamefont {Betzel}\ and\ \citenamefont
		{Bassett}(2017)}]{betzel2017generative}%
	\BibitemOpen
	\bibfield  {author} {\bibinfo {author} {\bibfnamefont {R.~F.}\ \bibnamefont
			{Betzel}}\ and\ \bibinfo {author} {\bibfnamefont {D.~S.}\ \bibnamefont
			{Bassett}},\ }\href@noop {} {\bibfield  {journal} {\bibinfo  {journal} {J R
				Soc Interface}\ }\textbf {\bibinfo {volume} {14}},\ \bibinfo {pages}
		{20170623} (\bibinfo {year} {2017})}\BibitemShut {NoStop}%
	\bibitem [{\citenamefont {Betzel}\ \emph
		{et~al.}(2016{\natexlab{b}})\citenamefont {Betzel}, \citenamefont
		{Avena-Koenigsberger}, \citenamefont {Go{\~{n}}i}, \citenamefont {He},
		\citenamefont {de~Reus}, \citenamefont {Griffa}, \citenamefont
		{V{\'{e}}rtes}, \citenamefont {Mi{\v{s}}ic}, \citenamefont {Thiran},
		\citenamefont {Hagmann}, \citenamefont {van~den Heuvel}, \citenamefont {Zuo},
		\citenamefont {Bullmore},\ and\ \citenamefont {Sporns}}]{Betzel2016b}%
	\BibitemOpen
	\bibfield  {author} {\bibinfo {author} {\bibfnamefont {R.~F.}\ \bibnamefont
			{Betzel}}, \bibinfo {author} {\bibfnamefont {A.}~\bibnamefont
			{Avena-Koenigsberger}}, \bibinfo {author} {\bibfnamefont {J.}~\bibnamefont
			{Go{\~{n}}i}}, \bibinfo {author} {\bibfnamefont {Y.}~\bibnamefont {He}},
		\bibinfo {author} {\bibfnamefont {M.~A.}\ \bibnamefont {de~Reus}}, \bibinfo
		{author} {\bibfnamefont {A.}~\bibnamefont {Griffa}}, \bibinfo {author}
		{\bibfnamefont {P.~E.}\ \bibnamefont {V{\'{e}}rtes}}, \bibinfo {author}
		{\bibfnamefont {B.}~\bibnamefont {Mi{\v{s}}ic}}, \bibinfo {author}
		{\bibfnamefont {J.~P.}\ \bibnamefont {Thiran}}, \bibinfo {author}
		{\bibfnamefont {P.}~\bibnamefont {Hagmann}}, \bibinfo {author} {\bibfnamefont
			{M.}~\bibnamefont {van~den Heuvel}}, \bibinfo {author} {\bibfnamefont
			{X.~N.}\ \bibnamefont {Zuo}}, \bibinfo {author} {\bibfnamefont {E.~T.}\
			\bibnamefont {Bullmore}}, \ and\ \bibinfo {author} {\bibfnamefont
			{O.}~\bibnamefont {Sporns}},\ }\href {\doibase
		10.1016/j.neuroimage.2015.09.041} {\bibfield  {journal} {\bibinfo  {journal}
			{NeuroImage}\ }\textbf {\bibinfo {volume} {124}},\ \bibinfo {pages} {1054}
		(\bibinfo {year} {2016}{\natexlab{b}})},\ \Eprint
	{http://arxiv.org/abs/1506.06795} {arXiv:1506.06795} \BibitemShut {NoStop}%
	\bibitem [{\citenamefont {Butts}(2009)}]{Butts2009}%
	\BibitemOpen
	\bibfield  {author} {\bibinfo {author} {\bibfnamefont {C.~T.}\ \bibnamefont
			{Butts}},\ }\href {\doibase 10.1126/science.1171022} {\bibfield  {journal}
		{\bibinfo  {journal} {Science}\ }\textbf {\bibinfo {volume} {325}},\ \bibinfo
		{pages} {414} (\bibinfo {year} {2009})},\ \Eprint
	{http://arxiv.org/abs/arXiv:1010.0725v1} {arXiv:arXiv:1010.0725v1}
	\BibitemShut {NoStop}%
	\bibitem [{\citenamefont {Rubinov}\ and\ \citenamefont
		{Sporns}(2010)}]{Rubinov2010}%
	\BibitemOpen
	\bibfield  {author} {\bibinfo {author} {\bibfnamefont {M.}~\bibnamefont
			{Rubinov}}\ and\ \bibinfo {author} {\bibfnamefont {O.}~\bibnamefont
			{Sporns}},\ }\href {\doibase 10.1016/j.neuroimage.2009.10.003} {\bibfield
		{journal} {\bibinfo  {journal} {NeuroImage}\ }\textbf {\bibinfo {volume}
			{52}},\ \bibinfo {pages} {1059} (\bibinfo {year} {2010})}\BibitemShut
	{NoStop}%
	\bibitem [{\citenamefont {Garcia}\ \emph {et~al.}(2018)\citenamefont {Garcia},
		\citenamefont {Ashourvan}, \citenamefont {Muldoon}, \citenamefont {Vettel},\
		and\ \citenamefont {Bassett}}]{garcia2018applications}%
	\BibitemOpen
	\bibfield  {author} {\bibinfo {author} {\bibfnamefont {J.~O.}\ \bibnamefont
			{Garcia}}, \bibinfo {author} {\bibfnamefont {A.}~\bibnamefont {Ashourvan}},
		\bibinfo {author} {\bibfnamefont {S.~F.}\ \bibnamefont {Muldoon}}, \bibinfo
		{author} {\bibfnamefont {J.~M.}\ \bibnamefont {Vettel}}, \ and\ \bibinfo
		{author} {\bibfnamefont {D.~S.}\ \bibnamefont {Bassett}},\ }\href@noop {}
	{\bibfield  {journal} {\bibinfo  {journal} {Proceedings of the IEEE}\
		}\textbf {\bibinfo {volume} {106}},\ \bibinfo {pages} {846 } (\bibinfo {year}
		{2018})}\BibitemShut {NoStop}%
	\bibitem [{\citenamefont {Giusti}\ \emph {et~al.}(2016)\citenamefont {Giusti},
		\citenamefont {Ghrist},\ and\ \citenamefont {Bassett}}]{giusti2016twos}%
	\BibitemOpen
	\bibfield  {author} {\bibinfo {author} {\bibfnamefont {C.}~\bibnamefont
			{Giusti}}, \bibinfo {author} {\bibfnamefont {R.}~\bibnamefont {Ghrist}}, \
		and\ \bibinfo {author} {\bibfnamefont {D.~S.}\ \bibnamefont {Bassett}},\
	}\href@noop {} {\bibfield  {journal} {\bibinfo  {journal} {J Comput
				Neurosci}\ }\textbf {\bibinfo {volume} {41}},\ \bibinfo {pages} {1} (\bibinfo
		{year} {2016})}\BibitemShut {NoStop}%
	\bibitem [{\citenamefont {Zomorodian}\ and\ \citenamefont
		{Carlsson}(2005)}]{Zomorodian2005}%
	\BibitemOpen
	\bibfield  {author} {\bibinfo {author} {\bibfnamefont {A.}~\bibnamefont
			{Zomorodian}}\ and\ \bibinfo {author} {\bibfnamefont {G.}~\bibnamefont
			{Carlsson}},\ }\href {\doibase 10.1007/s00454-004-1146-y} {\bibfield
		{journal} {\bibinfo  {journal} {Discrete and Computational Geometry}\
		}\textbf {\bibinfo {volume} {33}},\ \bibinfo {pages} {249} (\bibinfo {year}
		{2005})}\BibitemShut {NoStop}%
	\bibitem [{\citenamefont {Carlsson}(2009)}]{Carlsson2009}%
	\BibitemOpen
	\bibfield  {author} {\bibinfo {author} {\bibfnamefont {G.}~\bibnamefont
			{Carlsson}},\ }\href {\doibase 10.1090/S0273-0979-09-01249-X} {\bibfield
		{journal} {\bibinfo  {journal} {Bulletin of the American Mathematical
				Society}\ }\textbf {\bibinfo {volume} {46}},\ \bibinfo {pages} {255}
		(\bibinfo {year} {2009})},\ \Eprint {http://arxiv.org/abs/arXiv:1312.6184v5}
	{arXiv:arXiv:1312.6184v5} \BibitemShut {NoStop}%
	\bibitem [{\citenamefont {Sizemore}\ \emph {et~al.}(2018)\citenamefont
		{Sizemore}, \citenamefont {Phillips-Cremins}, \citenamefont {Ghrist},\ and\
		\citenamefont {Bassett}}]{sizemore2018importance}%
	\BibitemOpen
	\bibfield  {author} {\bibinfo {author} {\bibfnamefont {A.~E.}\ \bibnamefont
			{Sizemore}}, \bibinfo {author} {\bibfnamefont {J.}~\bibnamefont
			{Phillips-Cremins}}, \bibinfo {author} {\bibfnamefont {R.}~\bibnamefont
			{Ghrist}}, \ and\ \bibinfo {author} {\bibfnamefont {D.~S.}\ \bibnamefont
			{Bassett}},\ }\href@noop {} {\bibfield  {journal} {\bibinfo  {journal}
			{arXiv}\ }\textbf {\bibinfo {volume} {1806}},\ \bibinfo {pages} {05167}
		(\bibinfo {year} {2018})}\BibitemShut {NoStop}%
	\bibitem [{\citenamefont {Dotko}\ \emph
		{et~al.}(2016{\natexlab{a}})\citenamefont {Dotko}, \citenamefont {Hess},
		\citenamefont {Levi}, \citenamefont {Nolte}, \citenamefont {Reimann},
		\citenamefont {Scolamiero}, \citenamefont {Turner}, \citenamefont {Muller},\
		and\ \citenamefont {Markram}}]{Reimann2017}%
	\BibitemOpen
	\bibfield  {author} {\bibinfo {author} {\bibfnamefont {P.}~\bibnamefont
			{Dotko}}, \bibinfo {author} {\bibfnamefont {K.}~\bibnamefont {Hess}},
		\bibinfo {author} {\bibfnamefont {R.}~\bibnamefont {Levi}}, \bibinfo {author}
		{\bibfnamefont {M.}~\bibnamefont {Nolte}}, \bibinfo {author} {\bibfnamefont
			{M.}~\bibnamefont {Reimann}}, \bibinfo {author} {\bibfnamefont
			{M.}~\bibnamefont {Scolamiero}}, \bibinfo {author} {\bibfnamefont
			{K.}~\bibnamefont {Turner}}, \bibinfo {author} {\bibfnamefont
			{E.}~\bibnamefont {Muller}}, \ and\ \bibinfo {author} {\bibfnamefont
			{H.}~\bibnamefont {Markram}},\ }\href {\doibase 10.3389/fncom.2017.00048}
	{\bibfield  {journal} {\bibinfo  {journal} {Frontiers in Computational
				Neuroscience}\ }\textbf {\bibinfo {volume} {11}},\ \bibinfo {pages} {48}
		(\bibinfo {year} {2016}{\natexlab{a}})},\ \Eprint
	{http://arxiv.org/abs/1601.01580} {arXiv:1601.01580} \BibitemShut {NoStop}%
	\bibitem [{\citenamefont {Petri}\ \emph {et~al.}(2013)\citenamefont {Petri},
		\citenamefont {Scolamiero}, \citenamefont {Donato}, \citenamefont
		{Vaccarino},\ and\ \citenamefont {Lambiotte}}]{Petri2013}%
	\BibitemOpen
	\bibfield  {author} {\bibinfo {author} {\bibfnamefont {G.}~\bibnamefont
			{Petri}}, \bibinfo {author} {\bibfnamefont {M.}~\bibnamefont {Scolamiero}},
		\bibinfo {author} {\bibfnamefont {I.}~\bibnamefont {Donato}}, \bibinfo
		{author} {\bibfnamefont {F.}~\bibnamefont {Vaccarino}}, \ and\ \bibinfo
		{author} {\bibfnamefont {R.}~\bibnamefont {Lambiotte}},\ }\href {\doibase
		10.1371/} {\bibfield  {journal} {\bibinfo  {journal} {PLoS ONE}\ }\textbf
		{\bibinfo {volume} {8}} (\bibinfo {year} {2013}),\ 10.1371/}\BibitemShut
	{NoStop}%
	\bibitem [{\citenamefont {Sizemore}\ \emph
		{et~al.}(2017{\natexlab{b}})\citenamefont {Sizemore}, \citenamefont
		{Giusti},\ and\ \citenamefont {Bassett}}]{Sizemore2017}%
	\BibitemOpen
	\bibfield  {author} {\bibinfo {author} {\bibfnamefont {A.}~\bibnamefont
			{Sizemore}}, \bibinfo {author} {\bibfnamefont {C.}~\bibnamefont {Giusti}}, \
		and\ \bibinfo {author} {\bibfnamefont {D.~S.}\ \bibnamefont {Bassett}},\
	}\href {\doibase 10.1093/comnet/cnw013} {\bibfield  {journal} {\bibinfo
			{journal} {Journal of Complex Networks}\ }\textbf {\bibinfo {volume} {5}},\
		\bibinfo {pages} {245} (\bibinfo {year} {2017}{\natexlab{b}})},\ \Eprint
	{http://arxiv.org/abs/1512.06457} {arXiv:1512.06457} \BibitemShut {NoStop}%
	\bibitem [{\citenamefont {Horak}\ \emph {et~al.}(2009)\citenamefont {Horak},
		\citenamefont {Maleti},\ and\ \citenamefont {Rajkovi}}]{Horak2009}%
	\BibitemOpen
	\bibfield  {author} {\bibinfo {author} {\bibfnamefont {D.}~\bibnamefont
			{Horak}}, \bibinfo {author} {\bibfnamefont {S.}~\bibnamefont {Maleti}}, \
		and\ \bibinfo {author} {\bibfnamefont {M.}~\bibnamefont {Rajkovi}},\ }\href
	{\doibase 10.1088/1742-5468/2009/03/P03034} {\bibfield  {journal} {\bibinfo
			{journal} {Stat. Mech}\ } (\bibinfo {year} {2009}),\
		10.1088/1742-5468/2009/03/P03034}\BibitemShut {NoStop}%
	\bibitem [{\citenamefont {Bobrowski}\ and\ \citenamefont
		{Kahle}()}]{Bobrowski}%
	\BibitemOpen
	\bibfield  {author} {\bibinfo {author} {\bibfnamefont {O.}~\bibnamefont
			{Bobrowski}}\ and\ \bibinfo {author} {\bibfnamefont {M.}~\bibnamefont
			{Kahle}},\ }\href {\doibase 10.1007/s41468-017-0010-0} {\bibfield  {journal}
		{\bibinfo  {journal} {Journal of Applied and Computational Topology}\
		}\textbf {\bibinfo {volume} {1}},\ 10.1007/s41468-017-0010-0}\BibitemShut
	{NoStop}%
	\bibitem [{\citenamefont {Dotko}\ \emph
		{et~al.}(2016{\natexlab{b}})\citenamefont {Dotko}, \citenamefont {Hess},
		\citenamefont {Levi}, \citenamefont {Nolte}, \citenamefont {Reimann},
		\citenamefont {Scolamiero}, \citenamefont {Turner}, \citenamefont {Muller},\
		and\ \citenamefont {Markram}}]{Dotko2016a}%
	\BibitemOpen
	\bibfield  {author} {\bibinfo {author} {\bibfnamefont {P.}~\bibnamefont
			{Dotko}}, \bibinfo {author} {\bibfnamefont {K.}~\bibnamefont {Hess}},
		\bibinfo {author} {\bibfnamefont {R.}~\bibnamefont {Levi}}, \bibinfo {author}
		{\bibfnamefont {M.}~\bibnamefont {Nolte}}, \bibinfo {author} {\bibfnamefont
			{M.}~\bibnamefont {Reimann}}, \bibinfo {author} {\bibfnamefont
			{M.}~\bibnamefont {Scolamiero}}, \bibinfo {author} {\bibfnamefont
			{K.}~\bibnamefont {Turner}}, \bibinfo {author} {\bibfnamefont
			{E.}~\bibnamefont {Muller}}, \ and\ \bibinfo {author} {\bibfnamefont
			{H.}~\bibnamefont {Markram}},\ }\href {\doibase 10.3389/fncom.2017.00048}
	{\bibfield  {journal} {\bibinfo  {journal} {Frontiers in Computational
				Neuroscience}\ }\textbf {\bibinfo {volume} {1}},\ \bibinfo {pages} {1}
		(\bibinfo {year} {2016}{\natexlab{b}})},\ \Eprint
	{http://arxiv.org/abs/1601.01580} {1601.01580} \BibitemShut {NoStop}%
	\bibitem [{\citenamefont {Schiff}(2011)}]{Schiff2011}%
	\BibitemOpen
	\bibfield  {author} {\bibinfo {author} {\bibfnamefont {S.~J.}\ \bibnamefont
			{Schiff}},\ }\href@noop {} {\emph {\bibinfo {title} {{Neural Control
					Enginerring. The emerging intersection between control theory and
					neuroscience}}}}\ (\bibinfo  {publisher} {MIT Press},\ \bibinfo {year}
	{2011})\BibitemShut {NoStop}%
	\bibitem [{\citenamefont {Liu}\ \emph {et~al.}(2011)\citenamefont {Liu},
		\citenamefont {Slotine},\ and\ \citenamefont {Barab{\'{a}}si}}]{Liu2011}%
	\BibitemOpen
	\bibfield  {author} {\bibinfo {author} {\bibfnamefont {Y.~Y.}\ \bibnamefont
			{Liu}}, \bibinfo {author} {\bibfnamefont {J.~J.}\ \bibnamefont {Slotine}}, \
		and\ \bibinfo {author} {\bibfnamefont {A.~L.}\ \bibnamefont
			{Barab{\'{a}}si}},\ }\href {\doibase 10.1038/nature10011} {\bibfield
		{journal} {\bibinfo  {journal} {Nature}\ }\textbf {\bibinfo {volume} {473}},\
		\bibinfo {pages} {167} (\bibinfo {year} {2011})},\ \Eprint
	{http://arxiv.org/abs//www.nature.com/nature/journal/v473/n7346/abs/10.1038-nature10011-unlocked.html{\#}supplementary-information}
	{arXiv:/www.nature.com/nature/journal/v473/n7346/abs/10.1038-nature10011-unlocked.html{\#}supplementary-information
		[http:]} \BibitemShut {NoStop}%
	\bibitem [{\citenamefont {Bryson}(1996)}]{Bryson1996}%
	\BibitemOpen
	\bibfield  {author} {\bibinfo {author} {\bibfnamefont {A.~E.}\ \bibnamefont
			{Bryson}},\ }\href {\doibase 10.1109/37.506395} {\bibfield  {journal}
		{\bibinfo  {journal} {IEEE Control Systems}\ }\textbf {\bibinfo {volume}
			{16}},\ \bibinfo {pages} {26} (\bibinfo {year} {1996})}\BibitemShut {NoStop}%
	\bibitem [{\citenamefont {Yan}\ \emph {et~al.}(2017)\citenamefont {Yan},
		\citenamefont {Vertes}, \citenamefont {Towlson}, \citenamefont {Chew},
		\citenamefont {Walker}, \citenamefont {Schafer},\ and\ \citenamefont
		{Barabasi}}]{yan2017network}%
	\BibitemOpen
	\bibfield  {author} {\bibinfo {author} {\bibfnamefont {G.}~\bibnamefont
			{Yan}}, \bibinfo {author} {\bibfnamefont {P.~E.}\ \bibnamefont {Vertes}},
		\bibinfo {author} {\bibfnamefont {E.~K.}\ \bibnamefont {Towlson}}, \bibinfo
		{author} {\bibfnamefont {Y.~L.}\ \bibnamefont {Chew}}, \bibinfo {author}
		{\bibfnamefont {D.~S.}\ \bibnamefont {Walker}}, \bibinfo {author}
		{\bibfnamefont {W.~R.}\ \bibnamefont {Schafer}}, \ and\ \bibinfo {author}
		{\bibfnamefont {A.~L.}\ \bibnamefont {Barabasi}},\ }\href@noop {} {\bibfield
		{journal} {\bibinfo  {journal} {Nature}\ }\textbf {\bibinfo {volume} {550}},\
		\bibinfo {pages} {519} (\bibinfo {year} {2017})}\BibitemShut {NoStop}%
	\bibitem [{\citenamefont {Gu}\ \emph {et~al.}(2017)\citenamefont {Gu},
		\citenamefont {Betzel}, \citenamefont {Mattar}, \citenamefont {Cieslak},
		\citenamefont {Delio}, \citenamefont {Grafton}, \citenamefont {Pasqualetti},\
		and\ \citenamefont {Bassett}}]{gu2017optimal}%
	\BibitemOpen
	\bibfield  {author} {\bibinfo {author} {\bibfnamefont {S.}~\bibnamefont
			{Gu}}, \bibinfo {author} {\bibfnamefont {R.~F.}\ \bibnamefont {Betzel}},
		\bibinfo {author} {\bibfnamefont {M.~G.}\ \bibnamefont {Mattar}}, \bibinfo
		{author} {\bibfnamefont {M.}~\bibnamefont {Cieslak}}, \bibinfo {author}
		{\bibfnamefont {P.~R.}\ \bibnamefont {Delio}}, \bibinfo {author}
		{\bibfnamefont {S.~T.}\ \bibnamefont {Grafton}}, \bibinfo {author}
		{\bibfnamefont {F.}~\bibnamefont {Pasqualetti}}, \ and\ \bibinfo {author}
		{\bibfnamefont {D.~S.}\ \bibnamefont {Bassett}},\ }\href@noop {} {\bibfield
		{journal} {\bibinfo  {journal} {Neuroimage}\ }\textbf {\bibinfo {volume}
			{148}},\ \bibinfo {pages} {305} (\bibinfo {year} {2017})}\BibitemShut
	{NoStop}%
	\bibitem [{\citenamefont {Betzel}\ \emph
		{et~al.}(2016{\natexlab{c}})\citenamefont {Betzel}, \citenamefont {Gu},
		\citenamefont {Medaglia}, \citenamefont {Pasqualetti},\ and\ \citenamefont
		{Bassett}}]{betzel2016optimally}%
	\BibitemOpen
	\bibfield  {author} {\bibinfo {author} {\bibfnamefont {R.~F.}\ \bibnamefont
			{Betzel}}, \bibinfo {author} {\bibfnamefont {S.}~\bibnamefont {Gu}}, \bibinfo
		{author} {\bibfnamefont {J.~D.}\ \bibnamefont {Medaglia}}, \bibinfo {author}
		{\bibfnamefont {F.}~\bibnamefont {Pasqualetti}}, \ and\ \bibinfo {author}
		{\bibfnamefont {D.~S.}\ \bibnamefont {Bassett}},\ }\href@noop {} {\bibfield
		{journal} {\bibinfo  {journal} {Sci Rep}\ }\textbf {\bibinfo {volume} {6}},\
		\bibinfo {pages} {30770} (\bibinfo {year} {2016}{\natexlab{c}})}\BibitemShut
	{NoStop}%
	\bibitem [{\citenamefont {Pasqualetti}\ \emph {et~al.}(2014)\citenamefont
		{Pasqualetti}, \citenamefont {Zampieri},\ and\ \citenamefont
		{Bullo}}]{pasqualetti2014controllability}%
	\BibitemOpen
	\bibfield  {author} {\bibinfo {author} {\bibfnamefont {F.}~\bibnamefont
			{Pasqualetti}}, \bibinfo {author} {\bibfnamefont {S.}~\bibnamefont
			{Zampieri}}, \ and\ \bibinfo {author} {\bibfnamefont {F.}~\bibnamefont
			{Bullo}},\ }\href@noop {} {\bibfield  {journal} {\bibinfo  {journal} {IEEE
				Transactions on Control of Network Systems}\ }\textbf {\bibinfo {volume}
			{1}},\ \bibinfo {pages} {40} (\bibinfo {year} {2014})}\BibitemShut {NoStop}%
	\bibitem [{\citenamefont {Jeganathan}\ \emph {et~al.}(2018)\citenamefont
		{Jeganathan}, \citenamefont {Perry}, \citenamefont {Bassett}, \citenamefont
		{Roberts}, \citenamefont {Mitchell},\ and\ \citenamefont
		{Breakspear}}]{jeganathan2018fronto}%
	\BibitemOpen
	\bibfield  {author} {\bibinfo {author} {\bibfnamefont {J.}~\bibnamefont
			{Jeganathan}}, \bibinfo {author} {\bibfnamefont {A.}~\bibnamefont {Perry}},
		\bibinfo {author} {\bibfnamefont {D.~S.}\ \bibnamefont {Bassett}}, \bibinfo
		{author} {\bibfnamefont {G.}~\bibnamefont {Roberts}}, \bibinfo {author}
		{\bibfnamefont {P.~B.}\ \bibnamefont {Mitchell}}, \ and\ \bibinfo {author}
		{\bibfnamefont {M.}~\bibnamefont {Breakspear}},\ }\href@noop {} {\bibfield
		{journal} {\bibinfo  {journal} {NeuroImage Clinical}\ }\textbf {\bibinfo
			{volume} {In Press}} (\bibinfo {year} {2018})}\BibitemShut {NoStop}%
	\bibitem [{\citenamefont {Gu}\ \emph {et~al.}(2015)\citenamefont {Gu},
		\citenamefont {Pasqualetti}, \citenamefont {Cieslak}, \citenamefont
		{Telesford}, \citenamefont {Yu}, \citenamefont {Kahn}, \citenamefont
		{Medaglia}, \citenamefont {Vettel}, \citenamefont {Miller}, \citenamefont
		{Grafton},\ and\ \citenamefont {Bassett}}]{Gu2014}%
	\BibitemOpen
	\bibfield  {author} {\bibinfo {author} {\bibfnamefont {S.}~\bibnamefont
			{Gu}}, \bibinfo {author} {\bibfnamefont {F.}~\bibnamefont {Pasqualetti}},
		\bibinfo {author} {\bibfnamefont {M.}~\bibnamefont {Cieslak}}, \bibinfo
		{author} {\bibfnamefont {Q.~K.}\ \bibnamefont {Telesford}}, \bibinfo {author}
		{\bibfnamefont {A.~B.}\ \bibnamefont {Yu}}, \bibinfo {author} {\bibfnamefont
			{A.~E.}\ \bibnamefont {Kahn}}, \bibinfo {author} {\bibfnamefont {J.~D.}\
			\bibnamefont {Medaglia}}, \bibinfo {author} {\bibfnamefont {J.~M.}\
			\bibnamefont {Vettel}}, \bibinfo {author} {\bibfnamefont {M.~B.}\
			\bibnamefont {Miller}}, \bibinfo {author} {\bibfnamefont {S.~T.}\
			\bibnamefont {Grafton}}, \ and\ \bibinfo {author} {\bibfnamefont {D.~S.}\
			\bibnamefont {Bassett}},\ }\href {\doibase 10.1038/ncomms9414} {\bibfield
		{journal} {\bibinfo  {journal} {Nature Communications}\ }\textbf {\bibinfo
			{volume} {6}},\ \bibinfo {pages} {1} (\bibinfo {year} {2015})},\ \Eprint
	{http://arxiv.org/abs/1406.5197} {arXiv:1406.5197} \BibitemShut {NoStop}%
	\bibitem [{\citenamefont {Tang}\ \emph {et~al.}(2017)\citenamefont {Tang},
		\citenamefont {Giusti}, \citenamefont {Baum}, \citenamefont {Gu},
		\citenamefont {Pollock}, \citenamefont {Kahn}, \citenamefont {Roalf},
		\citenamefont {Moore}, \citenamefont {Ruparel}, \citenamefont {Gur},
		\citenamefont {Gur}, \citenamefont {Satterthwaite},\ and\ \citenamefont
		{Bassett}}]{Tang}%
	\BibitemOpen
	\bibfield  {author} {\bibinfo {author} {\bibfnamefont {E.}~\bibnamefont
			{Tang}}, \bibinfo {author} {\bibfnamefont {C.}~\bibnamefont {Giusti}},
		\bibinfo {author} {\bibfnamefont {G.~L.}\ \bibnamefont {Baum}}, \bibinfo
		{author} {\bibfnamefont {S.}~\bibnamefont {Gu}}, \bibinfo {author}
		{\bibfnamefont {E.}~\bibnamefont {Pollock}}, \bibinfo {author} {\bibfnamefont
			{A.~E.}\ \bibnamefont {Kahn}}, \bibinfo {author} {\bibfnamefont {D.~R.}\
			\bibnamefont {Roalf}}, \bibinfo {author} {\bibfnamefont {T.~M.}\ \bibnamefont
			{Moore}}, \bibinfo {author} {\bibfnamefont {K.}~\bibnamefont {Ruparel}},
		\bibinfo {author} {\bibfnamefont {R.~C.}\ \bibnamefont {Gur}}, \bibinfo
		{author} {\bibfnamefont {R.~E.}\ \bibnamefont {Gur}}, \bibinfo {author}
		{\bibfnamefont {T.~D.}\ \bibnamefont {Satterthwaite}}, \ and\ \bibinfo
		{author} {\bibfnamefont {D.~S.}\ \bibnamefont {Bassett}},\ }\href {\doibase
		10.1038/s41467-017-01254-4} {\bibfield  {journal} {\bibinfo  {journal}
			{Nature Communications}\ }\textbf {\bibinfo {volume} {8}} (\bibinfo {year}
		{2017}),\ 10.1038/s41467-017-01254-4},\ \Eprint
	{http://arxiv.org/abs/1607.01010} {arXiv:1607.01010} \BibitemShut {NoStop}%
	\bibitem [{\citenamefont {Wu-Yan}\ \emph {et~al.}(2017)\citenamefont {Wu-Yan},
		\citenamefont {Betzel}, \citenamefont {Tang}, \citenamefont {Gu},
		\citenamefont {Pasqualetti},\ and\ \citenamefont {Bassett}}]{Wu-Yan2017}%
	\BibitemOpen
	\bibfield  {author} {\bibinfo {author} {\bibfnamefont {E.}~\bibnamefont
			{Wu-Yan}}, \bibinfo {author} {\bibfnamefont {R.~F.}\ \bibnamefont {Betzel}},
		\bibinfo {author} {\bibfnamefont {E.}~\bibnamefont {Tang}}, \bibinfo {author}
		{\bibfnamefont {S.}~\bibnamefont {Gu}}, \bibinfo {author} {\bibfnamefont
			{F.}~\bibnamefont {Pasqualetti}}, \ and\ \bibinfo {author} {\bibfnamefont
			{D.~S.}\ \bibnamefont {Bassett}},\ }\href {http://arxiv.org/abs/1706.05117}
	{\bibfield  {journal} {\bibinfo  {journal} {Journal of Nonlinear Science}\ }
		(\bibinfo {year} {2017})},\ \Eprint {http://arxiv.org/abs/1706.05117}
	{arXiv:1706.05117} \BibitemShut {NoStop}%
	\bibitem [{\citenamefont {Cornblath}\ \emph {et~al.}(2018)\citenamefont
		{Cornblath}, \citenamefont {Tang}, \citenamefont {Baum}, \citenamefont
		{Moore}, \citenamefont {Roalf}, \citenamefont {Gur}, \citenamefont {Gur},
		\citenamefont {Pasqualetti}, \citenamefont {Satterthwaite},\ and\
		\citenamefont {Bassett}}]{cornblath2018sex}%
	\BibitemOpen
	\bibfield  {author} {\bibinfo {author} {\bibfnamefont {E.~J.}\ \bibnamefont
			{Cornblath}}, \bibinfo {author} {\bibfnamefont {E.}~\bibnamefont {Tang}},
		\bibinfo {author} {\bibfnamefont {G.~L.}\ \bibnamefont {Baum}}, \bibinfo
		{author} {\bibfnamefont {T.~M.}\ \bibnamefont {Moore}}, \bibinfo {author}
		{\bibfnamefont {D.~R.}\ \bibnamefont {Roalf}}, \bibinfo {author}
		{\bibfnamefont {R.~C.}\ \bibnamefont {Gur}}, \bibinfo {author} {\bibfnamefont
			{R.~E.}\ \bibnamefont {Gur}}, \bibinfo {author} {\bibfnamefont
			{F.}~\bibnamefont {Pasqualetti}}, \bibinfo {author} {\bibfnamefont {T.~D.}\
			\bibnamefont {Satterthwaite}}, \ and\ \bibinfo {author} {\bibfnamefont
			{D.~S.}\ \bibnamefont {Bassett}},\ }\href@noop {} {\bibfield  {journal}
		{\bibinfo  {journal} {arXiv}\ }\textbf {\bibinfo {volume} {1801}},\ \bibinfo
		{pages} {04623} (\bibinfo {year} {2018})}\BibitemShut {NoStop}%
	\bibitem [{\citenamefont {Menara}\ \emph {et~al.}(2018)\citenamefont {Menara},
		\citenamefont {Katewa}, \citenamefont {Bassett},\ and\ \citenamefont
		{Pasqualetti}}]{Menara2018}%
	\BibitemOpen
	\bibfield  {author} {\bibinfo {author} {\bibfnamefont {T.}~\bibnamefont
			{Menara}}, \bibinfo {author} {\bibfnamefont {V.}~\bibnamefont {Katewa}},
		\bibinfo {author} {\bibfnamefont {D.~S.}\ \bibnamefont {Bassett}}, \ and\
		\bibinfo {author} {\bibfnamefont {F.}~\bibnamefont {Pasqualetti}},\
	}\href@noop {} {\bibfield  {journal} {\bibinfo  {journal} {arXiv}\ }\textbf
		{\bibinfo {volume} {1706}},\ \bibinfo {pages} {05120} (\bibinfo {year}
		{2018})}\BibitemShut {NoStop}%
	\bibitem [{\citenamefont {Kim}\ \emph {et~al.}(2018)\citenamefont {Kim},
		\citenamefont {Soffer}, \citenamefont {Kahn}, \citenamefont {Vettel},
		\citenamefont {Pasqualetti},\ and\ \citenamefont {Bassett}}]{Kim2017}%
	\BibitemOpen
	\bibfield  {author} {\bibinfo {author} {\bibfnamefont {J.~Z.}\ \bibnamefont
			{Kim}}, \bibinfo {author} {\bibfnamefont {J.~M.}\ \bibnamefont {Soffer}},
		\bibinfo {author} {\bibfnamefont {A.~E.}\ \bibnamefont {Kahn}}, \bibinfo
		{author} {\bibfnamefont {J.~M.}\ \bibnamefont {Vettel}}, \bibinfo {author}
		{\bibfnamefont {F.}~\bibnamefont {Pasqualetti}}, \ and\ \bibinfo {author}
		{\bibfnamefont {D.~S.}\ \bibnamefont {Bassett}},\ }\href@noop {} {\bibfield
		{journal} {\bibinfo  {journal} {Nat Phys}\ }\textbf {\bibinfo {volume}
			{14}},\ \bibinfo {pages} {91} (\bibinfo {year} {2018})}\BibitemShut {NoStop}%
	\bibitem [{\citenamefont {Taylor}\ \emph {et~al.}(2015)\citenamefont {Taylor},
		\citenamefont {Thomas}, \citenamefont {Sinha}, \citenamefont {Dauwels},
		\citenamefont {Kaiser}, \citenamefont {Thesen},\ and\ \citenamefont
		{Ruths}}]{taylor2015optimal}%
	\BibitemOpen
	\bibfield  {author} {\bibinfo {author} {\bibfnamefont {P.~N.}\ \bibnamefont
			{Taylor}}, \bibinfo {author} {\bibfnamefont {J.}~\bibnamefont {Thomas}},
		\bibinfo {author} {\bibfnamefont {N.}~\bibnamefont {Sinha}}, \bibinfo
		{author} {\bibfnamefont {J.}~\bibnamefont {Dauwels}}, \bibinfo {author}
		{\bibfnamefont {M.}~\bibnamefont {Kaiser}}, \bibinfo {author} {\bibfnamefont
			{T.}~\bibnamefont {Thesen}}, \ and\ \bibinfo {author} {\bibfnamefont
			{J.}~\bibnamefont {Ruths}},\ }\href@noop {} {\bibfield  {journal} {\bibinfo
			{journal} {Front Neurosci}\ }\textbf {\bibinfo {volume} {9}},\ \bibinfo
		{pages} {202} (\bibinfo {year} {2015})}\BibitemShut {NoStop}%
\end{thebibliography}
\end{document}